\documentclass[12pt]{article}
\usepackage{rotating}
\usepackage{cleveref}
\usepackage{multirow}
\usepackage{amsmath,amssymb,graphicx,verbatim} 
\usepackage{cite}
\usepackage{latexsym} 
\usepackage{cancel}
\usepackage{hyperref}
\usepackage{enumerate} 
\usepackage[normalem]{ulem} 
\usepackage{array}
\graphicspath{{figs/}}
\usepackage{soul}
\usepackage{slashed}
\usepackage{caption}
\usepackage{subcaption}
\usepackage{longtable}

\pdfoutput=1
\usepackage{epsf}
\usepackage{pstricks} 
\usepackage{comment}

\newcommand{\centeron}[2]{{\setbox0=\hbox{#1}\setbox1=\hbox{#2}\ifdim
\wd1>\wd0\kern.5\wd1\kern-.5\wd0\fi \copy0
\kern-.5\wd0\kern-.5\wd1\copy1\ifdim\wd0>\wd1
                                   \kern.5\wd0\kern-.5\wd1\fi}}
\newcommand{\ltap}{\>\centeron{\raise.35ex\hbox{$<$}}
                           {\lower.65ex\hbox{$\sim$}}\>}
\newcommand{\gtap}{\>\centeron{\raise.35ex\hbox{$>$}}
                           {\lower.65ex\hbox{$\sim$}}\>}

\newcommand\ZZ{\hbox{\zfont Z\kern-.4emZ}}
\font\zfont = cmss10 

\newcommand{\figref}[1]{Fig.~\ref{fig:#1}}
\renewcommand{\eqref}[1]{Eq.~(\ref{eq:#1})}

\newcommand{\sref}[1]{Section \ref{s.#1}}

\newcommand{\tabref}[1]{Table~\ref{tab:#1}}

\newcommand{\ba}{\begin{array}}
\newcommand{\ea}{\end{array}}
\newcommand{\bpm}{\begin{pmatrix}}
\newcommand{\epm}{\end{pmatrix}}

\newcommand{\beq}{\begin{equation}}
\newcommand{\eeq}{\end{equation}}
\newcommand{\bit}{\begin{itemize}}
\newcommand{\eit}{\end{itemize}}
\newcommand{\ben}{\begin{enumerate}}
\newcommand{\een}{\end{enumerate}}
\newcommand{\bal}{\begin{align}} 
\newcommand{\eal}{\end{align}}
\newcommand{\f}{\frac}
\renewcommand{\t}{\tilde}

\def\bi{\begin{itemize}}
\def\ei{\end{itemize}}
\def\ben{\begin{enumerate}}
\def\een{\end{enumerate}}
\def\bc{\begin{center}}
\def\ec{\end{center}}
\def\bt{\begin{table}}
\def\et{\end{table}}
\def\btb{\begin{tabular}}
\def\etb{\end{tabular}}

\def\gev{\, {\rm GeV}}

\def\mass2{mass${}^2$}

\begin{document}
\begin{titlepage}
\begin{flushright}
\small{YITP-SB-17-25}
\end{flushright}

\vskip2.5cm
\begin{center}
\vspace*{5mm}
{\huge \bf Higgs-Precision Constraints on Colored Naturalness}
\end{center}
\vskip0.2cm

\begin{center}
{Rouven Essig$^{\,a}$, Patrick Meade$^{\,a}$, Harikrishnan Ramani$^{\,a}$, Yi-Ming Zhong$^{\,b}$}
\end{center}
\vskip 8pt

\begin{center}
{\it $^{a}$C. N. Yang Institute for Theoretical Physics\\ Stony Brook University, Stony Brook, NY 11794}\\
\vspace*{0.3cm}
{\it $^{b}$Physics Department\\ Boston University, Boston, MA 02215}\\
\vspace*{0.3cm}

\vspace*{0.1cm}

{\tt rouven.essig@stonybrook.edu, meade@insti.physics.sunysb.edu, hramani@insti.physics.sunysb.edu, ymzhong@bu.edu}
\end{center}

\vglue 0.3truecm

\begin{abstract}

The presence of weak-scale colored top partners 
is among the simplest solutions to the Higgs hierarchy problem and allows for 
a natural electroweak scale.  
We examine the constraints on generic colored top partners coming solely from their effect on the production and decay rates of the 
observed Higgs with a mass of 125~GeV.  
We use the latest Higgs precision data from the Tevatron and the LHC as of EPS 2017 to derive the current limits 
on spin-0, spin-1/2, and spin-1 colored top partners. 
We also investigate the expected sensitivity 
from the Run 3 and Run 4 of the LHC, as well from 
possible future electron-positron and proton-proton colliders, including the ILC, CEPC, FCC-ee, and FCC-hh.    
We discuss constraints on top partners in the Minimal Supersymmetric Standard Model and Little Higgs theories.  
We also consider various model-building aspects -- multiple top partners, modified couplings between the Higgs and Standard-Model particles, 
and non-Standard-Model Higgs sectors -- and evaluate how these weaken the current limits and expected sensitivities. 
By modifying other Standard-Model Higgs couplings, we find that the best way to hide low-mass top partners from current data is through modifications of the top-Yukawa coupling, 
although future measurements of top-quark-pair production in association with a Higgs 
will extensively probe this possibility.  
We also demonstrate that models with multiple top partners can generically avoid current and future Higgs precision measurements.  
Nevertheless, some of the model parameter space can be probed with precision measurements at future electron-positron colliders 
of, for example, the $e^+e^-\to Zh$ cross section.
\end{abstract}

\end{titlepage}

\tableofcontents

\section{Introduction}
\label{s.intro} \setcounter{equation}{0} \setcounter{footnote}{0}

The discovery of a Higgs-like particle at Run~I of the LHC~\cite{Aad:2012tfa,Chatrchyan:2012xdj} without any accompanying Beyond 
the Standard Model (BSM) particles has brought into sharp focus the hierarchy problem and the naturalness of the Weak scale. 
A minimal ingredient of any natural model is a mechanism to soften the quadratic divergences of the Higgs mass-squared parameter, $m_h^2$, that 
appear when computing quantum corrections in the Standard Model (SM).   
Since models of naturalness characteristically predict additional particles near the Weak scale, an urgent question 
is how these particles could have escaped detection at the LHC.  
Various possibilities exist, ranging from particle spectra that are hidden from direct searches to models 
of ``neutral" naturalness~\cite{Chacko:2005pe,Burdman:2006tz}.  

Currently, two symmetry mechanisms are known that can account for a light Higgs naturally and satisfy experimental constraints --- supersymmetry (SUSY) and having the Higgs arise as a pseudo-Nambu-Goldstone boson (PNGB) of a larger global symmetry.\footnote{
Conformal symmetry could yield a light scalar~\cite{Bardeen:1995kv}, but we will not consider this further, since no complete model exists.
} 
Models of naturalness contain new BSM ``partner'' states, which are related to the SM particles by these symmetries.   
While the partner states cancel the quadratic divergence of $m_h^2$, the Higgs mass is now quadratically dependent on the mass scale of the 
partner states.  
A natural theory thus requires that these states be near the Weak scale.  
Moreover, the largest contribution to $m_h^2$ in the SM comes from the top quark, $t$, since it has the largest coupling to the Higgs.  
This implies that the largest correction among partner particles comes from the partner of the top quark, motivating searches for top partners. 

The partner particles affect the rates for producing the Higgs boson at colliders as well as the decay rates of Higgs bosons to SM states.  
Precision measurements of Higgs properties can thus constrain the partner particles.  
In this paper, we focus solely on the impact of top-partners on Higgs precision physics.  To carry out this program we have to choose first whether the 
symmetry that relates the top partners to the top itself commutes with $SU(3)$-color of the SM.  In addition, the spin of the top partner dictates the 
symmetry structure that is needed to cancel the quadratic divergences.  In this paper, we choose to focus on colored top partners of spin-0, 1/2, and 1, which represents all models of naturalness other than those that fall under the rubric of ``neutral naturalness".   This is a sensible choice, since the top is colored and color does not play a priori a central role in naturalness. Moreover, current Higgs precision measurements, which are the focus of this paper, 
are not the best probe of ``neutral naturalness"~\cite{Craig:2013xia,Curtin:2015bka}. 

The generic predictions of colored top partners from naturalness have been studied in many contexts. In particular, both 
indirect searches and more model-dependent direct searches for colored top partners have been proposed and undertaken at the LHC.     
Direct searches can offer a powerful way to search for particular top partners, but specific partner-mass 
spectra~\cite{Fan:2011yu, Curtin:2014zua} or additional BSM physics can easily hide the top-partner signal from specific 
searches without affecting the ``naturalness'' of the model.  
For example, in the context of SUSY, direct production bounds on the colored top partners (``stops'') can be as high as 
$m_{\tilde t} \gtrsim \mathcal{O}(800)$~GeV~\cite{ATLAS:2016jaa, ATLAS:2016kts,
CMS:2016mwj, CMS:2016xva,CMS-PAS-SUS-16-016}, but assume that the stops decay to energetic SM final states and missing transverse energy. 
Instead, if the stop decays to a top quark and neutralino, $\t \chi^0_1$, with $m_{\t t}-m_{\t \chi^0_1}\approx m_t$, the searches become much more challenging. In this ``compressed'' region, the top quark and neutralino are collinear in the lab frame; the stop-pair production will thus be similar to the top-pair production, and no significant missing transverse energy is observed~\cite{Han:2012fw, Hagiwara:2013tva, An:2015uwa, Cheng:2016mcw}.   
Similar difficult regions emerge when $m_{\t t}-m_{\t \chi^0_1}\approx m_W$ or $m_{\t t}-m_{\t \chi^0_1}\approx 0$.   Direct searches can be complicated further by additions to the Minimal Supersymmetric Standard Model (MSSM) such as Stealth SUSY~\cite{Fan:2011yu,Fan:2012jf}.  As such, although direct searches are powerful there still is room for colored top partners to evade these searches, which argues for employing also alternative search methods. 

Indirect searches for top partners from Higgs precision measurements have been carried out, for example, in the context of SUSY, 
see e.g.~\cite{Arvanitaki:2011ck,Carmi:2012yp,Blum:2012ii, Farina:2013ssa,Fan:2014txa,Fan:2014axa, Englert:2014uua,Henning:2014gca,Bae:2015nva} and references therein. 
Since the stops couple to the Higgs, they can significantly affect the loop-induced Higgs-gluon-gluon ($hgg$) and 
Higgs-photon-photon ($h\gamma\gamma$) couplings.   
Higgs precision data can thus constrain low-mass stops independent of their production or decay modes, although heavier stops remain 
unconstrained as they decouple from the $hgg$ and $h\gamma\gamma$ loops as~$\mathcal{O}(1/m_{\tilde{t}}^2)$. 
For example, the earliest Higgs data constrained models of electroweak baryogenesis in the Minimal Supersymmetric Standard Model (MSSM), 
which require a light stop (independent of naturalness)~\cite{Curtin:2012aa, Cohen:2011ap, Cohen:2012zza, Carena:2012np}. 
Direct and indirect searches are thus complementary.  

In this paper, we update the bounds on colored top partners using the latest available Higgs precision data, including data up to  
EPS 2017~\cite{EPS2017}.  
We also provide projections for expected future data from the LHC Run~3 (300 fb$^{-1}$) and LHC Run~4 (3 ab$^{-1}$), as well as from possible future 
colliders, namely the Future Circular Collider (FCC-ee and FCC-hh), Circular Electron Positron Collider (CEPC), and 
the International Linear Collider (ILC).
This updates and extends related previous work on spin-0 (SUSY)~\cite{Arvanitaki:2011ck,Carmi:2012yp,Blum:2012ii, Farina:2013ssa,Fan:2014txa,Fan:2014axa, Englert:2014uua,Henning:2014gca,Bae:2015nva}, spin-1/2~\cite{Reuter:2012sd, Berger:2012ec, Yang:2014mba, Han:2014qia}, and 
spin-1~\cite{Collins:2014pba}.

We also investigate the robustness of the current and projected indirect constraints by allowing for non-SM Higgs couplings, invisible 
Higgs decays, and exotic Higgs decays.  We describe which of these possibilities 
are best at hiding the effects of top partners and weakening the constraints on their masses.  
We discuss briefly how these additional deviations could be implemented in realistic models, which allows for further work that focuses on the 
most ``natural" models that explain electroweak symmetry breaking (EWSB).   
In principle there will also be bounds from direct constraints or other precision tests of colored top partners, which are often 
complementary. However, as we show, there are remarkably powerful statements that can be made through Higgs precision measurements alone. 

The rest of the paper is organized as follows.  
In Section~\ref{s.divtunelowenergy}, we review the interplay between fine-tuning and the masses of the colored top partners. In Section~\ref{s.precision}, we discuss the methodology for setting constraints on top partners, as well as how to reduce the sensitivity 
of Higgs precision data to colored top partners through modifications that affect the Higgs cross sections and decay rates. 
Section~\ref{s.data} describes the Higgs data that we use to calculate our constraints and projections, and also details our analysis method.  
Section~\ref{s.theory} discusses canonical top partner models and some extensions.  
In Section~\ref{results}, we present our results, before concluding in Section~\ref{s.conclusion}.  In Appendix~\ref{blindspot}, we review how Higgs precision measurements, only relevant at future colliders, such as the $e^+e^-\rightarrow Zh$ cross section or $pp\to hh$ production can provide complementary probes.
Appendices \ref{loopcoupling}, \ref{validity}, and \ref{datasets} contain additional information, including the detailed data sets used in our 
paper, as well as a validation cross-check of our results. 

\section{Naturalness and Higgs Couplings}
\label{s.divtunelowenergy}
To understand the hierarchy problem and possible symmetry-based solutions, we utilize the one-loop Coleman-Weinberg  (CW) potential with a hard UV cutoff, $\Lambda$, and examine the contributions to the Higgs-mass term in the potential in the mass eigenbasis.  The form of the Coleman-Weinberg potential for the Higgs is~\cite{Farina:2013ssa}
\begin{equation}
V_\text{CW}=\frac{1}{64\pi^2}\sum_i  (-1)^{F_i} N_{f,i} \left(2M_i^2 \Lambda^2+ M_i^4 \log{\frac{M_i^2}{\Lambda^2}}\right) \,,
\label{eq:VCW}
\end{equation}
where $i$ runs over all particles in the Higgs loop diagrams,  $N_{f, i}$ is the number of flavors of particle $i$, $F_i$ is the fermion number, and $M_i$ is the field-dependent mass taking the form 
\begin{equation}
 M_i^2=\mu_i^2 + a_i h^2\,,
 \label{eq:massform}
 \end{equation}
 where $a_i$ is given by the particle's effective coupling to the Higgs, and $\mu_i$ represents a possible bare mass for the particle
whose origin is not from the Higgs mechanism. The origin of the hierarchy problem comes from quadratic divergences that appear when  computing the shift in the Higgs mass at one-loop, 
\begin{equation}
\label{eq:delmh}
\delta m_h^2=\frac{d^2 V_\text{CW}}{d h^2}\simeq  \frac{1}{32\pi^2} \sum_i (-1)^{F_i} N_{f,i} \left(a_i \Lambda^2+2 \mu_i^2 a_i  \ln \frac{\Lambda^2}{M_i^2}\right).
\end{equation}
A necessary condition to ``solve" the hierarchy problem is then given by 
\begin{equation} 
 \sum_i (-1)^{F_i} N_{f,i} a_i =0\,.
 \label{eq:cancel}
 \end{equation}
This imposes certain relationships among the Higgs couplings that must be preserved by a symmetry. If the particles responsible for the cancellation are charged under the SM gauge symmetries, as in our case, this immediately has implications for Higgs precision physics, 
since their couplings to the Higgs are related to SM couplings of the Higgs through~\eqref{delmh}. 

To predict the impact of the new physics on Higgs phenomenology, we also need its overall mass scale.  This is dictated by the sub-leading terms in \eqref{delmh}, since cancellation of the quadratic divergences do not automatically eliminate $\log \Lambda$ divergences 
as well.\footnote{Note that even if one eliminates the $\log\Lambda$ divergences, which could be done with, for example, a conformal symmetry, 
there would likely be problems with achieving electroweak symmetry breaking.} 
This results in the usual logarithmic dependence on the cutoff in theories that solve the hierarchy problem, and a quadratic sensitivity to the mass of the BSM states
\begin{equation}
\delta m_h^2\simeq  \frac{1}{16\pi^2} \sum_i (-1)^{F_i} N_{f,i} \mu_i^2 a_i  \ln \frac{\Lambda^2}{M_i^2}\,.
\label{eq:delmhlog}
\end{equation}
Since the top has by far the largest SM coupling to the Higgs, the top-partner mass scale is the most critical among the BSM masses.  
For a natural theory, all $\mu_i$ for the top partners should be $\mathcal{O}(m_\text{weak})$, and masses heavier than this require tuning for successful EWSB. 

On the other hand, lowering top-partner masses as required by fine-tuning considerations increases the visibility of the partners at colliders. Top partners that share the top's color and electrical charge affect the loop-induced $hgg$ and $h\gamma\gamma$ couplings. The qualitative behavior can be immediately understood by considering the low-energy Higgs theorem~\cite{Ellis:1975ap, Shifman:1979eb} , which relates the mass of charged particles to their contribution to these couplings. For a heavy particle that receives some or all of its mass from the Higgs mechanism, the effective coupling is proportional to
\begin{equation}
\frac{v^2}{M_{\hat{t}}^2} \frac{\partial M_{\hat{t}}^2}{\partial v^2} \frac{h}{v} G^{\mu \nu}G_{\mu \nu}\,,
\end{equation}
where $v\simeq 246$~GeV is the Higgs vacuum expectation value (VEV), $M_{\hat{t}}^2$ is the appropriately evaluated mass-squared matrix 
for the top partner, and $G^{\mu\nu}$ is 
the gluon field strength; a similar equation holds for the electromagnetic field-strength. When the partial derivative is an $\mathcal{O}(1)$ constant, we can see the $\propto 1/M_{\hat{t}}^2$ dependence drives the phenomenology. Therefore in a natural theory the 
largest contributions to Higgs observables arise from colored top partners. We shall also encounter cases where, due to fortuitous cancellations, the partial derivative evaluates to a very small number. However the colored top partners still contribute to Higgs wave-function renormalization (WFR), which leads to deviations from the SM prediction for Higgs associated production.
Broadly, improving Higgs precision measurements without seeing deviations from the SM expectations results in a more fine-tuned theory, since it requires larger top-partner masses. Thus  Higgs phenomenology and naturalness are inexorably tied together for colored top partners, and it provides an important constraint independent of direct searches.

\section{Higgs Precision Constraints \& Colored Top Partners}

\label{s.precision}

In this section, we describe our formalism and strategy to constrain colored top-partner models through Higgs precision physics.  
We discuss the generic features of these models that are most constrained by current data.  
Moreover, we identify those model features that are best at hiding top partners from current Higgs precision data alone, thereby 
reducing tension with naturalness.  
Finally, we describe how upcoming data from the LHC, future proton-proton, and future precision electron-positron colliders 
will affect these model features. 

Higgs precision data can constrain BSM models mainly if these models modify the coupling of the Higgs 
to SM particles or contain new decay modes for the Higgs.  
Modifications to the Higgs couplings can affect the Higgs partial widths and production modes, while new decay modes affect only 
the partial widths.  In our attempt to be model-agnostic, we do not investigate off-shell decays of the Higgs to top partners, but we do include the contributions to the Higgs partial width of on-shell decays for $M_{\hat{t}}<m_h/2$.  This provides an important constraint at low masses.  We will also consider the possibility that new decay modes can help hide colored top partners from 
Higgs precision measurements. 

\subsection{Definitions for non-Standard Model Higgs couplings}

As we have emphasized in the introduction, a generic prediction of colored top partners is a modification of a certain set of Higgs couplings.  
Since the SM Higgs fits the data well, we parameterize modifications to the most important tree-level SM Higgs couplings as
\beq 
r_j=c_{hjj}/c_{hjj}^\text{SM}\,,
\eeq 
where $j=t$, $b$, $V$, or $\tau$, $c_{hjj}$ is the coupling of the Higgs to the state $j$, and $c_{hjj}^{\rm SM}$ is the coupling of the SM 
Higgs to the state $j$; the SM value is $r_j^\text{SM}=1$.\footnote{For simplicity, we assume that the Higgs boson couples equally 
to the $W$- and $Z$-bosons, i.e.~$r_V\equiv r_W=r_Z$, as otherwise electroweak precision tests would be far more constraining 
than Higgs precision data for the foreseeable future.}  

However, this definition is not sufficient for BSM particles, since $c_{hjj}^{\rm SM}$ is not defined. 
We present a more general definition that works for both SM and BSM particles, which is derived directly from the mass $M_j$
for a mass-eigenstate $j$ . 
The 125 GeV Higgs, $h$, can  in principle be a linear combination of Higgses $H_i$ with vacuum expectation values (VEVs), $v_i$, 
that supply some or all of 
the mass to $j$.\footnote{$h=\sum_i R_i H_i$ such that $\langle h | H_i \rangle = R_i$.}  The ratio of the $hjj$ coupling to its SM value is then given by
\begin{equation}\label{eq:rj-def}
r_j=\sum_i\langle h | H_i \rangle \frac{v}{v_i}\frac{\text{d} \log [M_j^2]}{\text{d} \log [v_i^2]}\,. 
\end{equation}
In the SM there is only a single Higgs and therefore this reduces to $r_j=1$ as desired. Note that this is similar to the ``$\kappa$ framework" in~\cite{Heinemeyer:2013tqa} and in practice they are the same except for how new contributions are included in the Higgs width.  
Here we use the precise definition given in \eqref{rj-def}.  

Apart from BSM physics affecting tree-level couplings, the loop-induced couplings $hgg$ and $h\gamma\gamma$ 
play a particularly important role in constraining colored and electrically charged top partners.  
These partners appear at the same order in perturbation theory as SM processes.  
Moreover, the $hgg$ coupling controls the dominant production mechanism in the SM, and the $h\gamma\gamma$ coupling 
controls one of the most sensitive decay channels.  
When these loop particles are heavy and can be integrated out, the effective vertex is given by the low-energy 
Higgs theorems~\cite{Ellis:1975ap,Shifman:1979eb}. 
The one-loop result for the $hgg$ coupling, including finite mass effects, is given by
\beq
\label{eq:lowmed1}
\mathcal{L} \supset-\frac{1}{4} c_G \frac{h}{v} G_{\mu\nu}^a G^{\mu\nu a}\,,
\eeq
where 
\beq
c_G = \frac{\alpha_s}{12\pi}\sum_j N_{c,j} C_2(R) r_j \mathcal{A}_j(\tau_j)\,. 
\eeq
Here $N_{c,j}$ is the number of colors, $C_2(R)$ is the quadratic Casimir, and $\mathcal{A}_j\equiv \mathcal{A}^{s_j}(\tau_j)$  are the loop 
functions defined in Appendix~\ref{loopcoupling}, which depend on the spin $s_j$ and $\tau_j\equiv m_h^2/4m_j^2$, where $m_j$ are the 
eigenvalues of $M_j$.  
For electrically charged states, a similar operator with the electromagnetic field strength replacing the gluon field strength can be 
used to calculate the Higgs coupling to two photons.  

These definitions allow us to express the modifications to the effective $hgg$ coupling from a colored top-partner, $\hat t$, 
and any accompanying BSM physics that affects the tree-level couplings $r_j$ as
\beq
r_G \equiv \frac{c_G}{c_G^{\text{SM}}} =\f{r_t \mathcal A_t+r_b \mathcal A_b+r_{\hat t} \mathcal A_{\hat t}+\delta r_G}{\mathcal A_t+\mathcal A_b}\,,
\label{eq:myrg}
\eeq
where $\delta r_G$ captures the presence of other colored BSM (non-top-partner) particles.
For modifications to the effective $h\gamma\gamma$ coupling, we have 
\beq
r_\gamma \equiv \frac{c_\gamma}{c_\gamma^{\text{SM}}} =\f{\sum_{j = W, t, b, \tau} r_j Q_j^2 \mathcal{A}_j   + r_{\hat{t}} Q_{\hat{t}}^2 \mathcal{A}_{\hat{t}}+\delta r_\gamma} {\sum_{j =  W, t, b,  \tau } Q_j^2 \mathcal{A}_j}\,,
\label{eq:myrgamma}
\eeq
where $\delta r_\gamma$ captures other (non-top-partner) particles carrying electrical charge. 

It is now useful to define a new variable $\mathcal{N}_{\hat{t}}$ such that 
\beq
\mathcal{N}_{\hat{t}} \equiv \frac{ r_{\hat t}\mathcal A_{\hat t}}{r_t \mathcal A_{t}}\,.
\label{eq:rGhatt}
\eeq
This serves also to eliminate the $\langle h | H_i \rangle$ dependence of the top-partner contribution, as it is common to both the top-quark 
and top partner. 
We can then re-express $r_G$ as
\beq
r_G =\f{r_t \mathcal A_t (1+\mathcal{N}_{\hat{t}}) +r_b \mathcal A_b+\delta r_G}{\mathcal A_t+\mathcal A_b}\,.
\label{eq:myrg2}
\eeq

Finally, changing the couplings of the Higgs to SM particles affects the partial widths and hence the total width possibly as well.  To parametrize the effects of this shift we define 
\beq\label{eq:rh}
r_h \equiv 1+\sum_{j=G, \gamma, V, b, \tau}(|r_j|^2-1)B^\text{SM}_{h\to jj}\,,
\eeq
where the SM branching ratios are given in e.g.~\cite{Higgsbr:2016}.
While this is a redundant definition, it is useful, because in addition to colored top partners there may also be new decay modes for the 
Higgs, which would either be a contribution to the invisible width of the Higgs or an exotic decay channel.  
The difference between invisible and exotic decays will occur only whether we include direct searches for invisible Higgs decays.  
In particular, it is much easier to constrain an invisible decay rather than an arbitrary exotic decay.   
We also explicitly include the partial width into the top partners, which is nonzero only for low top-partner masses.  
We do not specify how the top partners decay, so this possibility is only constrained by how it affects the Higgs branching fractions of the  
other states. 
The total decay width of the Higgs, $\Gamma_\text{tot}$, is then given by
\beq
\Gamma_\text{tot} = r_h \Gamma^\text{SM}_\text{tot} + \Gamma_\text{exo}+\Gamma_\text{inv}+\Gamma_{\hat{t}\hat{t}}\,\Theta(m_h/2-m_{\hat{t}})\,,
\label{eq:myinv}
\eeq
where $\Gamma^\text{SM}_\text{tot}$, $\Gamma_\text{exo}$, $\Gamma_\text{inv}$, and $\Gamma_{\hat{t}\hat{t}}$ are decay widths of the Higgs to SM particles, 
exotic final states, invisible final states, and top partner(s), respectively. This relation can be re-parameterized as
\begin{eqnarray}
\label{eq:rinv}
r_\text{exo}&\equiv & \f{\Gamma_\text{exo}}{\Gamma_\text{tot}^\text{SM}}=\frac{r_h B_\text{exo}}{1-B_\text{exo}-B_\text{inv}-B_{\hat{t}\hat{t}}\,\Theta(m_h/2-m_{\hat{t}})}\,, \\
 r_\text{inv}&\equiv & \f{\Gamma_\text{inv}}{\Gamma_\text{tot}^\text{SM}}=\frac{r_h B_\text{inv}}{1-B_\text{exo}-B_\text{inv}-B_{\hat{t}\hat{t}}\,\Theta(m_h/2-m_{\hat{t}})}\,,
\end{eqnarray}
where the branching ratios are defined as $B_\text{inv}\equiv \Gamma_\text{inv}/\Gamma_\text{tot}$, $B_\text{exo}\equiv \Gamma_\text{exo}/\Gamma_\text{tot}$ and $B_{\hat{t}\hat{t}}\equiv \Gamma_{\hat{t}\hat{t}}/\Gamma_\text{tot}$. 

\subsection{How to Constrain and Hide Top Partners}
\label{subsec:avoiding-constraints}

With the definitions given in the previous sub-section, it is straightforward to understand where the strongest constraints on top partners come from.  To constrain top partners, we use, for example, the signal strengths reported by ATLAS and CMS for particular final states of the Higgs.  
These are given by
\begin{align}
\mu_{f}\equiv{}& \f{\sigma_\text{prod}^\text{BSM}\times B^\text{BSM}_{h\to ff}}{\sigma_\text{prod}^\text{SM} \times B^\text{SM}_{h\to ff}}\,,~~~~~  
\mu_{\text{inv}}\equiv{} \f{\sigma_\text{prod}^\text{BSM}\times B^\text{BSM}_{h\to \text{inv}}}{\sigma_\text{prod}^\text{SM}}\,,
\label{eq:mu}
\end{align}
where for the SM $\mu^\text{SM}_f=1$ and $\mu_{\text{inv}}^\text{SM}=0$.  
Given that a particular final state may come from a variety of different production modes, we must also take into account the weighting of the production modes, $\xi_{G, V, T}$, which give the relative strength of contributions to Higgs production from  gluon fusion (ggF), vector-boson fusion plus Higgs associated production (VBF+VH), and top-quark-pair production in association with a Higgs (ttH), respectively. 
Similarly, for a particular production channel with multiple final states, we weight the decay modes (see Appendix~\ref{datasets} 
for more details).

In the limit that all SM particles couple to the Higgs with their SM tree-level values ($r_j=1$), and assuming there are no exotic/invisible Higgs decays, the largest shift to Higgs phenomenology appears in $r_G$ (see \eqref{myrg2}), 
\beq
r_G \sim (1+\mathcal{N}_{\hat{t}})\,.
\eeq
In this limit, 
\beq
\mu_f \sim \left( |r_G|^2 \xi_{G} +  \xi_{V}+\xi_{t}\right) \frac{1}{1+(|r_G|^2-1)B^\text{SM}_{h\to gg}+\cdots}\,,
\eeq
which will give $\mu_f \gg 1$ for many channels if they have a large contribution from gluon fusion.  This then implies a bound on $\mathcal{N}_{\hat{t}}$, which can be translated into a bound on the mass of the top-partner and a constraint on naturalness.
However, many models have modified Higgs-SM couplings from an extended Higgs sector or from non-renormalizable contributions, so 
that other shifts, $r_j$, must be taken into account, especially for larger top-partner masses when $\mathcal{N}_{\hat{t}}$ becomes small.  
In models with multiple top partners, the contribution to $r_G$ from the different top partners can cancel amongst themselves, a 
possibility we will investigate in Section~\ref{s.theory}.

Ignoring the possibility of multiple top partners for now, it is useful to see how best to alleviate a shift in $r_G$ from its SM value.  From \eqref{myrg}, naively the simplest way to return $r_G$ to its SM value would be by appropriately adding an equal and opposite contribution of $\delta r_G$ from some new physics contribution.  For instance in SUSY, one could add vector-like matter with a large bare mass, and interactions with the Higgs that could give the requisite $\delta r_G$ to offset the stop contribution.  Because of the large bare mass, they would not be seen in direct searches nor in other Higgs precision observables.  
Nevertheless we will not investigate this option further as it requires additional fine tuning for the new sector to cancel the inherent change in $r_G$ without a symmetry, and additionally the extra sector would also contribute more significantly to the naturalness problem throughout most of its parameter space. 

Next, we investigate how to alleviate changes in $r_G$ through other coupling changes for the Higgs.  If gluon fusion was the {\em only} way the LHC produced the Higgs, it would be straightforward to change the total width of the Higgs to offset this with a $r_\text{exo/inv}$ contribution to attempt to hide this shift.   However, since gluon fusion is not the only production channel, this will not reduce the constraints significantly, as there are currently strong constraints on all production mechanisms except for ttH. 
Instead, a shift in one or more SM couplings is needed to offset the contribution of a colored top partner to $r_G$. 

From \eqref{myrg2}, in the limit that the top quark and top partner dominate the contributions to $r_G$, we have 
\beq\label{eq:mylim}
r_G\sim r_t (1+\mathcal{N}_{\hat{t}})\,.
\eeq
We thus see that a natural way to hide the shift from $\mathcal{N}_{\hat{t}}$ is by adjusting $r_t$. 
While other coupling modifications are possible, they would require a parametrically larger shift from their SM values than a shift in 
$ht \bar{t}$ coupling, which is the largest and among the least-constrained couplings.  
In particular, $r_t$ is currently only constrained independently from measurements where ttH is the dominant production mode. 
Whether $r_t$ can be modified from its SM value, is a model-dependent question.  In particular, the spin of the top partner is correlated with the sign of $\mathcal{N}_{\hat{t}}$.   For spin-0 partners, $\mathcal{N}_{\hat{t}}$ is positive, which implies that $r_t$ must be smaller than 1.  For spin-1/2 partners, $\mathcal{N}_{\hat{t}}$ is negative, implying that $r_t>1$ is desired;  however, in models it is usual to have $r_t <1$.   
We will comment more on particular model building aspects in future sections. 

Adjusting $r_t$ is currently the best mechanism for hiding the effects of a colored top partner and only measurements of the ttH coupling at the ILC and FCC-hh will constrain $r_t$ at a percent level. This points to this channel as the best possible mode for indirect hints of top partners.  
However, it also implies that other mechanisms are needed to avoid Higgs precision constraints on top partners if no deviations are found. 
After $r_t$, changing $r_b$ is most promising, since it also enters into the loop functions.  As stated earlier, $r_{\rm inv}$ can offset some of the increase from $r_G$ but only at the expense of affecting other channels as well.   With this in mind, we will investigate the correlations of each possible shift in correlation with $\mathcal{N}_{\hat{t}}$ to determine the best mechanism for hiding top partners in current and in future datasets.  We will also investigate whether existing models like the MSSM can be effective in hiding top partners from Higgs precision data, or if further model building is needed.

\section{Data Sets and Fitting Procedure}\label{s.data}

\subsection{Current and future proton collider data}
Here we review the Higgs-signal-strength data sets used in our analyses.   
Explicit values and detailed references are included in Appendix~\ref{datasets}. 
 Note that the constraints on $h\to {\rm invisible}$ requires special treatment for the current and projected data, as we discuss below.
 \begin{enumerate}
 \item {\bf Current Limits}: 
 This data set consists of existing Higgs measurements from ATLAS and CMS Run~1 (7 and 8 TeV), Run 2 (13 TeV), and the Tevatron 
 (which is only marginally relevant for the $b\bar b$-channel).  It includes data up to EPS 2017~\cite{EPS2017}. We denote the observed signal strength as $\mu$, and $1\sigma$ upper and lower error bars as $\sigma^\text{up}$ and $\sigma^\text{down}$ respectively.
  
 \item {\bf Current Expected Sensitivities}: 
Several existing measurements of the Higgs-signal-strengths differ slightly from their SM values.  
This is expected from statistical fluctuations, but could also be a sign of new physics.  
It is thus useful to compare the current constraints on top partners with the expected constraints assuming the 
existing measurements would have been in perfect agreement with the SM values.  
Moreover, the expected constraints provide a good benchmark with which to compare the projected sensitivities from future Run 3 and Run 4
data sets (see below).  

We construct the current expected sensitivities from the same Higgs measurements as used to derive the ``Current Limits'' above, 
but with the signal strengths set to their SM values, $\mu_\text{exp}^\text{obs}=\mu^\text{SM}$ ($\mu_{f}^\text{SM}=1$, $\mu_{\text{inv}}^\text{SM}=0$).  We take the $1\sigma$ error bar to be the average of the original asymmetric $1\sigma$ error bars, i.e.~$\sigma_\text{exp}^\text{up/down}=(\sigma^\text{up}+\sigma^\text{down})/2$. 

 \item {\bf LHC Run 3 Data (projected, 300~fb$^{-1}$)}: 
To derive the prospective constraints on top partners by the end of the LHC Run 3, 
we use the projected sensitivities on the Higgs signal strengths 
for ATLAS Run~3 from~\cite{atlas2014:pro}.  The projections are based on simulations of various search channels and 
assume a center-of-mass energy of 14~TeV and an integrated luminosity of 300~fb$^{-1}$. 
The theoretical uncertainties are assumed to be the same as today.  
CMS also has several projections~\cite{CMS-projection}, but for simplicity, we assume that they will analyze the same search channels 
as ATLAS and obtain identical results as in~\cite{atlas2014:pro}.  
We thus calculate projected sensitivities for the combined ATLAS$+$CMS data set (referred to as ``LHC Run 3''), which 
consists of 600~fb$^{-1}$.  
We take the production-channel weights from~\cite{Bechtle:2014ewa}. 
Note that current LHC Run~2 measurement have already included several search channels that are not listed 
in~\cite{atlas2014:pro}, such as measuring $hb\bar{b}$ through ttH~\cite{CMS:2016zbb, ATLAS:2016awy}. 
Our projections are thus conservative. 
 
\item {\bf  LHC Run 4 Data (HL-LHC) (projected, 3~ab$^{-1}$)}: 
We also derive prospective constraints on top partners by the end of the LHC Run~4 (high-luminosity run), again using the 
ATLAS prescription and the same caveats discussed above for Run~3~\cite{atlas2014:pro}.  
This data set is similar to the LHC Run~3 data, but the integrated luminosity is set instead to 3~ab$^{-1}$ for each experiment 
(i.e.~for a total of 6~ab$^{-1}$). 
We assume that the theoretical systematic uncertainties remain unchanged as the integrated luminosity grows from 300 fb$^{-1}$ to 3 ab$^{-1}$ in~\cite{atlas2014:pro} (i.e.~they are assumed to be the same as today). 
This is likely a pessimistic assumption, but it is conservative. 

\item{\bf Proton-proton beams at the Future Circular Collider (FCC-hh)}: 
We project sensitivities for FCC-hh (100 TeV, 30 ab$^{-1}$) based on the Higgs-coupling data from FCC-ee 
(see Sec.~\ref{subsec:data-future} below) but assuming a measurement of the $ht \bar{t}$ coupling, $r_t$. 
This coupling is expected to be measured at a statistical limited level of 1\%~\cite{FCChh:2017}, an  
improvement compared to the expected FCC-ee measurement of 13\% (see Table~\ref{tab:future}). 

 \end{enumerate}
 
Current and projected $h\to{\rm invisible}$ search data is treated differently in our analysis compared to the other data, as we now describe.  
Most of the published LHC results only show upper limits for $\mu_{\text{inv}}\equiv\sigma_\text{prod}\times B_\text{inv}/\sigma_\text{prod}^\text{SM}$ at 95\% confidence level (CL) (denoted as $\sigma^{95\%}_\text{inv}$) rather than a likelihood scan with respect to $\mu$.
In the absence of the likelihood curve, we do not know the best-fit and $1\sigma$ values.  
In these cases, we set the ``observed'' $\mu$ to be 0 and translate the 95\% CL upper limit into $1\sigma$ CL uncertainty ($\sigma^\text{up/down}_\text{inv}=\sigma^{95\%}_\text{inv}/\sqrt{3.84}$). 
For our prospective constraints, we use projected $\text{VBF}\to h\to\text{invisible}$ data from~\cite{CMS-DP-2016-064}, which is 
based on $\sqrt{s}=13$ TeV with up to 3~ab$^{-1}$ of data (similar projections are also found in \cite{Brooke:2016vlw}).  
We choose the scenario in~\cite{CMS-DP-2016-064} in which the experimental systematic uncertainties and the theoretical systematic uncertainties stay the same as the integrated luminosity increases. 
This projects $\sigma^{95\%}_\text{inv} = 21\%$ for LHC Run 3 Data and $\sigma^{95\%}_\text{inv} = 20\%$ for LHC Run 4 Data. 

The Higgs may also have exotic decays~\cite{Curtin:2013fra}, distinct from purely invisible decays.  
Exotic decays of the Higgs are an important window to new physics and several searches have been conducted by ATLAS and CMS~\cite{ATLAS:2015bra, ATLAS:2015xda, Aad:2015uaa, Aad:2015oqa, Aad:2015sva, Khachatryan:2015vta, Aad:2015txa, Aad:2015pla, Aad:2015bua, CMS:2015iga, Khachatryan:2015nba, Orimoto:2015ueg, CMS:2016cqw, CMS:2016cel, Aaboud:2016oyb}. 
However, many possibilities exist, which have not yet all been constrained.  
The bound on the total Higgs width~\cite{Caola:2013yja,Dixon:2013haa} from direct measurements does not provide a 
strong constraint on arbitrary exotic Higgs decays. 
Instead, as discussed in Sec.~\ref{subsec:avoiding-constraints}, exotic-decay modes are constrained as they 
would also modify the signal strengths.  
As we will see, while they can help to hide spin-0 top partners, they do not help in hiding spin-1/2 and spin-1 
top partners. 
 
We construct the overall $\chi^2$ fitting function of all the search channels as
\beq\label{eq:chisqFunction}
\chi^2=\sum_{f, \text{inv}} \f{(\mu_f-\mu_f^\text{obs})^2}{\sigma^2_{f}}\quad \text{with}\quad  \sigma_{f}=\left\{ 
\begin{array}{ll}
\sigma^\text{up}_{f}, & \mu_f \geq \mu^\text{obs}_f \\
\sigma^\text{down}_{f}, & \mu_f < \mu^\text{obs}_f
\end{array}\,,
\right.
\eeq
where $f$ (inv) runs over all the (invisible) search channels. In Appendix~\ref{validity}, we show good agreement between our results, obtained using the above $\chi^2$, and the results obtained using 
the more involved method adopted by \texttt{HiggsSignals}~\cite{Bechtle:2013xfa}.
 
\subsection{Future lepton collider data 
}\label{subsec:data-future}  
Future lepton colliders provide new opportunities to constrain the Higgs sector. Here we focus on three proposed projects and compare their reaches to those of the current and future proton colliders (see Table~\ref{tab:future} for the expected precision on the Higgs couplings): 
 \begin{enumerate}
  \item {\bf International Linear Collider (ILC)}~\cite{Behnke:2013xla}. Projected Higgs-signal-strengths on individual search channels for ILC do not exist yet. However, there are combined Higgs-coupling fits to the 11-parameter set consisting of $r_W$, $r_Z$, $r_b$, $r_G$, $r_\gamma$, $r_\tau$, $r_c$, $r_\mu$, $r_t$, $\Gamma_\text{tot}$, and $B_\text{inv}$~\cite{Fujii:2015jha}. We choose the sensitivities of  the ``Full Data Set" of ILC (250~GeV, 2~ab$^{-1}$ $\oplus$ 350 GeV, 200 fb$^{-1}$ $\oplus$ 550 GeV, 4 ab$^{-1}$) for our projections. 
  
 \item {\bf Circular Electron Positron Collider (CEPC)}~\cite{CEPC-SPPCStudyGroup:2015csa}. Similar to ILC, there are only sensitivities from combined Higgs-coupling fits to the 10-parameter set: $r_W$, $r_Z$, $r_b$, $r_G$, $r_\gamma$, $r_\tau$, $r_c$, $r_\mu$, $\Gamma_\text{tot}$, and $B_\text{inv}$~\cite{CEPC-SPPCStudyGroup:2015csa}. 
We use the CEPC (240~GeV, 10~ab$^{-1}$) expected sensitivities to derive our projections. 
  
 \item {\bf Electron-positron beams at the Future Circular Collider  (FCC-ee)}~\cite{dEnterria:2016sca}. FCC-ee also only provides sensitivities from combined Higgs-coupling fits, to the following 11-parameters:  
 $r_W$, $r_Z$, $r_b$, $r_G$, $r_\gamma$, $r_\tau$, $r_c$, $r_\mu$, $r_t$, $\Gamma_\text{tot}$, and $B_\text{inv}$~\cite{dEnterria:2016sca,FCChh:2017}. Although FCC-ee, running at 350 GeV, cannot directly measure $r_t$ like ILC, it could constrain $r_t$ indirectly in the $e^+ e^-\to t\bar{t}$ channel through virtual Higgs-exchange. We use the FCC-ee (240~GeV, 10~ab$^{-1}$ $\oplus$ 350~GeV, 2.6~ab$^{-1}$) expected sensitivities to derive our projections.  
 The improvements in the sensitivities of FCC-ee compared to the ILC and CEPC are due to the increased integrated luminosity, 
more interaction points, and a better electron beam energy resolution~\cite{dEnterria:2016sca}. 
 \end{enumerate}
Other future lepton colliders, such as the Compact Linear Collider~\cite{Lukic:2016equ}, yield similar constraints on colored top partners. 
We do not include them below. 

We interpret the projected sensitivities in Table~\ref{tab:future} 
as 1$\sigma$-error bars for a particular coupling, $\sigma_{r_i}$, and construct a $\chi^2$-function 
as
\beq\label{eq:futurelepton}
\chi^2 =\sum_{i = W, Z, b, G, \gamma, \tau, c , \mu, t}  \frac{(r_i-1)^2}{\sigma_{r_i}^2}+\f{(r_h+r_\text{inv}+r_\text{exo}-1)^2}{\sigma_{\Gamma_\text{tot}^2}}+\f{(r_\text{inv}-0)^2}{\sigma_{r_\text{inv}}^2},
\eeq
where $r_h$, $r_\text{inv}$, and $r_\text{exo}$ are given by \eqref{rh} and \eqref{rinv}, respectively. 
Since we focus only on the parameters most relevant for deriving the constraints on top partners, we set $r_W=r_Z=r_V$ and    
$r_\mu=r_\tau$, and construct $r_G$ and $r_\gamma$ from $r_t$,\footnote{For CEPC, we set $r_t=1$ in Eqs.~(\ref{eq:myrg}), (\ref{eq:myrgamma}), and (\ref{eq:futurelepton}).} $r_b$, $r_V$, $r_\tau$, and $r_{\hat t}$ according to Eqs.~(\ref{eq:myrg}) 
and (\ref{eq:myrgamma}), respectively.  
In many cases discussed below, we set $r_c=1$, since it is not relevant for constraining top partners.  
However, in some cases (e.g., in the MSSM or in 2HDM models), we set $r_c=r_t$, and thus a precise measurement of $r_c$ will 
allow strong constraints to be set on top-partner models in which $r_t$ is affected. 
For invisible decays, we use \eqref{rinv} and translate $\sigma_{B_\text{inv}}$ from the 95\% CL upper limit given in \tabref{future} 
using the relation $\sigma_{B_\text{inv}}= \sigma^{95\%}_{B_\text{inv}}/\sqrt{3.84}$.

\section{Canonical Top Partner Models and Extensions}
\label{s.theory}

We outline now the models of three specific classes of colored top partners -- spin-0, 1/2, and 1 -- that we study.  The symmetries that enforce the cancellation of quadratic divergences, see \eqref{cancel}, will be different in the various cases, and therefore the basic moving parts of a model and their predictions for Higgs phenomenology are different.   We briefly comment on their generic prediction for $\mathcal{N}_{\hat{t}}$ defined by \eqref{rGhatt} and the extensions that can reduce the overall contribution to $r_G$.

\subsection{Spin-0}
\label{s.spin-0}

For spin-0 colored top partners (without loss of generality we will refer to them as stops), enforcing~\eqref{cancel} requires a symmetry between the fermionic tops of the SM and scalar particles.   Supersymmetry is the only known symmetry that can have such a relation, and in the minimal incarnations that can incorporate the SM there will be two stops, a partner for the right-handed and left-handed top quarks. 
Moreover, since the top-quark is part of an $SU(2)$ doublet with the bottom-quark, SUSY will also require a left-handed bottom-partner.  
We assume that other partner particles are heavy, which in any case does not spoil naturalness.  
 
\begin{figure}[htbp]
   \centering
  \includegraphics[width=0.25\textwidth]{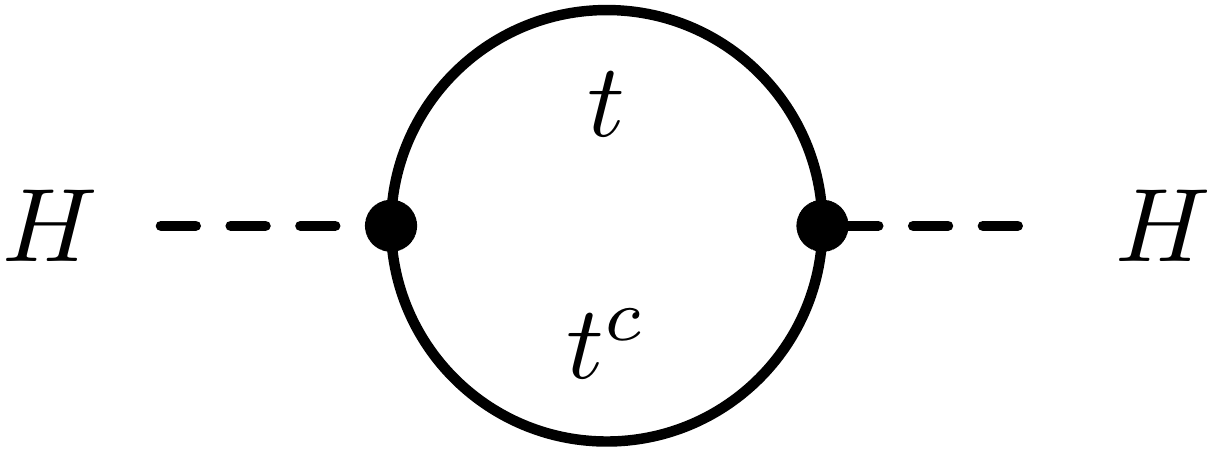}~\includegraphics[width=0.2\textwidth]{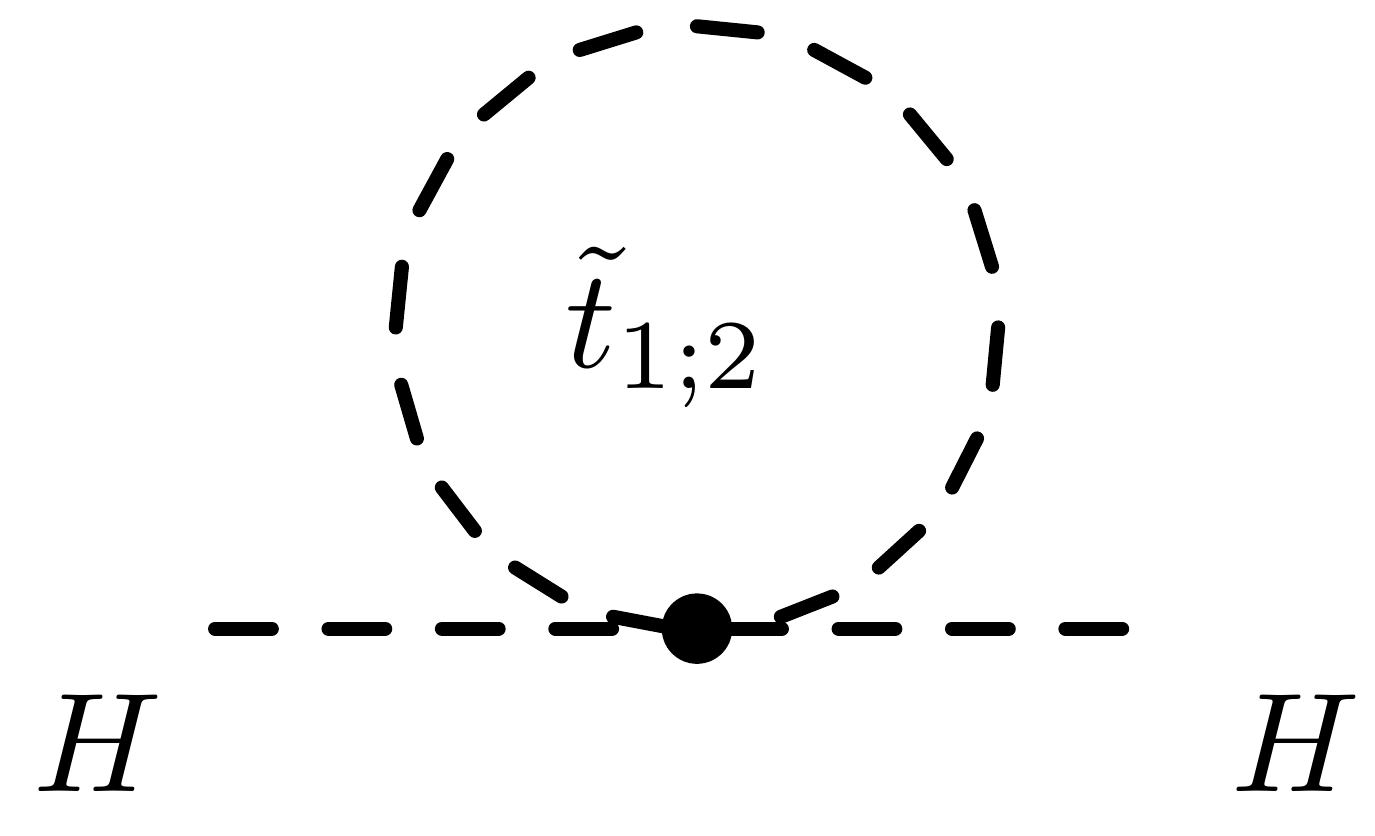}
   \caption{Diagrams relevant to the cancellation of the top loop with spin-0 partners: the one-loop diagram for the SM top (left) and the two stops (right).}
   \label{fig:topcancel1}
\end{figure}

In this limit we have the equality $a_{\t t}=a_t=\lambda_t^2$ in \eqref{massform}, and due to scalars and fermions contributing with opposite signs to the Higgs-loop integrals, \eqref{cancel} is automatically satisfied. 
The usual diagrammatic presentation of the cancellation of quadratic divergences is shown in \figref{topcancel1}. 

Even though \eqref{cancel} is satisfied for a natural theory, the precise structure of the model and the Higgs sector is model-dependent.   
In the case of SUSY, the Higgs sector must be enlarged at least to a two-Higgs-doublet model (2HDM) because of the symmetry that enforces~\eqref{cancel}.  This has implications for the Higgs phenomenology, since the Higgs observed at the LHC must then be a linear combination of the fields within the 2HDM (or additional Higgs states if there are more).  For simplicity, we will start with the minimal 2HDM required for stops, which is a type-II model as in the MSSM, and take the decoupling limit.  
In the decoupling limit, the Higgs couples to the SM fields as the SM Higgs (i.e., $r_t$ and the other Higgs-SM couplings are 1).  
We will comment below in Sections~\ref{subsec:MSSMlike} and \ref{subsec:Spin0extendedHiggs} 
on changes that occur from extended Higgs sectors. 

We now first review the structure of the stop 
masses and their couplings to the Higgs, and then explain how Higgs precision 
measurements constrain the stops.  
We then discuss how the stop constraints are affected when including the left-handed sbottom.  

After EWSB the stop-mass matrix is given by
\beq
\begin{pmatrix} m^2_{Q_3}+m_t^2 + D_L^t &m_t X_t  \\ m_t X^*_t  &  m^2_{U_3}+m_t^2 +D^t_R \end{pmatrix}\,,
\label{eq:stop1}
\eeq
where $m_{Q_3}$ and $m_{U_3}$ are the soft SUSY breaking masses of the left- and right-handed stops, 
respectively.\footnote{In the MSSM, the off-diagonal mixing parameter can be written as $X_t= {A_t}-{\mu}\,{\cot \beta}$,  
where $A_t$ is a soft-SUSY breaking parameter and $\mu$ is a supersymmetric mass term for the Higgs doublets; 
for a review see~\cite{Martin:1997ns}.} 
$D_L^t$ and $D_R^t$ are $D$-terms
\beq
D_L^t =\left(\f{1}{2}-\f{2}{3}\sin^2 \theta_W\right) m_Z^2 \cos 2\beta,\quad D_R^t =\frac{2}{3}\sin^2 \theta_W m_Z^2 \cos 2\beta\,, 
\eeq
where $\theta_W$ is the Weinberg angle and $ \tan \beta\equiv v_2/v_1$ is the ratio of the two Higgs VEVs. 
The stop matrix has two eigenvalues, $m_{\t t_1}$ and $m_{\t t_2}$. 
We order the eigenvalues such that $m_{\t t_1}$ contains mostly $m_{Q_3}$ and $m_{\t t_2}$
contains mostly $m_{U_3}$. The eigenvalues $m_{\t t_1}$ and $m_{\t t_2}$ satisfy the relation
\beq
|m_{\t t_1}^2-m_{\t t_2}^2|=\sqrt{(m^2_{Q_3}-m^2_{U_3}+D_L^t-D_R^t)^2+4m_t^2X_t^2}\,.
\label{eq:diagonalstops}
\eeq
Since $m_{Q_3}$, $m_{U_3}$, $D_L^t$, and $D_R^t$ are real, the maximum value for $X_t$ is given by 
\beq
|X^{\text{max}}_t| =\f{|m_{\t t_1}^2-m_{\t t_2}^2|}{2m_t}\,.
\label{eq:Xtmax}
\eeq

To calculate the effect of the stops on Higgs-precision measurement, we need to know the couplings between the Higgs, $\t t_1$, and 
$\t t_2$. 
In the decoupling limit, these are given by 
\begin{align} 
g_{h \t t_1 \t t_1} ={}&\f{2}{v} \left (m_t^2 - \frac{m_t^2 X_t^2}{m_{\t t_2}^2-m_{\t t_1}^2}+D_{11}  \right)\,,\label{eq:ght1t1}\\
g_{h \t t_2 \t t_2} ={}&\f{2}{v} \left (m_t^2 + \frac{m_t^2 X_t^2}{m_{\t t_2}^2-m_{\t t_1}^2}+D_{22} \right)\,, \label{eq:ght2t2}\\
g_{h \t t_1 \t t_2}={}& \f{m_t}{v} X_t \left(\cos{2\theta_t}+D_{12}\right)\,, \label{eq:ght1t2}
\end{align}
where
\begin{align}
D_{11} ={}&\cos 2\beta m_Z^2 \left[\frac{1}{4}+c_{2\theta_t}(\f{1}{4}-\frac{2}{3}\sin^2 \theta_W)\right]\,,\\
D_{22} ={}&\cos 2\beta m_Z^2 \left[\frac{1}{4}-c_{2\theta_t}(\frac{1}{4}+\frac{2}{3}\sin^2 \theta_W)\right]\,,\\
D_{12} ={}&-  \frac{4 }{m_{\t t_2}^2-m_{\t t_1}^2} m_Z^2 \cos 2\beta \left(\frac{2}{3}\sin \theta_W^2-\frac{1}{4}\right)\,,\\
\cos{2\theta_t} ={}&\sqrt{1- \left(\frac{2 m_t X_t}{m_{\t t_2}^2-m_{\t t_1}^2}\right)^2}\,.
\end{align}

We can now discuss how the stops affect Higgs precision measurements.  
First, the stops can contribute to the $hgg$ coupling, $\mathcal{N}_{\tilde{t}}$ (see \eqref{rGhatt}) as  
\beq\label{eq:Nt-spin-0}
\mathcal{N}_{\tilde{t}}  = \frac{{\mathcal A}^{\text{s}=0}(m_h^2/4 m_{\t t_1}^2)}{{\mathcal A}^{\text{s}=1/2} (m_h^2/4 m_t^2)} \frac{g_{h \t t_1 \t t_1}}{m_{\t t_1}^2}\f{v}{2}+  \frac{{\mathcal A}^{\text{s}=0}(m_h^2/4 m_{\t t_2}^2)}{{\mathcal A}^{\text{s}=1/2} (m_h^2/4 m_t^2)} \frac{g_{h \t t_2 \t t_2}}{m_{\t t_2}^2}\f{v}{2}\,.
\eeq
Second, stops with mass below $m_h/2$ do not only appear in such loop processes, 
but also provide an exotic decay-channel for the Higgs, 
contributing to the total Higgs width (recall that we do not specify the stop decay channels or investigate off-shell decays in this paper).  
The tree-level Higgs-decay width to two stops is given by
\beq
\Gamma (h \to \t t_i \t t_j) = \frac{3}{16\pi m_h} g^2_{h \t t_i \t t_j} \left[1- \frac{2(m_{\t t_i}^2+m_{\t t_j}^2)}{m_h^2}-\frac{(m_{\t t_i}^2 -m_{\t t_j}^2)^2}{m_h^4}\right]^{1/2} \Theta\left(m_h - m_{\t t_i}-m_{\t t_j}\right)\,,
\eeq
so that the total width for the Higgs decays into stops is 
\beq\label{eq:gamma-h-stop}
\Gamma_\text{tot} (h \to \t t \t t) \equiv \Gamma (h \to \t t_1 \t t_1)+ \Gamma (h \to \t t_2 \t t_2)+ \Gamma (h \to \t t_1 \t t_2)\,.
\eeq

In Section \ref{results}, we will present the explicit exclusions from all the various contributions.  However it is useful, as in Section~\ref{subsec:avoiding-constraints}, to develop an intuition for the shape of the exclusion curves.  Given that the two main sources are contributions to $\mathcal{N}_{\tilde{t}}$ and the width, it is helpful to look at their approximate expressions and understand where their contributions are extremized.  In the limit $m_{\t t_{1,2}}\gg m_h/2$, we get
\beq
\mathcal{N}_{\tilde{t}} \approx  \f{v}{8} \left(\frac{g_{h\t t_1 \t t_1}}{m_{\t t_1}^2}+\frac{g_{h \t t_2 \t t_2}}{m_{\t t_2}^2}\right). 
\label{eq:largemt}
\eeq
It is a good approximation to neglect the $D$-terms, so that $\mathcal{N}_{\tilde{t}}$ becomes 
\beq
\mathcal{N}_{\tilde{t}} \approx \f{1}{4}\left(\f{m_t^2}{m_{\t t_1}^2}+\f{m_t^2}{m_{\t t_2}^2}-\f{m_t^2 X_t^2}{m_{\t t_1}^2 m_{\t t_2}^2}\right)\,. 
\label{eq:softHiggs}
\eeq
(We note that this expression can be also obtained from the low-energy Higgs theorem discussed in Section~2~\cite{Ellis:1975ap, Shifman:1979eb, Dermisek:2007fi, Blum:2012ii, Carmi:2012in, Fan:2014txa}.)
\eqref{softHiggs} depends only on the stop masses and the mixing.  Moreover, $m_{\t t_1}$ and $m_{\t t_2}$ have symmetric contribution to Higgs precision measurements.\footnote{In Appendix~\ref{blindspot}, we discuss how stops are constrained from the future Higgs 
precision probe $e^+ e^- \to Zh$; we will see that $\t t_1$ and $\t t_2$ contribute differently to this process.} 
The lowest allowed value for $X_t$ from \eqref{softHiggs} is given by  
\beq
|X^{\text{min}}_t|^2\approx\f{m_t^2\left(m_{\t t_1}^2+m_{\t t_2}^2\right)- 4 m_{\t t_1}^2 m_{\t t_2}^2 (\mathcal{N}_{\tilde{t}})^\text{fit;max}}{m_t^2}\,,
\label{eq:Xtmin}
\eeq
where $(\mathcal{N}_{\tilde{t}})^\text{fit;max}$ is the upper limit allowed from Higgs precision data. Combined with \eqref{Xtmax}, 
a given set of $m_{\t t_1}$ and $m_{\t t_2}$ are ruled out if $|X^{\text{min}}_t| > |X^{\text{max}}_t|$. 
The resulting constraints are strongest in the degenerate limit ($m_{\t t_1}=m_{\t t_2}\equiv m_{\t t}$);  from \eqref{softHiggs}, the 
constraint is given by 
\beq
m_{\t t} \ge  \f{m_t}{\sqrt{2 \ (\mathcal{N}_{\tilde{t}})^\text{fit;max}}}\,. 
\label{eq:deg}
\eeq
As a result, smaller $(\mathcal{N}_{\tilde{t}})^\text{fit;max}$ leads to stronger constraints on $m_{\t t}$, while $(\mathcal{N}_{\tilde{t}})^\text{fit;max}<0$ completely rules out the degenerate direction. 

The non-degenerate direction is less constrained. $X^{\text{max}}_t$ increases as the difference between $m_{\t t_1}$ and $m_{\t t_2}$ increases, see \eqref{Xtmax}, and it becomes easier to find specific values of $X_t = X^{\text{blind}}_t$ that allow 
$\mathcal{N}_{\tilde{t}}$ to vanish (this was referred to as the stop blind-spot in~\cite{Fan:2014axa}). 
$X^{\text{blind}}_t$ is given by 
\begin{equation}
X^{\text{blind}}_t= ({m_{\t t_1}^2+m_{\t t_2}^2})^{1/2}\,.  
\end{equation}
For $m_{\t t_1}\to 0$, one finds that $m_{\t t_2}=|X^{\text{min}}_t| \le |X^{\text{max}}_t|=m^2_{\t t_2}/ (2 m_t)$ is always satisfied for 
$m_{\t t_2} \gtrsim 2 m_t$ (similarly for $m_{\t t_2}\to 0$).
This means that Higgs precision measurements that constrain only $\mathcal{N}_{\tilde{t}}$ are not sufficient to probe this region. 

Back to the full expression, the vanishing of $\mathcal{N}_{\tilde{t}}$ in the non-degenerate direction occurs since 
$g_{h\t t_1 \t t_1}/m^2_{\tilde{t}_1}=-g_{h\t t_2 \t t_2}/m^2_{\tilde{t}_2}$, see \eqref{largemt}. 
However, in this limit, at least for $m_{\t t_{1}} < m_h/2$ and/or $m_{\t t_{2}} < m_h/2$, the Higgs can also decay to the stops.  
The Higgs decay width to stops does not vanish for the same choice of parameters as does $\mathcal{N}_{\t t}$, and one might 
naively conclude that any stop lighter than $m_h/2$ will be ruled out.  
However, in the limit that the other stop is sufficiently heavy, the coupling of the lighter stop to the Higgs becomes small.  
We can see this by integrating out the heavy stop~\cite{Fan:2014axa} to obtain the following effective lighter stop-Higgs coupling: 
\beq
\mathcal{L}=\frac{2 m_t^2}{v^2} \left(1-\frac{X_t^2}{m^2_{\t t_h}-m^2_{\t t_l}}\right)|H_u|^2 |\t t_l|^2\,. 
\eeq
where $H_u$ represents a up-type Higgs doublet, $\t t_h$ and $\t t_l$ stands for the heavier and lighter stops respectively. 
This vanishes for $X_t=X^{\text{blind}}_t$ in the non-degenerate limit.  
In this case, the lighter stop only couples very weakly to the Higgs, 
while the heavier stop is too heavy to affect the Higgs precision measurements.  
Higgs precision measurements thus cannot alone rule out the possibility of a very light stop entirely (see also Section~\ref{results} and Fig.~\ref{fig:stopstop}). 

In the particular limit we are studying here, there also will be a left-handed sbottom in the spectrum.  In the full MSSM there is a right-handed sbottom and the full couplings of the sbottom will resemble those of Eqns.~(\ref{eq:ght1t1})--(\ref{eq:ght1t2}).  
We leave a full accounting of third-generation squarks to future work, 
and focus instead on the natural SUSY limit in which the right-handed sbottom is decoupled.  
The sbottom-eigenstate mass can be written in terms of the stop parameters as 
\beq\label{eq:sbottom-mass}
m_{\t b_1}^2 = \frac{1}{2}\left(1+\cos 2\theta_t\right) m_{\t t_1}^2 +\frac{1}{2}\left(1-\cos 2\theta_t\right) m_{\t t_2}^2 - m_t^2 - m_W^2 \cos 2 \beta + m_b^2\,.
\eeq
The Higgs-sbottom-sbottom coupling is given by 
\begin{align}\label{eq:sbottom-h}
g_{h \t b_1 \t b_1}\approx{}&\f{2}{v}\left\{m_b^2 + m_Z^2 \cos 2 \beta \left[-\f{1}{2} + \frac{2}{3}\sin^2 \theta_W \right]\right\}\,.
\end{align}
The sbottom can contribute to both $\mathcal{N}_{\t b}$ and $\Gamma (h \to \t b_1 \t b_1)$, with expressions 
similar to the ones given for the stops above.  
From \eqref{sbottom-h}, we see that there are two contributions to $g_{h \t b_1 \t b_1}$: the first is suppressed by the 
small bottom quark mass, while the $D$-term contribution generically gives a large coupling of $\mathcal{O}(v)$. 
We thus see that it is useful to investigate two limits: one in which the sbottoms do not contribute at all and one in which the 
sbottoms contribute with a large coupling given by the $D$-term. 
Below, we will investigate these two cases with the following specific parameter choices:
\begin{eqnarray}
\tan\beta\simeq 1 & \implies & g_{h \t b_1 \t b_1}=0  \label{eq:gnb1b1=0}\,, \\
{\rm large}~\tan\beta & \implies & g_{h \t b_1 \t b_1}\simeq\frac{m_Z^2}{v} \left(1-\frac{2}{3}\ \sin^2\theta_W \right)\,. \label{eq:gnb1b1-D-term} 
\end{eqnarray}
We will show the results for these two cases in Section~\ref{results} (see Fig.~\ref{fig:stopstop}).  
Note that we also ensure that $m_{\t b_1}$ in \eqref{sbottom-mass} is real for all viable choices of 
$m_{\t t_1}$, $m_{\t t_2}$, and $X_t$ when we calculate the current constraints and projected sensitivities in Section~\ref{results}.  

Finally, we note that the process $e^+e^- \to Z h$ will also generically be affected by stops~\cite{Craig:2014una}, and 
could avoid some of the blind spots of the other measures.  However, since it is only relevant for a future high-precision $e^+e^-$ collider, 
we defer its discussion to Appendix~\ref{blindspot}.

\vskip 0.5 cm

Our discussion thus far is sufficient to talk about the bounds on stops in any model.  
However, as mentioned above, there are at least two Higgs doublets instead of a single doublet as in the SM, which 
can impact the phenomenology.   
Furthermore, in a concrete model such as the MSSM, there will be a number of relationships that relate the top-sector with other sectors through the particular structure of EWSB.  We next discuss how the relations imposed by concrete models affect their ability to accommodate light stops.  We then outline the most promising model building directions 
to hide spin-0 colored top partners. 

\subsubsection{Concrete model: MSSM}
\label{subsec:MSSMlike}

In this subsection, we restrict ourselves to the EWSB structure of the MSSM. After EWSB, there are two CP-even Higgs bosons, and we identify the lightest of these as the 125 GeV SM-like Higgs boson for the rest of our discussion. The Higgs couplings in a MSSM(-like) model are described by two parameters: the rotation angle of the Higgs mass matrix, $\alpha$,\footnote{In MSSM models $-\pi/2\leq \alpha \leq 0$ while in a general type-II 2HDM $-\pi/2\leq \alpha < \pi$.} and the ratio of the two Higgs VEVs, $\tan \beta$. The modifications of tree-level Higgs couplings are
\beq
r_c=r_t=\f{\cos \alpha}{\sin \beta},\quad r_b=r_\tau=-\f{\sin \alpha}{\cos \beta},\quad r_V=\sin(\beta-\alpha),
\label{eq:threeparameters}
\eeq
which can be recast into a more convenient form in terms of $r_t$ and $\tan \beta$
\begin{equation}
r_c=r_t, \quad r_b=r_\tau=\sqrt{1+(1-r_t^2)\tan^2 \beta},\quad r_V=\f{r_t \tan^2 \beta +\sqrt{1+(1-r_t^2)\tan^2 \beta}}{1+\tan^2\beta}\,.
\label{eq:rtrbrv}
\end{equation}

As discussed in \sref{precision}, the most powerful way to hide top partners is to modify $r_t$ directly. The 2HDM naturally allows for modifications of top Yukawa to non-SM values, and to hide stops, we would require $r_t < 1$ given $\mathcal{N}_{\tilde{t}}>1$.  However, in the MSSM,  \eqref{rtrbrv} indicates that $r_b$ and $r_V$ depends on $r_t$, which is not well constrained by current data. The $\tan \beta \rightarrow 0$ limit however removes the $r_t$ dependence of $r_b$ and $r_V$, and fixes them to 1 allowing lower values of $r_t$ without changing other Higgs observables. There is a lower limit of $\tan \beta=2.2$ (obtained at one-loop, ignoring threshold corrections) in order to retain perturbativity of Yukawa couplings at the GUT scale. Smaller values of $\tan \beta$ would necessarily require new physics below the 
GUT scale but even $\tan \beta < 1$ is in principle possible. Requiring the top Yukawa to be smaller than $4\pi$ at 10~TeV (100~TeV) leads to $\tan \beta \ge 0.63~(0.8)$. 
In the opposite limit of large $\tan \beta$, $r_t, r_V \to 1$, independent of $\tan\beta$. We also emphasize that the equality $r_c=r_t$ still holds in all these limits. Precise measurement of $r_c$ at future lepton colliders will therefore indirectly constrain $r_t$.
\subsubsection{Extended Higgs sectors}\label{subsec:Spin0extendedHiggs}
The restrictive relation \eqref{rtrbrv}  could in principle be eased by extending the MSSM Higgs sector. The simplest extension is to add a new scalar singlet as in the Next-to-Minimal Supersymmetric Standard Model (NMSSM). However this yields a uniform reduction in all the couplings~\cite{Ellwanger:2009dp,Maniatis:2009re}. This is not sufficient for hiding top partners because the tradeoff for reducing the contribution to $r_G$ reduces also all other SM Higgs couplings, in particular $r_V$.  Introducing additional Higgs doublets could break the relationship of $r_V$ with other couplings and could thus better hide top partners from Higgs precision measurements.  Additional modifications to the Higgs sector that break the $r_c=r_t$ relation while avoiding flavor constraints is another interesting direction to pursue. We leave a detailed investigation to future work. 

\subsection{Spin-1/2}
\label{spinhalf}

\begin{figure}[t]
   \centering
  \includegraphics[width=0.3\textwidth]{feyn_2}~\includegraphics[width=0.3\textwidth]{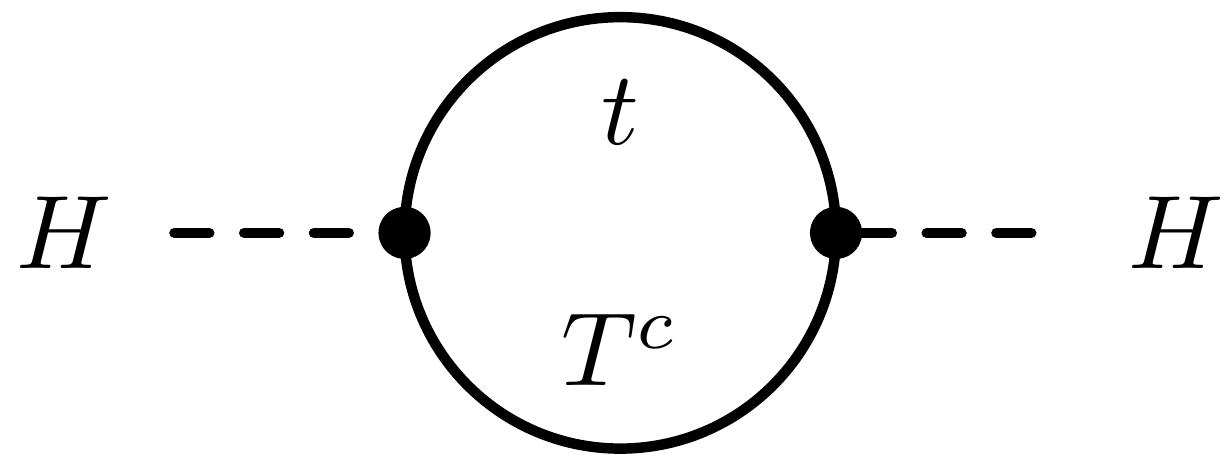}~ \includegraphics[width=0.22\textwidth]{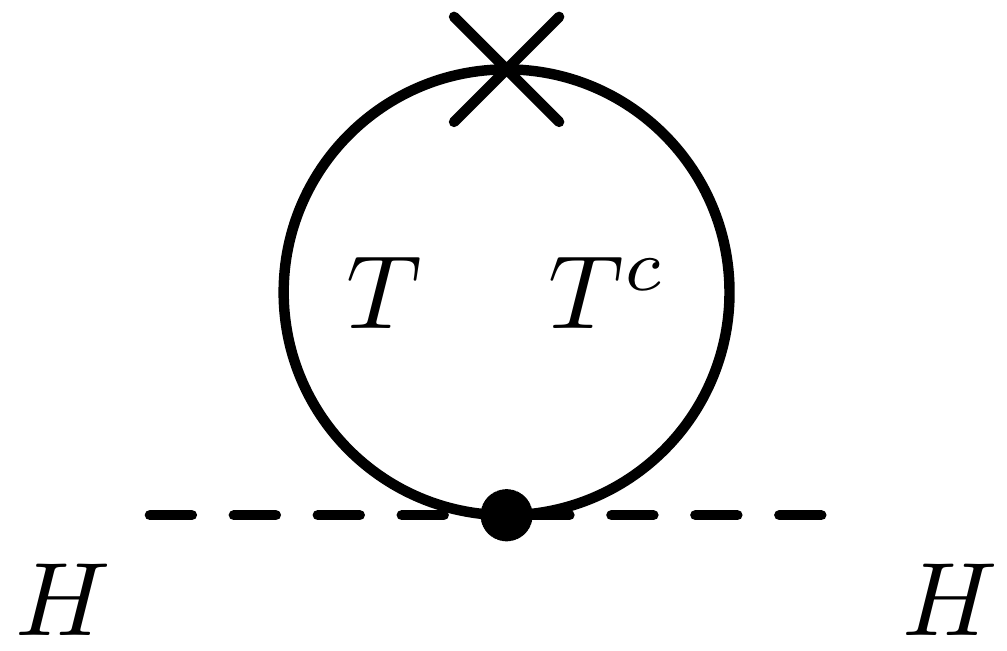} 
   \caption{Diagrams involved in the cancellation of top loop in a spin 1/2 top partner model.  The original one-loop diagram of SM top (left), the one-loop diagram with $HTt^c$ interaction (middle), and the one-loop diagram with a dimension-five $h^2 T T^c$ coupling and a $T$ mass insertion (right).}
   \label{fig:topcancel}
\end{figure}

Spin-1/2 top partners appear in Little Higgs  (LH) theories (see e.g. \cite{ArkaniHamed:2002qx, ArkaniHamed:2002qy, Kaplan:2003uc,Schmaltz:2010ac}) and Composite Higgs (CH) models (for recent reviews see e.g. \cite{Bellazzini:2014yua, Csaki:2015hcd}). In these theories, the Higgs is a pseudo-Nambu-Goldstone-boson (PNGB) of a larger symmetry that is collectively broken, ensuring the cancellation of one-loop quadratic divergences.  This is a different symmetry realization than spin-0 that ensures the cancellation in \eqref{cancel}, and thus the diagrammatic cancelation also is different in the low-energy effective field theory (EFT).  For fermionic top partners, the cancelation occurs because of a higher-dimension interaction between the top-partner and Higgs, unlike the spin-statistics cancellation with renormalizable terms for spin-0. For instance, if the fermionic top partner, $T$, is a singlet under $SU(2)$,  one can add a dimension-five operator $h^2 T T^c$  in addition to the allowed renormalizable interactions.  Diagramatically a cancellation can occur as shown in \figref{topcancel}.  The collective symmetry breaking ensures the couplings of the various terms are appropriately related to preserve the cancelation. Since the Higgs is realized as a PNGB it can be parametrized by an EFT expansion with the Higgs field residing in a nonlinear-sigma-model (NLSM) field and an expansion scale $f$ with cutoff $\Lambda \sim 4\pi f$. 

 Rather than investigating a complete model, we focus on the physics of the fermionic top partner's cancellation of quadratic divergences.  We start with the simplest spin-1/2 top partner extension, a singlet fermionic top partner, $T$ under the EW gauge group. The Lagrangian of the top sector takes the form,
\begin{equation}
\mathcal L_\text{top} 
=  (T, t)\, M\, \begin{pmatrix} T^c \\ t^c \end{pmatrix}+\text{h.c.}\,,
\label{eq:topyukawa}
\end{equation}
where $M$ is a $2\times 2$ mixing matrix of the top/top-partner, and $t^c$ and $T^c$ are the right-handed top/top-partner conjugates.  We assign $t$ and $t^c$ with $SU(2)_{L}$ charge $(t, t^c)=(\bf{2, 1})$ as in the SM.
 The top Lagrangian, before EWSB, in the mass eigenbasis up to $\mathcal O ({1}/{f^2})$ is then restricted to be
\beq
\mathcal L_\text{top}  =  (T, t)\begin{pmatrix} M_1 -a H^2 /f & -b H^2  /f \\ c \left(H - c' H^2/f^2 \right) & d H \left(1- d' H^2/f^2\right) \end{pmatrix}\, \begin{pmatrix} T^c \\ t^c \end{pmatrix}+\text{h.c.}\,,
\label{eq:top1}
\eeq
where \{$a$, $b$, $c$, $d$, $c'$, $d'$\} are dimensionless real coefficients obtained from the expansion of a NLSM field and $M_1$ is a bare mass that can exist for singlets. $H^2$ is the shorthand notation for $H^\dagger H$.  The mass matrix given in \eqref{top1} shows that $T$ is massive and $t$ remains massless before EWSB. The cancellation of diagrams in~\figref{topcancel} requires the following relationship to be satisfied
\begin{equation}\label{eq:equality}
2 a M_1 /f= c^2+d^2\,.
\end{equation}

After EWSB, $H$ gets a VEV, which generates additional mixing in the top/top-partner sector, requiring further rotation to switch to the mass eigenbasis. This gives electroweak-scale masses to the top partner and top of
\begin{equation}
m_T=M_1\left[1-\frac{v^2}{f^2}\left(\frac{f^2}{M_1^2}\right)\frac{d^2}{2} +\mathcal{O}\left(\frac{v^4}{f^4}\right)\right],\quad
m_t=  v d\left[1+\frac{v^2}{f^2}\frac{\epsilon_t}{2}+\mathcal{O}\left(\frac{v^4}{f^4}\right)\right]
\label{eq:mtafter}
\end{equation}
with $\epsilon_t=\left[ -4 a^2 c^2 d+4 a b c (c^2+d^2)+2 d^2 d'
   (c^2+d^2)^2 \right]/(c^2+d^2)^2$. 
In the limit of a heavy top/top-partner, $\mathcal{N}_{T}$ defined in \eqref{rGhatt} is given 
by 
\begin{equation}
\mathcal{N}_{T}=-\frac{m_t^2}{m_T^2}\,,
\label{eq:rTG}
\end{equation}
up to $\mathcal{O}({v^2}/{f^2})$ corrections. The relation is very similar to the degenerate stop case in \eqref{deg}. Furthermore this is a negative-definite quantity. Given a lower limit on $\mathcal{N}_{T}$ ($-1<\mathcal{N}_{T}<0$) from Higgs precision measurements, we can use \eqref{rTG} to constrain $m_T$. 

\subsubsection{Concrete models: Little Higgs models with one Higgs doublet}\label{subsubsec:concrete-models-spin-1/2}

Two classes of concrete LH models exist in literature:
the Simplest Little Higgs (SLH) models~\cite{Kaplan:2003uc,Schmaltz:2004de,Schmaltz:2010ac} and the Littlest Little Higgs (LLH) models~\cite{ArkaniHamed:2002qx, ArkaniHamed:2002qy}.
 In the $SU(3)$ SLH and the $SU(5)$ LLH, there is only a single Higgs doublet.  
A generic feature of these models is that $r_t\leq 1$. 
Since we need to have $r_t > 1$ to compensate for the negative definite $\mathcal{N}_{T}$, constraints on top partners from Higgs precision data cannot be weakened by adjusting $r_t$.  

\subsubsection{Concrete models: Little Higgs models with two Higgs doublets}\label{subsubsec:spin-1/2-2HDM}

If the Higgs sector is extended to a 2HDM, as in the $SU(4)$ SLH model, then $r_t>1$ is possible.  
In contrast to the MSSM case, we can explore also other types 
of 2HDM, but these cannot weaken appreciably constraints on top partners. 
For type-III (lepton-specific), this limitation is due to the restriction $r_t=r_b$, so that a precise measurement of $hb\bar{b}$, which restricts $r_b$, also indirectly constrains $r_t$. For type-IV (flipped), a similar limitation arises from $r_t=r_\tau$.  For type-I, both limitations exist. For type-II models, there is no restriction between $r_t$ and $r_b$ or $r_\tau$. It can be tuned independently and hence is helpful to hide the top partner. In all these cases, it is important to note that $r_t=r_c$, which results in competitive indirect constraints on $r_t$ through a precise measurement of $r_c$ at future lepton colliders. In the sections below, we will focus on LH with type-II 2HDM.

\subsubsection{Top-partners with additional resonances}\label{subsubsec:spin-1/2-2HDM-resonances}

The generic prediction of $r_t<1$ is tied to the NLSM nature of the Higgs fields, however this is just a low-energy EFT description.  In principle one could imagine that a strongly coupled UV complete description of the theory contained additional resonances similar to QCD where the Higgs as ``pions" would come along with the $\rho$ mesons and other resonances.  The Yukawa coupling of the top-quark to the Higgs in this low-energy EFT theory should then be thought of as an effective form factor.  Similar to QCD if these additional resonances were introduced they could come with different signs and change the generic relation of $r_t<1$.  However, this is not a standard prediction of the low-energy theory and would require a more complete model~\cite{Peskin:2017} to investigate the constraints.  Nevertheless, this is an interesting model building direction for the near future given it is the strongest avenue for maintaining naturalness in the basic top-partner sector.

\subsubsection{Extended fermionic top partners sectors}\label{subsec:better-models}
Another promising direction for naturalness in fermionic top partner models is to extend this sector itself with additional top partners.
The Lagrangian in \eqref{topyukawa} can be extended trivially to 
multiple top partners with degenerate top-partner masses and couplings.\footnote{If $N$ top partners have identical masses and Higgs couplings, $g_{hTT}$, the product $N g_{hTT}$ needs to be kept invariant to cancel the Higgs mass loop. The same factor occurs inside the gluon fusion loop and therefore all our arguments in the previous section remain valid.} We investigate now the effects of non-degeneracy in masses and couplings in the case of two spin-1/2 top partners.  For this case, we consider the mass matrix to the same order in $1/f$ given by

\begin{equation}
M=\begin{pmatrix} M_1 -  a_{11} H^2 /{f}  & 0 &0 \\  0 &  M_2-{a_{22}}H^2/{f}  & 0  \\ 0 & 0& a_{33} H \end{pmatrix}\,,
\end{equation}
where again the $a$'s are dimensionless coefficients from a NLSM field expansion as before. In general off-diagonal terms would also be present and require a symmetry to be forbidden. This generality would cause increased mixing between the SM top and top-partners making our choice more conservative.

The cancellation of Higgs mass loops requires the relation
\begin{equation}
2(M_1 a_{11}+M_2 a_{22})=f a^2_{33}\,,
\end{equation}
and consequently the total contribution from top partners to $hgg$ is given by
\begin{equation}
\mathcal{N}_{T}=-m_t^2 \left(\frac{\rho }{{m_{T_1}^2}}+\frac{1-\rho}{{m_{T_2}^2}}\right)\,, 
\label{eq:2toprg}
\end{equation}
where 
\begin{equation}
\rho \equiv \frac{2 M_1 a_{11}}{f a^2_{33}} 
\end{equation}
defines the ``fraction" of the cancellation coming from the $T_1$ loop. 
It is interesting to note that for $\rho> 1$, $T_2$ and the SM $t$ yield the same-sign contribution to the quadratic divergence of Higgs mass, which is cancelled entirely by $T_1$.  This scenario, if realizable in a complete model, would allow for tuning $r_G$ to the SM value without affecting other Higgs precision data. This happens when

\beq
\frac{\rho}{m^2_{T_1}}=-\frac{1-\rho}{m^2_{T_2}},
\label{eq:cancellaF}
\eeq
for $\rho$ in \eqref{2toprg}. This allows for a stealth region to avoid Higgs precision measurement, which is unavailable with a single spin-1/2 top partner, and is 
reminiscent of the stop blind spot. 
However, unlike the stop blind spot, the parameter space that is open at low masses for spin-1/2 top partners is very 
small (see Section \ref{subsec:multiple-spin-1/2}).  We thus do not explicitly investigate if Higgs decays to spin-1/2 top partners can 
constrain it further.  
Instead, we investigate in Appendix~\ref{blindspot} how future precision measurements of the $Zh$ cross section can 
probe the stealth region. 

\subsection{Spin-1}
\label{spinone}

For a spin-1 top partner, the cancellation between the top  and its partner again relies on the two being in the same multiplet of a larger symmetry.  As in the case of the spin-0 top partner, the only symmetry that can do this is SUSY~\cite{Haag:1974qh}.  However, this immediately presents a challenge: the top lives in a vector multiplet as a ``gaugino", which should be in a real representations of a gauge symmtetry.  Ref.~\cite{Cai:2008ss} proposes  a way around: the gauge symmetry is enlarged beyond the SM and broken in a way such that the heavy gauge bosons transform in other representations of the residual unbroken SM symmetries.   The original Cai-Cheng-Terning (CCT) model~\cite{Cai:2008ss} includes a breaking of  $SU(5)\rightarrow SU(3)\times SU(2)\times U(1).$\footnote{Not to be confused with the SM symmetries, which result from a subsequent symmetry breaking step when mixed with other gauge groups.} 
This generates massive $X,Y$ gauge bosons
that can be identified as the spin-1 partners of the left-handed top.\footnote{The usual problems of $X, Y$ gauge bosons are avoided because this model is not solely an $SU(5)$ and the gauge symmetry has to be enlarged.}  This in turn requires  the left-handed  top to be a gaugino of the enlarged gauge symmetry. 

The structure of a spin-1 top partner makes the connection between Higgs precision and colored naturalness more tenuous because many other particles must be introduced to generate the correct interactions.  
For instance, the $h t \bar{t}$ interaction must arise from a gaugino interaction of the form 
\beq
 g \lambda^{a} \phi^* T^a \psi\,.
\eeq
This in turn requires: (1) the Higgs must be inside a representation of the larger gauge symmetry and (2) the Higgsino needs to be identified as the right-handed top to generate a Yukawa type SM interaction.   This then dictates that the gauge coupling of the enlarged gauge symmetry is identified with the top Yukawa coupling. 
Those requirements necessarily introduce other interactions that contribute to the Higgs mass tuning. In~\cite{Cai:2008ss}, the Higgsino partners in the multiplet of the top-quark and the gauge bosons of the $SU(2)$ give a quadratic contribution to the Higgs potential proportional to the top Yukawa and their masses.

Nevertheless, it is still meaningful to ask what is the correction to the Higgs couplings from a spin-1 top partner alone, as for instance in~\cite{Collins:2014pba}.  
Determining this will give us a conservative lower estimate on tuning as there is necessarily additional large tuning coming from the same interaction term but with particles not included here.\footnote{This can be contrasted to the spin-0 or spin-1/2 scenarios, where additional tunings to the Higgs mass exist other than from the top partners, but they are controlled by other interactions.}  
Keeping this in mind, we investigate a particular implementation, the CCT model~\cite{Cai:2008ss, Collins:2014pba}. In this model, the symmetry group in the UV is $SU(5)\times SU(3)\times SU(2) \times U(1)_H \times U(1)_V$. We refer the detailed description of the field content to~\cite{Cai:2008ss}. The relevant terms in the Lagrangian that lead to Higgs coupling are,
\begin{equation}
\mathcal{L}\supset |D_{\mu}\bar{H}|^2 +\sqrt{2}\hat{g_5} \bar{H}^* T^a \lambda_a \widetilde{\bar H}\,,   
\end{equation}
where $D_{\mu}=d_{\mu}-i \hat{g_5} A^a_{\mu} T_a$, $\bar{H}=(t_R,H_d)$, $\t H=(\t t_R, \t H_d)^T$, and $\lambda^a$ are gauginos corresponding to the broken $SU(5)$. After the UV symmetry group breaks to the SM gauge groups, we are left with multiple heavy gauge bosons (and gauginos), which consist of heavy gluons $G'$,  $\vec{Q}$ are top-partners transforming like the SM left-handed top, heavy $SU(2)$ gauge boson $W'$, and $U(1)$ gauge bosons, $B'$ and $B''$.
For Higgs precision studies, it turns out that \emph{only}  $\vec{Q}$ and $W'$  are relevant. Their color and electrical charge are $(N_{c, \vec Q}, Q_{\vec Q})=(3, 2/3)$ and $( N_{c, W'}, Q_{W'})=(1,1)$.
After EWSB, we get the Higgs from $H_d=(0,(v_d+h)/\sqrt{2})^T$ (note that it is a down-type Higgs doublet $H_d$,  not a up-type Higgs doublet $H_u$ as in a spin-0 model). The relevant Lagrangian is
\begin{equation}
\mathcal{L}\supset \f{1}{2} \hat{g_5^2}h^2 \vec{Q}^2+\f{1}{2} \hat{g_5^2}h^2W'^+ W'^-+\hat{g_5} h t_R t_R^* + \text{h.c.}\,,
\end{equation}
where  $\hat{g_5} h t_R t_R^*$ is identified with the top Yukawa, and hence $r_t= \hat g_5\approx \sqrt{1+\tan^2 \beta}>1$.
 The $h^2\vec{Q}^2$ interaction modifies both $hgg$ and $h\gamma\gamma$ couplings, while $h^2 W'^+ W'^-$ affects only the latter. In principle, $m_{\vec Q}$ and $m_{W'}$ are uncorrelated. However, they should be around the same energy scale due to the their common origin. For simplicity, we enforce $m_{\vec Q} = m_{W'}$ in our fits. For $\mathcal{N}_{\vec Q}$ in \eqref{rGhatt}, we obtain 
\begin{equation}
\mathcal{N}_{\vec Q}=-\frac{1}{\cos \beta}\frac{21}{4}\frac{m_t^2}{m_Q^2}\,,
\label{eq:CCTrG}
\end{equation}
where $\tan \beta$ is the usual MSSM VEV ratio.   The large ${21}/{4}$ prefactor in \eqref{CCTrG} comes from the spin-1 loop function, see Appendix~\ref{loopcoupling}. Therefore, there are stricter limits on vector top partners compared to other top partners. While novel, this model requires a plethora of additional particles resulting in tuning penalties as well as large deviations in Higgs phenomenology. For this reason, we do not investigate further extensions.


\section{Results and Discussions}
\label{results}

In this section, we present the exclusion limits for various top-partner scenarios and their extensions. 
We derive current and projected bounds on top-partner masses by first considering the minimal case in which 
all Higgs-couplings are SM-like except that the top partners (and bottom partners, if present) 
can contribute to the $hgg$ and $h\gamma\gamma$ loops, i.e.~they affect $r_G$ and $r_\gamma$;  moreover, we include Higgs 
decays to top partners (and bottom partners) when allowed.   
We then additionally modify other Higgs couplings from their SM values, and numerically evaluate which modifications are most effective 
at hiding the top partners from Higgs precision studies.  
Finally, we consider constraints and projections for canonical models (like the MSSM) and extensions. 

As discussed in Section~\ref{s.data}, we use the existing results from the LHC and Tevatron to derive the current constraints, 
and we derive projected sensitivities based on projected Higgs coupling measurements at the future LHC Runs 3 and 4, 
the proposed electron-positron colliders ILC, CEPC, and FCC-ee, and the proposed proton-proton collider FCC-hh. 

\subsection{Constraints on top partners that only affect $hgg, h\gamma\gamma$ loops}\label{subsec:t-hat-only}

\subsubsection{Spin-0}
We begin with the spin-0 scenario discussed in Section~\ref{s.spin-0}.  Recall that we assume that there are two 
spin-0 particles, which we call stops, $\t t_1$ and $\t t_2$ and a light left-handed sbottom $\t b_1$; any other partner particles have been decoupled.  
(This limit is similar to the natural SUSY limit, although we do not include Higgsinos.) 
The stop mass eigenstate $\t t_1$ is mostly left-handed (i.e., $m_{\t t_1}$ contains mostly $m_{Q_3}$), 
while $\t t_2$ is mostly right-handed (i.e., $m_{\t t_2}$ contains mostly $m_{U_3}$). 

We first assume that the top partners are the only BSM contributions to the Higgs couplings $r_G$ and $r_\gamma$, and can 
contribute to $r_{\rm exo}$ through possible exotic Higgs decay to stops and sbottoms. 
The other Higgs couplings are fixed to their SM values, i.e., $r_t=r_b=r_\tau=r_V=1$ and $r_\text{inv}=\delta r_\gamma=\delta r_G=0$. 
The $\chi^2$-function in Eq.~(\ref{eq:chisqFunction}) is then only a function of $\mathcal{N}_{\hat t}$ (\eqref{rGhatt} 
or \eqref{Nt-spin-0}), $\Gamma_\text{tot} (h \to \t t \t t)$ (\eqref{gamma-h-stop}), and, possibly, $\Gamma(h \to \t b_1 \t b_1)$. 
To calculate the excluded parameter space in the $m_{\t t_2}$ versus $m_{\t t_1}$ plane, we proceed as follows.  
First, we fix $\tan\beta$.  Then for a given value of $m_{\t t_1}$ and $m_{\t t_2}$, we let $X_t$ take on values 
from 0 up to $X_t^{\rm max}$ in \eqref{Xtmax}.  For each value of $X_t$, we calculate $g_{h \t t_1 \t t_1}$,  
$g_{h \t t_1 \t t_2}$, $g_{h \t t_2 \t t_2}$,  
$g_{h \t b_1 \t b_1}$,  $m_{\t b_1}$, 
with which we then calculate $\mathcal{N}_{\tilde{t}}$, 
$\mathcal{N}_{\tilde{b}}$, 
$\Gamma_\text{tot} (h \to \t t \t t)$, $\Gamma (h \to {\t b_1} {\t b_1})$, and check that $m_{\t b_1}$ is real. 
This determines $r_G$, $r_\gamma$, and $r_{\rm exo}$, 
which are used, together with the Higgs precision data described in Section~\ref{s.data}, as inputs to the $\chi^2$-fitting procedure.  
If no value of $X_t$ can be found for which the $\chi^2$ is satisfactory, the chosen $m_{\t t_1}$ and $m_{\t t_2}$ values are disfavored. 

\begin{figure}[t!]
   \centering
   \includegraphics[width=0.48\textwidth]{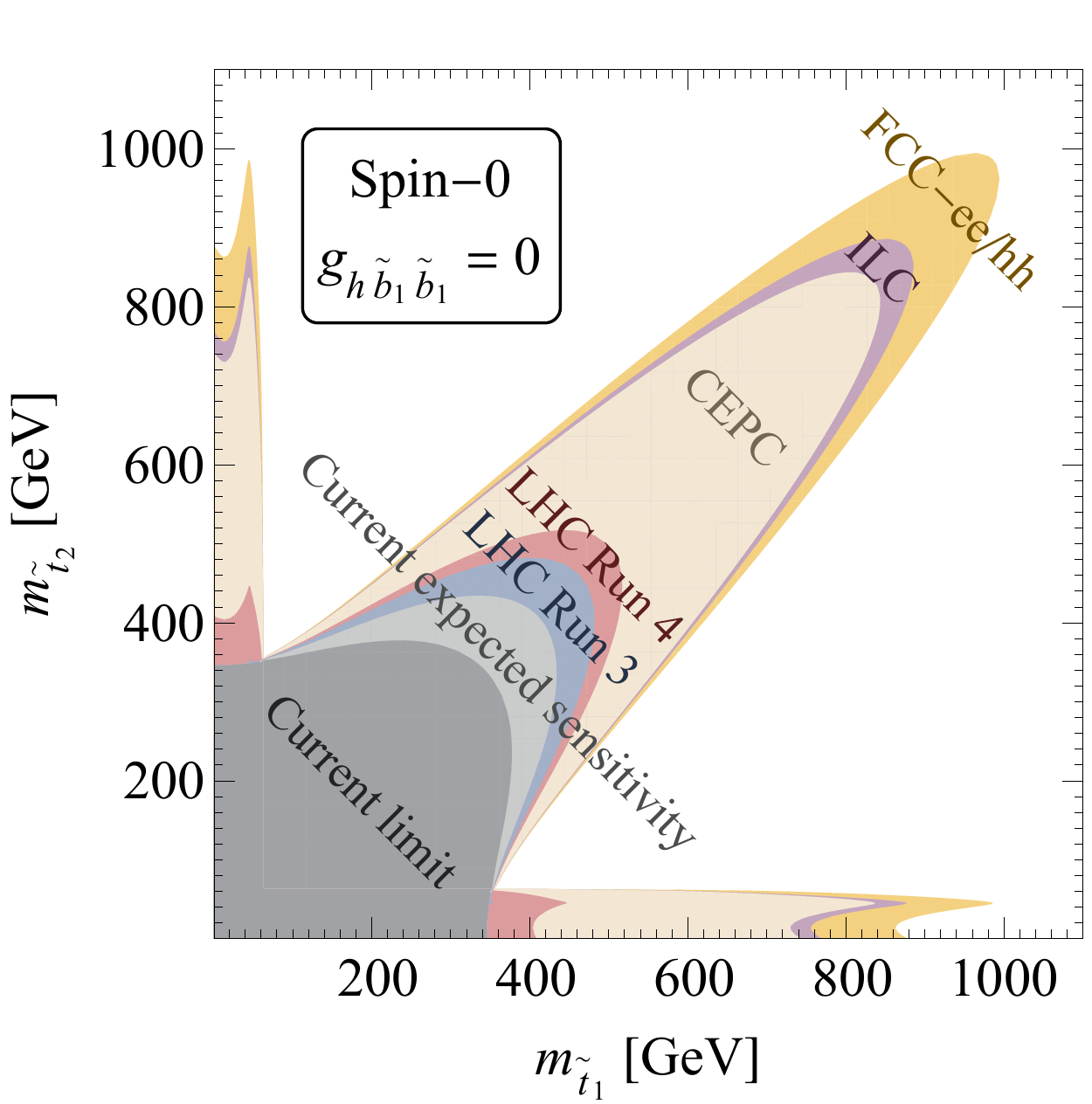}
~~~    \includegraphics[width=0.48\textwidth]{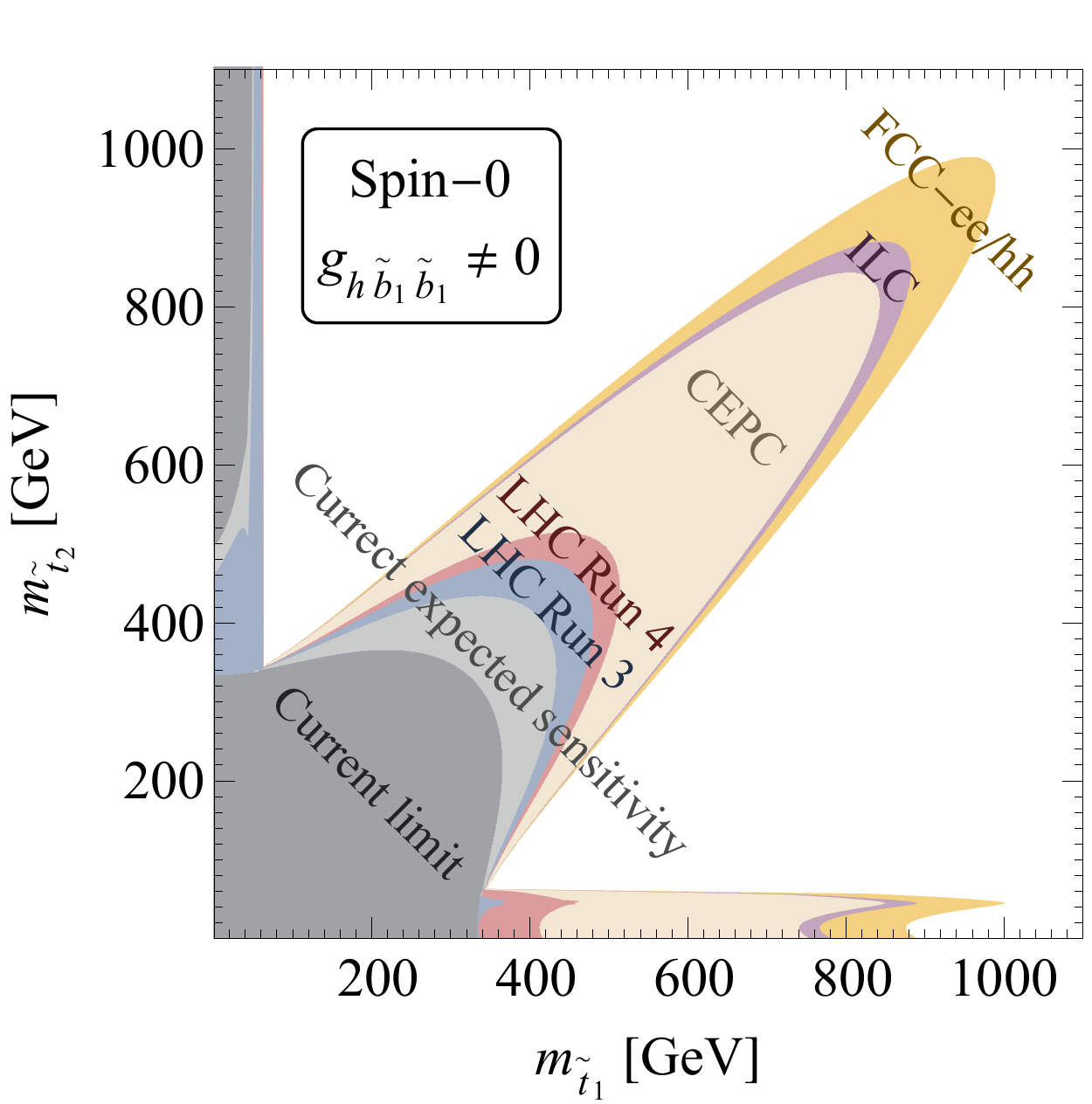}
   \caption{
   Excluded parameter space and expected sensitivities at the $2\sigma$ CL~of current (gray) and future data (various colors) 
   for spin-0 top-partners in the $m_{\t t_2}$ versus $m_{\t t_1}$ plane.  
   In the {\it left} plot, we assume $\tan \beta \simeq 1$ and $h \t b_1 \t b_1$ coupling vanishes (\eqref{gnb1b1=0}), while in the {\it right} plot, 
   $\tan \beta$ is large 
   to maximize the $D$-term contributions in the stop and sbottom sector (\eqref{gnb1b1-D-term}). 
   We assume that top partners are the only BSM contributions to the Higgs couplings and can 
   contribute to exotic Higgs decay through $h \to \t t \t t$ and, possibly, $h \to \t b_1 \t b_1$.  
   The other Higgs couplings are fixed to their SM values. 
   For both plots, we require $m_{\t b_1}$ to be real in the allowed region. 
}
   \label{fig:stopstop}
\end{figure}

In \figref{stopstop} we show the resulting excluded parameter space (dark gray region) from current LHC and Tevatron data.  
The expected sensitivity from the current data is shown in light gray, while the expected sensitivities 
from future collider data is shown in various colors.   
We consider two choices for $\tan\beta$ that are representative of the possible range for the phenomenology: 
\figref{stopstop} (left) shows the constraints without any sbottom contribution 
(i.e.~$\tan\beta\simeq 1$, $g_{h  \t b_1 \t b_1} = 0$, \eqref{gnb1b1=0}), 
while \figref{stopstop} (right) shows the constraints when sbottom contributions are included 
(i.e.~large $\tan\beta$, with $g_{h  \t b_1 \t b_1}$ given by the $D$-term contribution, \eqref{gnb1b1-D-term}).  
Note that even for the left plot, although we set $g_{h  \t b_1 \t b_1} = 0$, we require that the choice of stop-sector masses and mixing 
allow the left-handed sbottom to be real, see \eqref{sbottom-mass}.

As anticipated in Section \ref{s.spin-0}, the lower bounds on the masses are strongest for $m_{\t t_1} = m_{\t t_2}$ 
and weaker for split masses.  
The constraints and projections along the degenerate direction for high masses arise dominantly from the presence of the two stops in 
the $hgg$ and $h\gamma\gamma$ loops.  Comparing the two plots in this region, we see that the $D$-term contribution in the stop 
mass matrix \eqref{stop1} and in the Higgs-stop-stop couplings Eqs.~(\ref{eq:ght1t1})--(\ref{eq:ght1t2}), as well as including the 
sbottom contribution, only slightly extends the constraints and projections at the $\mathcal{O} (1\%)$ level. 
When one of the stops becomes lighter than half the Higgs mass, constraints arise from $h \to \t t \t t$ (left plot) and from 
$h \to \t t \t t$ and $h \to \t b_1 \t b_1$ (right plot).  
If one of the stops becomes heavy, the coupling of the Higgs to the lighter stop with mass below $m_h/2$ becomes small and the 
Higgs decay to the lighter stop vanishes.  
However, in the presence of a light left-handed sbottom (corresponding to a light left-handed stop, $\t t_1$), the Higgs decay width to sbottoms is large; 
while the current data is unable to rule out the $m_{\t t_2}<m_h/2$ region entirely, future LHC Run 3 data can sufficiently constrain 
exotic Higgs decays to probe this region completely.  

\begin{figure}[t]
   \centering
   \includegraphics[width=\textwidth]{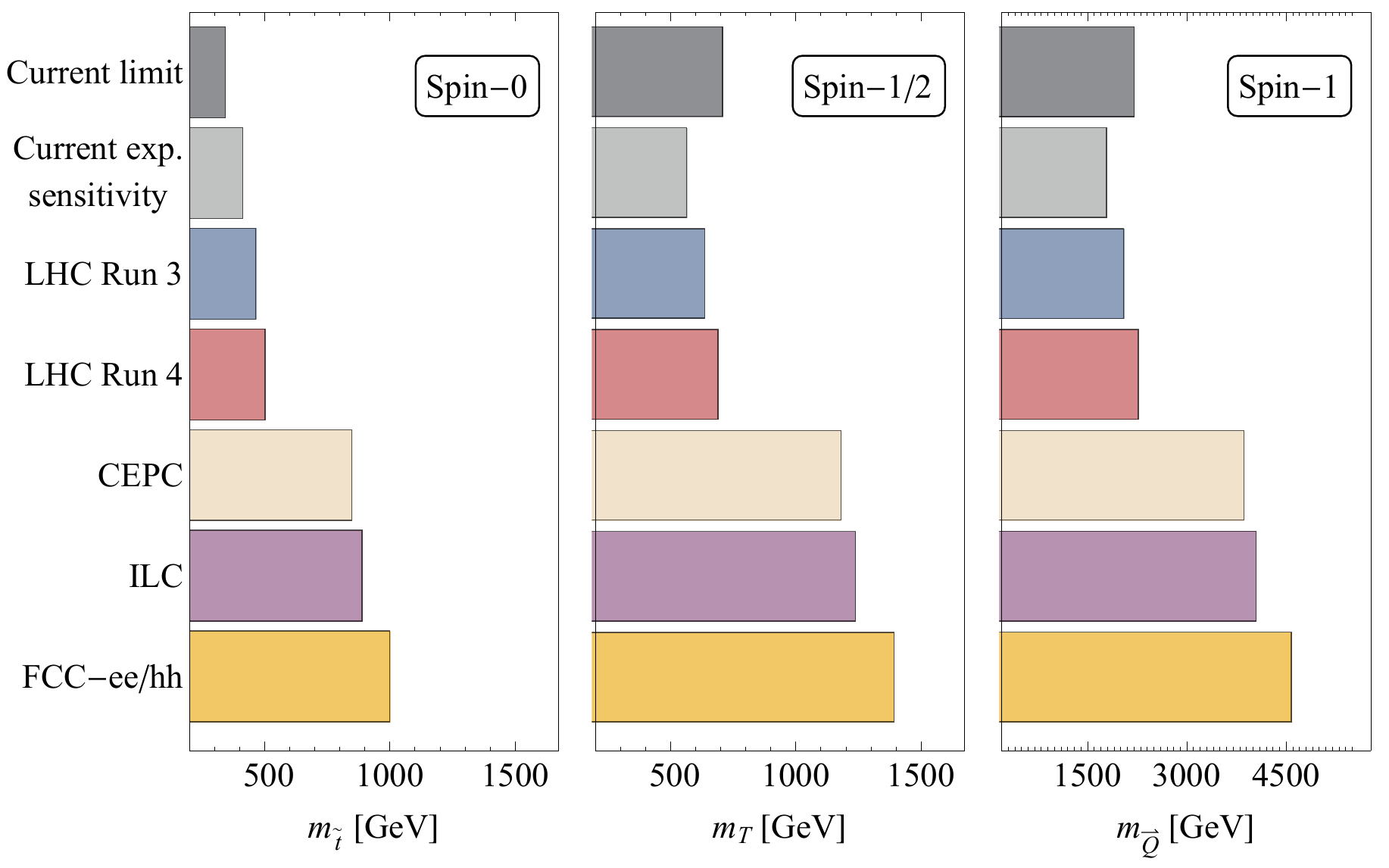} 
   \caption{
   Excluded parameter space and expected sensitivities at the $2\sigma$ CL~of current and future data 
   for spin-0 (left), spin-1/2 (middle), and spin-1 (right) top-partners.  
   We assume that the two spin-0 top partners are degenerate in mass, $m_{\t t_1} = m_{\t t_2} \equiv m_{\t t}$. 
   We assume that top partners contribute only in the $hgg$ and $h\gamma\gamma$ loops, there 
   are no modifications of the Higgs couplings to other SM particles, and there are no exotic or invisible Higgs decays.  
   The parameter space excluded by current LHC and Tevatron data is shown in dark gray, while the expected sensitivity of the current data 
   is shown in light gray.  
   Future LHC runs and the proposed future colliders (ILC, CEPC, and FCC-ee/hh) are shown in various colors.     
   }
   \label{fig:spinzerohalfone}
\end{figure}

\subsubsection{Comparison of Constraints between Spin-0, Spin-1/2, and Spin-1}

To compare constraints on spin-0 particles with constraints on spin-1/2 and spin-1, we focus on the degenerate direction for 
spin-0 ($m_{\t t_1} = m_{\t t_2}$), because our canonical spin-1/2 and spin-1 models only have a single top partner.  
Recall that along the high-mass spin-0 degenerate direction, the contributions from the left-handed sbottom and 
from stop $D$-terms only matter at 
a few-percent level. 
For the remainder of Section 6, we set $g_{h  \t b_1 \t b_1} = 0$, but require that the choice of stop-sector masses and mixing allow 
the left-handed sbottom to be real, see Section~\ref{s.spin-0} (note that we include $D$-term contributions in the stop-sector, i.e., large $\tan\beta$).

In \figref{spinzerohalfone} we show the current constraints and expected sensitivities 
for degenerate spin-0 (left), spin-1/2 (middle), and spin-1 (right) top-partners.  
The current constraints from Tevatron and LHC data for these different spin-states are about 350~GeV,  700~GeV, and 
2.2~TeV, respectively. 
The LHC Run 4 is expected to improve on these by a few hundred GeV, while the possible future ILC, CEPC, and FCC-ee/hh 
are expected to improve by another few hundred GeV for spin-1/2 and by almost 2 TeV for spin-1. 
These projected sensitivities probe similar parameter space to current direct searches, but of course with fewer 
model assumptions. Due to the current data preferring $r_G>1$, the current constraints are weaker (stronger) for spin-0 (spin-$\frac{1}{2}$ and spin-1) models compared to their expected sensitivities. 

The constraints on spin-1 top partners are much stronger than for spin-0 and spin-1/2 states.  
The tuning from the spin-1 state alone is already significant given that the current limit on the top-partner mass is already 
approaching 2 TeV.  Moreover, as discussed in 
Section~\ref{spinone}, a contribution to the tuning should be included from the other scalars and vectors that are required in 
spin-1 top-partner models.   
We will thus not consider spin-1 top partners further, focusing instead on how best to hide spin-0 and spin-1/2 partners from 
Higgs precision measurements.  

\subsection{Constraints on top-partners with modified SM Higgs couplings}\label{subsec:constraints-with-nonSM}

\begin{figure}[t]
   \centering
   \includegraphics[width=0.66\textwidth]{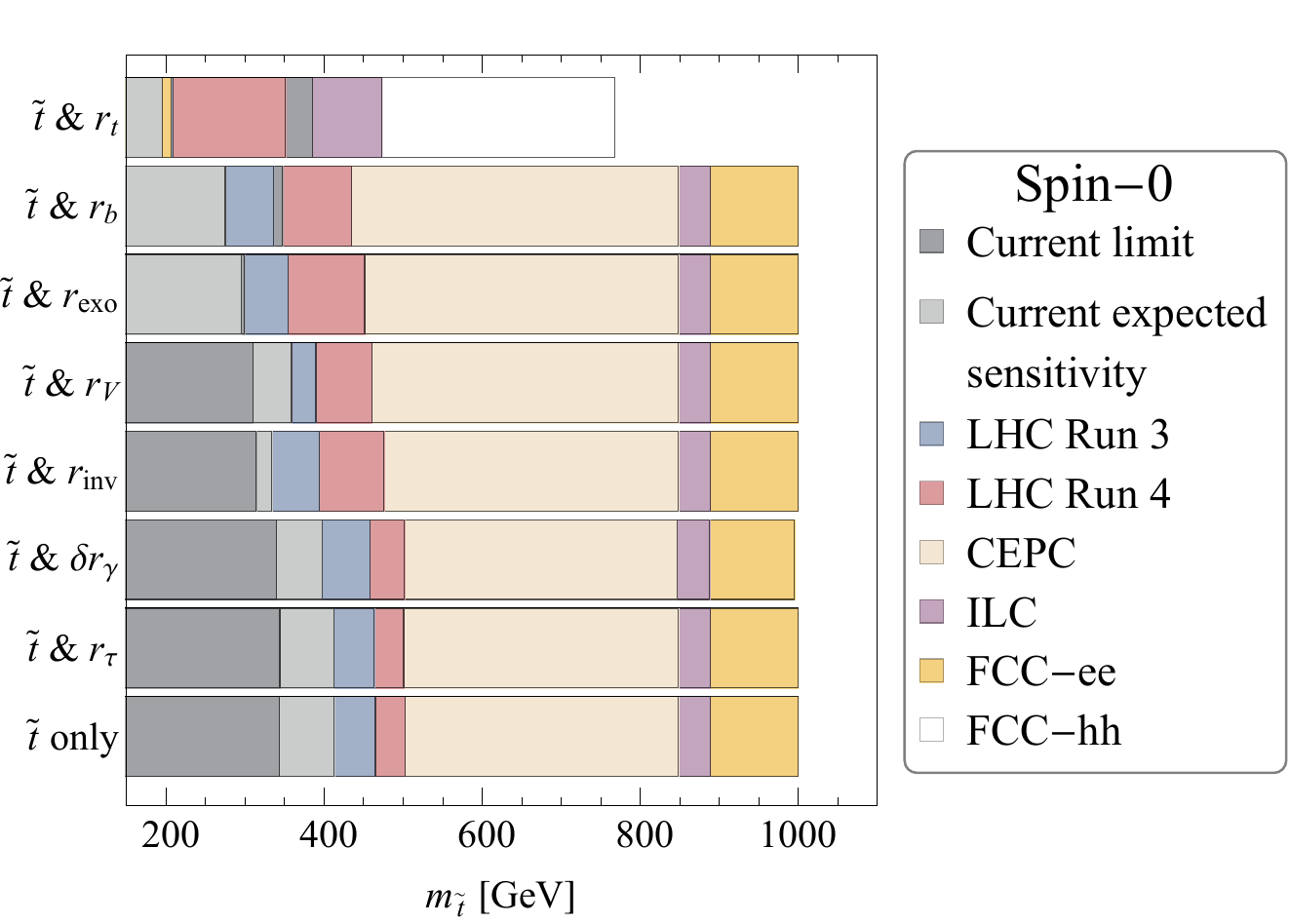} 
   \caption{Excluded parameter space and expected sensitivities at the $2\sigma$ C.L. on degenerate spin-0 top partner masses, $ m_{\t t_1} = m_{\t t_2} \equiv m_{\t t} $, 
   from various joint-fits of current and future data.  
   We assume here that in addition to top partners contributing in the $hgg$ and $h\gamma\gamma$ loops, there 
   is one additional modification to the couplings as indicated by the description on the left axis; for example, for ``${\t t} $ \& $r_t$", the 
   top-partner contributes to the $hgg$ and $h\gamma\gamma$ loops and $r_t$ is allowed to vary from its SM value, while all other 
   $r_j$ are fixed to their SM value.
Note that the current limit shaded in dark gray is naively stronger 
for ``$\t t ~\& ~r_t$'' and ``$\t t ~\&~ r_b$'' than the expected sensitivity of the future LHC Run 3 and/or Run 4 data (see text for details).  
}
   \label{fig:stopgeneral}
\end{figure}

In addition to the top-partners contributing to the $hgg$ and $h\gamma\gamma$ loops, the Higgs couplings to SM particles could 
also be modified from their SM values.  
In this section, we numerically quantify which modifications are most efficient at absorbing the top-partner-loop contributions. 
We allow for one Higgs coupling, $r_i \in \{r_t$, $r_b$, $r_\tau$, $r_V$,  $r_\text{exo}$, $r_\text{inv}$, $\delta r_\gamma$\}, 
to differ from its SM value, while fixing all other couplings to their SM values. 
To obtain the $2\sigma$~CL regions for the top-partner masses, we adjust their masses, 
while marginalizing over $r_i$, until their contributions to $hgg$ and $h\gamma\gamma$ expressed in 
terms of the variable $\mathcal{N}_{\hat t}$ in \eqref{rGhatt} gives the appropriate $\Delta \chi^2$.   
\
\begin{figure}[t]
   \centering
   \includegraphics[width=0.66\textwidth]{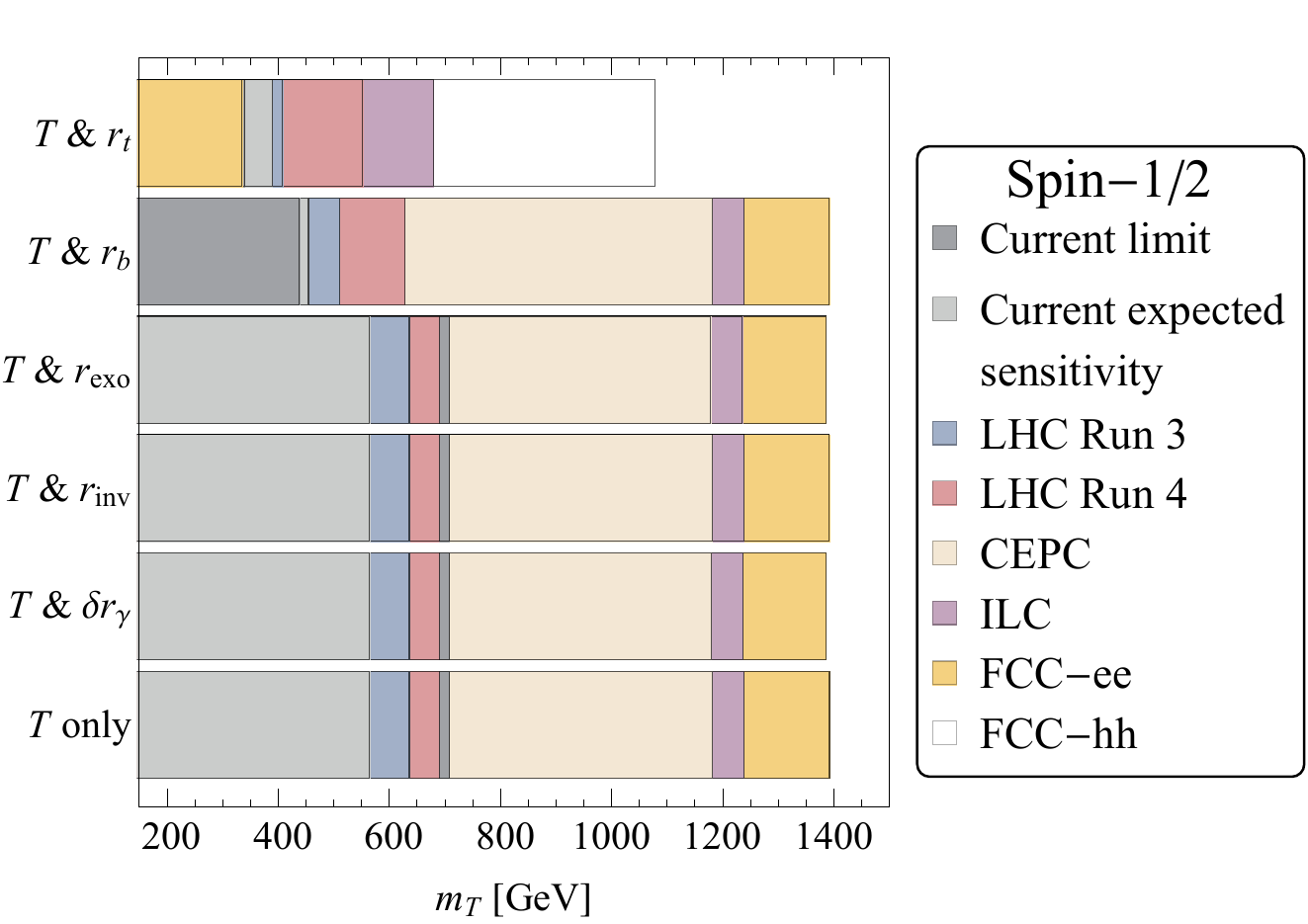} 
   \caption{Similar to \figref{stopgeneral}, but here showing excluded parameter space and expected sensitivities at the $2\sigma$ C.L. on spin-1/2 top partner mass, $m_T$, from various joint-fits of current and future data. 
      }
   \label{fig:LHgeneral}
\end{figure}

The results for spin-0 and spin-1/2 scenarios are shown in \figref{stopgeneral} and \figref{LHgeneral}, respectively. 
Not surprisingly, the additional degree of freedom helps in reducing the current top-partner bounds and projected sensitivities. 
As anticipated in Section~\ref{s.precision}, allowing for a non-SM $h t \bar{t}$ coupling is currently the best way to hide a top partner. 
Future LHC, ILC, and FCC-hh data will measure $h t \bar{t}$ production precisely improve constraints on top parters.

For the spin-0 ``$\t t$ \& $r_t$" and   ``$\t t$ \& $r_b$" scenario (see \figref{stopgeneral}), 
the current data naively excludes larger degenerate stop masses than the expected sensitivity of the data from the LHC Run 3 and 
Run 4. 
This is because some of the current Higgs data prefers $r_t = 1.18$ and $r_b = 0.89$ that are far away from 1.
Given~\eqref{myrg2}, i.e.,
\beq
r_G = 1.05 r_t (1 +\mathcal N_{\hat t})  + (- 0.05+0.07 i) r_b,
\eeq
a negative $(\mathcal{N}_{\t t})^\text{fit}$ is favored to remove $r_t > 1$ or $r_b < 1$. 
We find that the $2\sigma$ CL upper limit, $(\mathcal{N}_{\t t})^\text{fit;max}$, while not negative, is a small positive number.  
This leads to a relatively large bound on degenerate stop mass compared to the LHC Run 4 and Run 4 expected sensitivities, 
which, by definition, assume a measured $r_t=r_b=1$. 
It is thus better to compare future expected sensitivities to the current expected sensitivity indicated by the light-gray shaded region. 
We note that the current limit is weaker than the current expected sensitivity for the spin-1/2 ``$T$ \& $r_t$" scenario. This is because $\mathcal{N}_{T}$ is negative-definite from LH theories ($\mathcal{N}_{T}=-{m_t^2}/{m_T^2}$), and a 
smaller $m_T$ is preferred to cancel $r_t > 1$ or $r_b<1$. 

Besides varying the $h t \bar{t}$ coupling, varying the $hb\bar{b}$ coupling is the second most effective way to hide both spin-0 and spin-1/2 
top partners.  Exotic Higgs decays, invisible Higgs decays, and $hVV$ 
also help to hide top-partners 
for the spin-0 scenario, although to a lesser extent than varying either $r_t$ or $r_b$.  
For the spin-1/2 scenario, exotic or invisible decays do not help.  
The reason is that spin-1/2 top partners can only suppress the $hgg$ coupling, with exotic or invisible decays suppressing the signal 
strength further.  For spin-0 partners, the $hgg$ coupling can also be enhanced, in which case additional non-standard Higgs 
decay modes help hide the partners. 
Finally particles contributing to $h\gamma\gamma$, or non-SM $h\tau^+\tau^-$ decays, barely help to hide the spin-0 top partners. 

Nevertheless, as seen in \figref{stopgeneral} and \figref{LHgeneral}, those hiding methods are no longer effective for future colliders. Generally speaking, future colliders can measure precisely the 
$hb\bar{b}$ and $hVV$ couplings as well as the invisible and exotic Higgs-decay width. 
Therefore the modifications to SM Higgs couplings are very constrained, and the exclusion limits on top-partner masses are 
almost identical to the top-partner only cases.

Allowing several Higgs couplings to vary simultaneously would further weaken current constraints, but the effect would still be dominated by $r_t$.  
We thus consider concrete models next.

\subsection{Constraints on canonical models and extensions}  

\subsubsection{Spin-0 top partners in the MSSM}\label{subsubsec:mssm}

\begin{figure}[t!]
   \centering
 \includegraphics[width=0.52\textwidth]{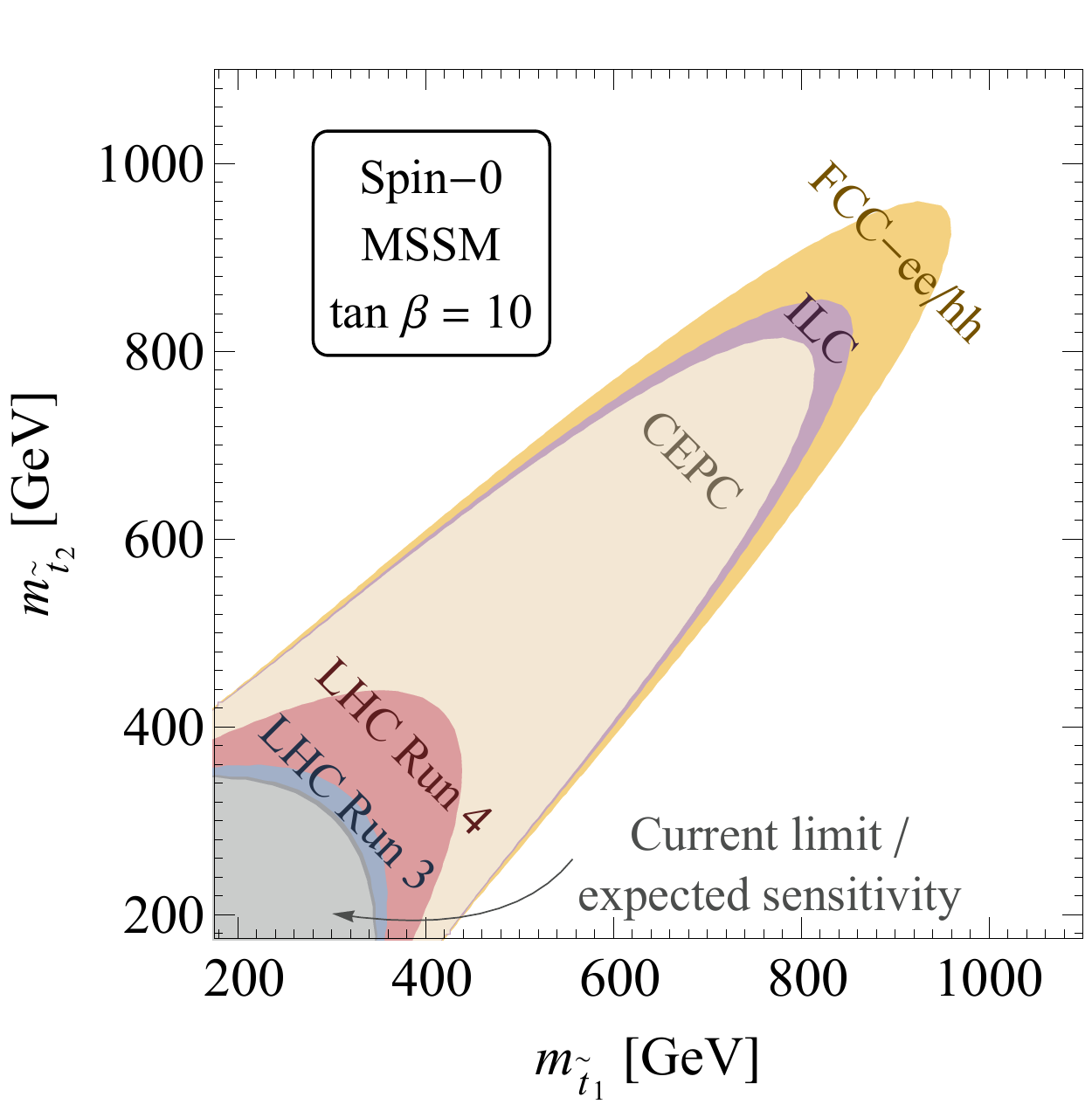} 
   \caption{
   Excluded parameter space and projected expected sensitivities at the $2\sigma$ C.L.~of current and future data, respectively,  
   for stops in the MSSM in the $m_{\t t_2}$ versus $m_{\t t_1}$ plane.  
   The parameter space formally excluded by current LHC and Tevatron data is shown in dark gray. It mostly overlaps with the current expected sensitivity in light gray.  
   Future LHC runs and the proposed future colliders (ILC, CEPC, and FCC-ee/hh) are shown in various colors. 
   }   \label{fig:stopstop2}
\end{figure}

We consider now constraints on stops in the context of the MSSM.  
We have seen that allowing $r_t<1$, i.e.~lowering the $h t \bar{t}$-coupling, currently provides the best way to hide spin-0 top partners. 
In the MSSM, we can lower $r_t$ by lowering $\tan\beta$. 
However, this forces $r_b$, $r_\tau$, and $r_V$ to vary in a correlated way, see \eqref{rtrbrv}.  
Since these parameters are well-constrained by current and expected future data, 
it is more difficult to hide stops in the MSSM than in models in which only $r_t$ is being varied.  
We also note that we set $r_t=r_c$, so that future measurements of $r_c$ indirectly constrain $r_t$.   

\begin{figure}[t]
   \centering
   \includegraphics[width=0.66\textwidth]{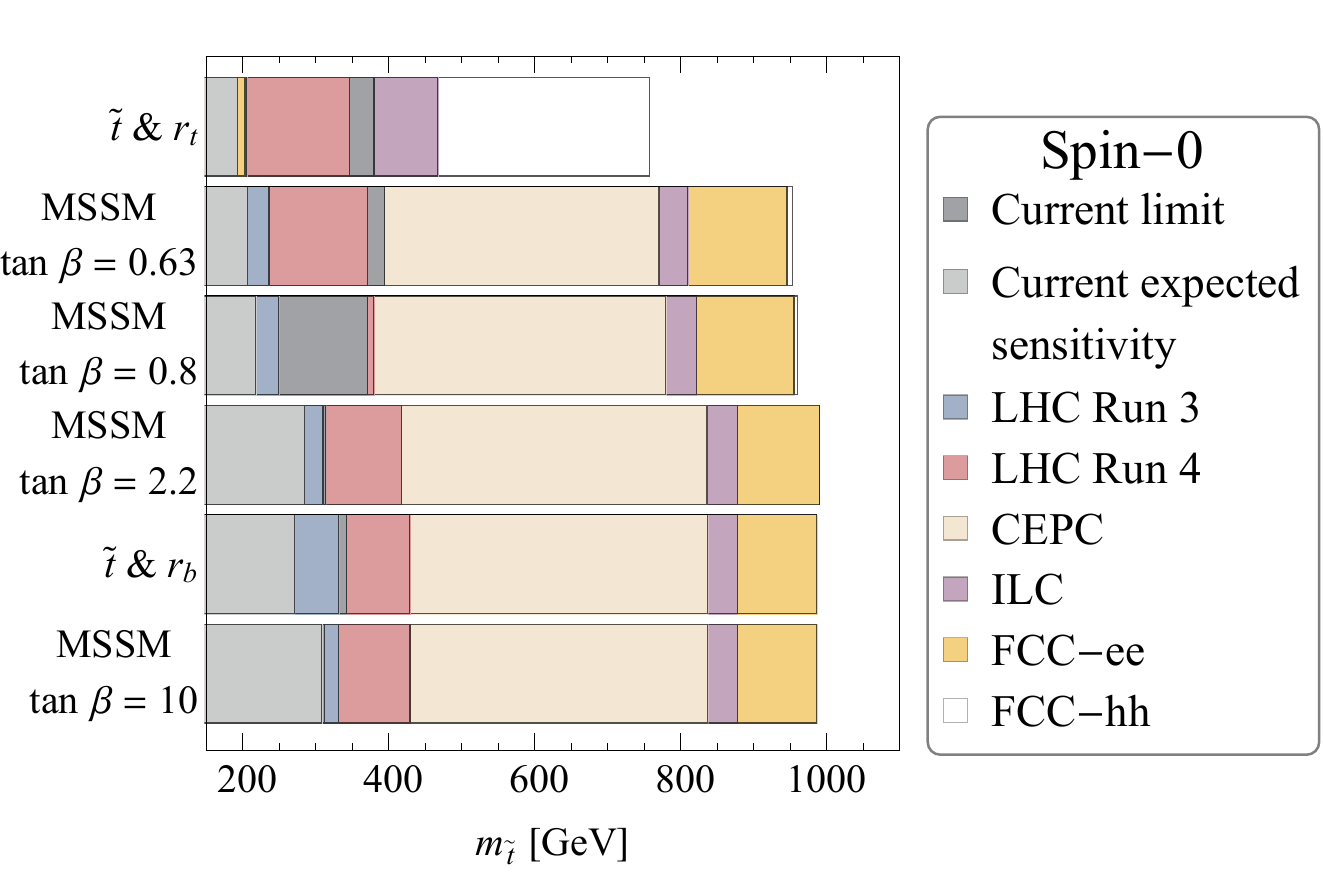} 
   \caption{
   Excluded parameter space and expected sensitivities at $2\sigma$ C.L. on degenerate spin-0 top partner mass, 
   $ m_{\t t_1} = m_{\t t_2}\equiv m_{\t t} $, in the 
   MSSM with various $\tan \beta$ of current and future data.  In the plots, we again show the (non-MSSM) ``${\t t}$ \& $r_t$" and ``${\t t}$ \& $r_b$" 
   results from \figref{stopgeneral} as a reference.  
   Note that the anomalously strong limits in the ``current data" fit (dark gray) are due to the current data favoring a 
   minimum with $r_t >1$ and $r_b <1$.}
   \label{fig:stop-mssm}
\end{figure}

We show in \figref{stopstop2} the constraints on the two stop-mass eigenstates in the MSSM, setting $\tan\beta=10$.  
We see again that the constraint is much stronger for degenerate stop masses compared to the case with large stop mixing, and more
generally is very similar to the case in which only the top-partner contributions to the $hgg$ and $h\gamma\gamma$ loops are included, 
see \figref{stopstop} (left). Note that like in Fig.~\ref{fig:stopstop} (left), we set the sbottom-sbottom-Higgs coupling to zero, but ensure that the sbottom mass is real. 
The anomalously strong current limits (dark gray) are due to the current data favoring a 
minimum with $r_b<1$; this is for the same reason as discussed in Section~\ref{subsec:constraints-with-nonSM} and seen in \figref{stopgeneral}.  

In \figref{stop-mssm}, we show current constraints and projected sensitivities on stops in the MSSM for various $\tan\beta$. Note that the constraints are significantly stronger compared to ``$\t t$ \& $r_t$" mostly due to the $r_t=r_c$ relation in the MSSM. Furthermore, we clearly see that lower values of $\tan\beta$ provide weaker projected sensitivities. This is because for small $\tan\beta$ we can vary $r_t$ without affecting $r_b$ or $r_V$, which are measured very well. 
We cannot lower $\tan\beta$ too much without the top Yukawa reaching a Landau pole near the weak scale. 

It is worth comparing the MSSM scenario with $\tan \beta=10$ in Fig.~\ref{fig:stop-mssm} and the ``$\t t$ \& $r_b$" scenario from \figref{stopgeneral} and repeated in Fig.~\ref{fig:stop-mssm}.  
Naively they are very similar given that $\tan\beta$ is large, which fixes $r_t=r_V=1$, but leaves $r_b$ as free parameter.  
In the ``$\t t$ \& $r_b$" scenario, $r_b$ is constrained by $h b\bar b$ only.  In the MSSM scenario, it is additionally restricted by $h\tau^+\tau^-$ measurements due to the type-II 2HDM structure that restricts $r_b=r_\tau$.   The current data prefers $r_b<1$, but the additional restriction $r_b=r_\tau$ together with the current data preferring a value for $r_\tau$ just slightly above 1, forces $r_b$ to be closer to 1.  This results in weaker exclusions in the MSSM scenario on $m_{\t t}$.  When $r_b =1$ (as is assumed when calculating the current and 
future projected sensitivities), the additional restriction is less important when comparing the two scenarios. 

\subsubsection{Spin-1/2 top partners with one Higgs doublet} 

In LH models with only a single Higgs doublet, such as the $SU(3)$ SLH and $SU(5)$ LLH models, we have $r_t\le 1$.  
As discussed in Section~\ref{subsubsec:concrete-models-spin-1/2}, since $\mathcal{N}_{T}$ is negative-definite, 
$r_t<1$ does not help to hide the top partner compared to what is shown in Fig.~\ref{fig:LHgeneral}, 
which marginalizes over all values of $r_t$. 
These theories thus prefer $r_t = 1$, and \figref{spinzerohalfone} shows the resulting constraints on the mass of the top partners up 
to 1.4~TeV at FCC-ee/hh. 

\subsubsection{Spin-1/2 top partners in type-II 2HDM} 

\begin{figure}[t]
   \centering
   \includegraphics[width=0.66\textwidth]{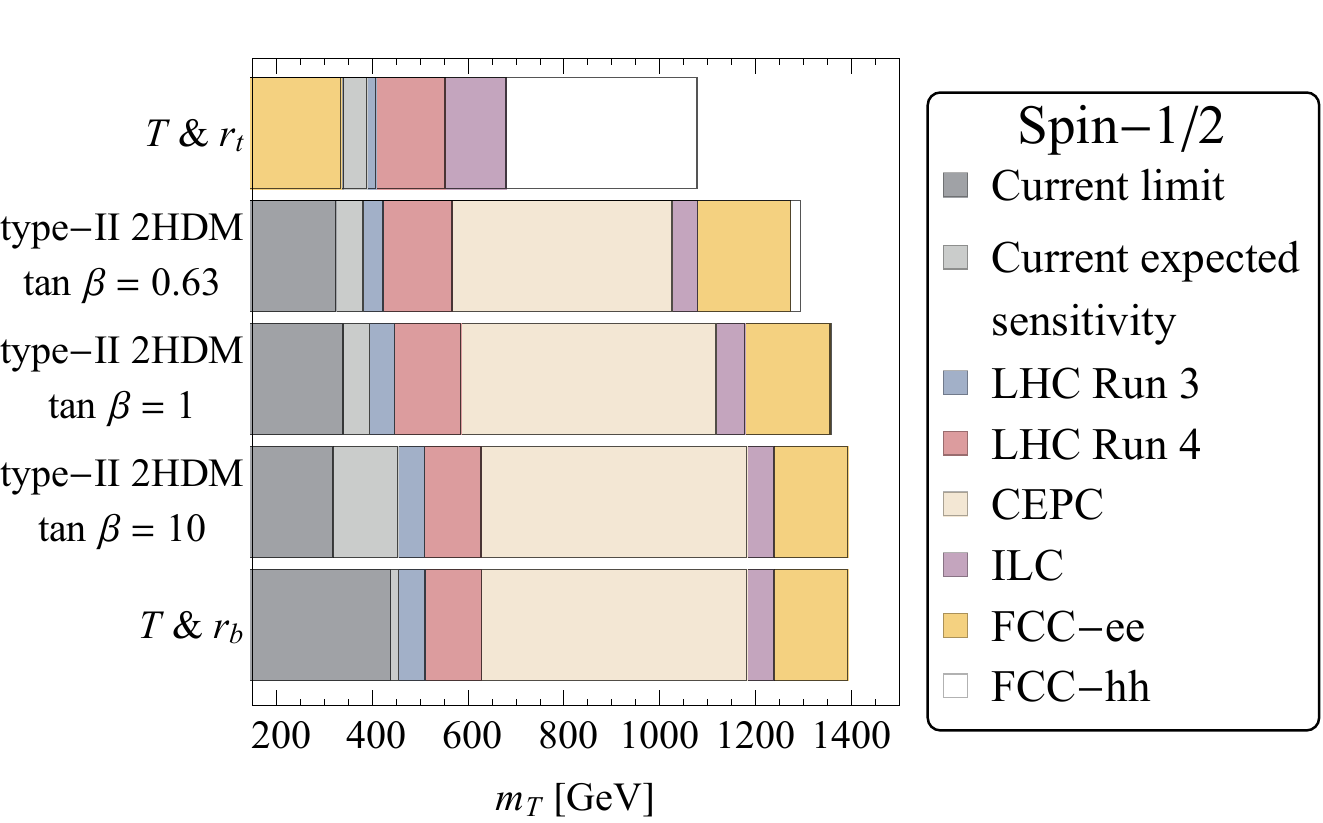} 
   \caption{\ 
  Excluded parameter space at $2\sigma$ C.L. on spin-1/2 top-partner mass, $m_T$, in type-II 2HDM with various $\tan \beta$ of current and future data (various colors).  In the plots, we again show the (non-2HDM) ``${T} $ \& $r_t$" and  ``${T}$ \& $r_b$" scenarios from 
  \figref{LHgeneral} for ease of comparison.  
   }
   \label{fig:2HDMparam}
\end{figure}

Similarly to the MSSM case in Section~\ref{subsubsec:mssm}, we can now consider spin-1/2 top partners with a 
Higgs sector given by a 2HDM model (such as the $SU(4)$ SLH).  
We focus on a 2HDM type-II model, since this allows for the weakest constraints on top partners as discussed in 
Section~\ref{subsubsec:spin-1/2-2HDM}.  
The results are given in \figref{2HDMparam} for various values of $\tan\beta$.  To simplify comparisons with the cases in 
Section~\ref{subsec:constraints-with-nonSM}, we again show the results for ``$T$ \& $r_b$'' and ``$T$ \& $r_t$''.  
As for the MSSM case, we see that lower values of $\tan\beta$ help in hiding the spin-1/2 top partners. 

\subsubsection{Extended spin-1/2 top partner sectors}\label{subsec:multiple-spin-1/2}
\begin{figure}[t!]
   \centering
   \includegraphics[width=0.45\textwidth]{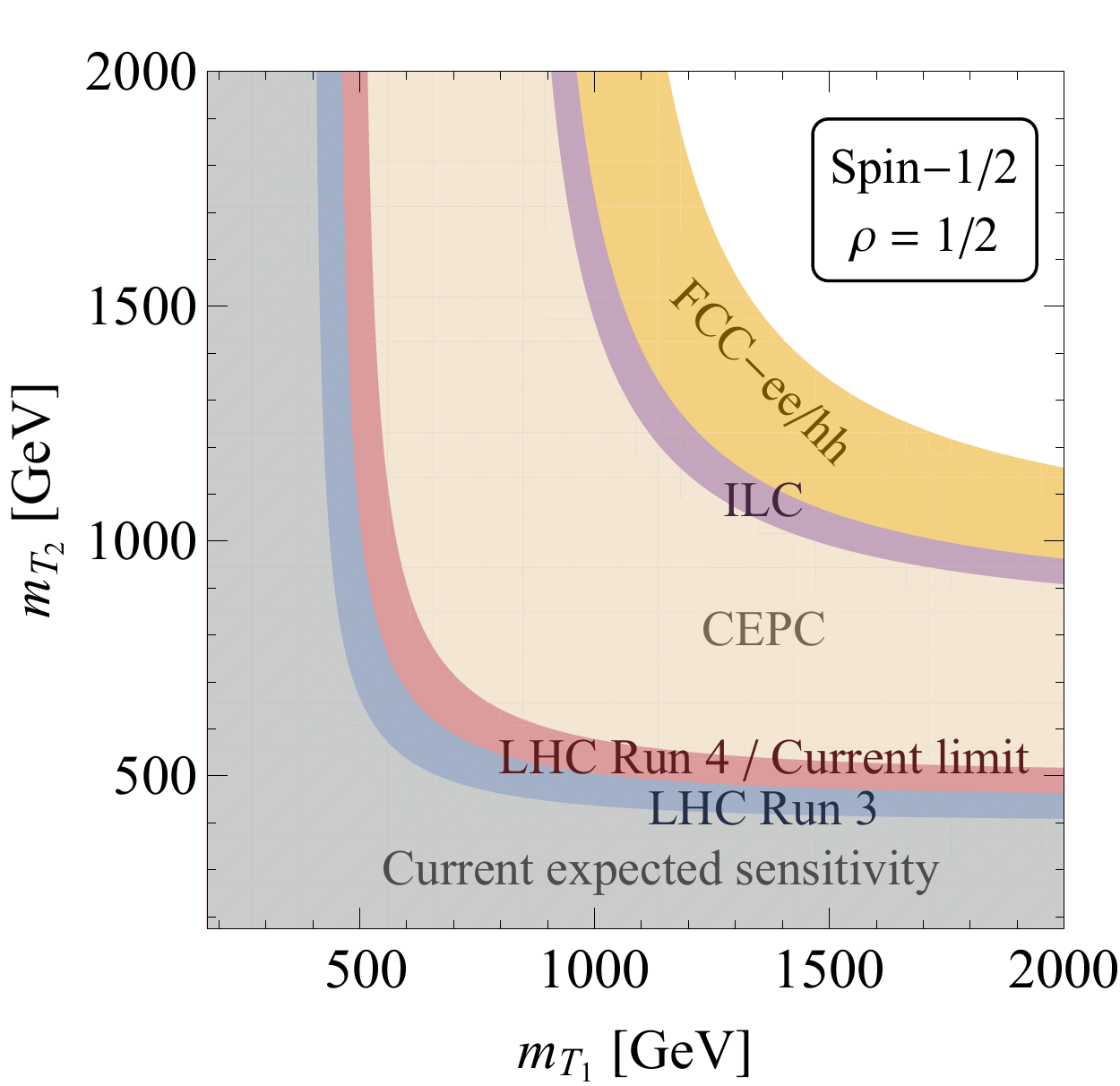}~\includegraphics[width=0.45\textwidth]{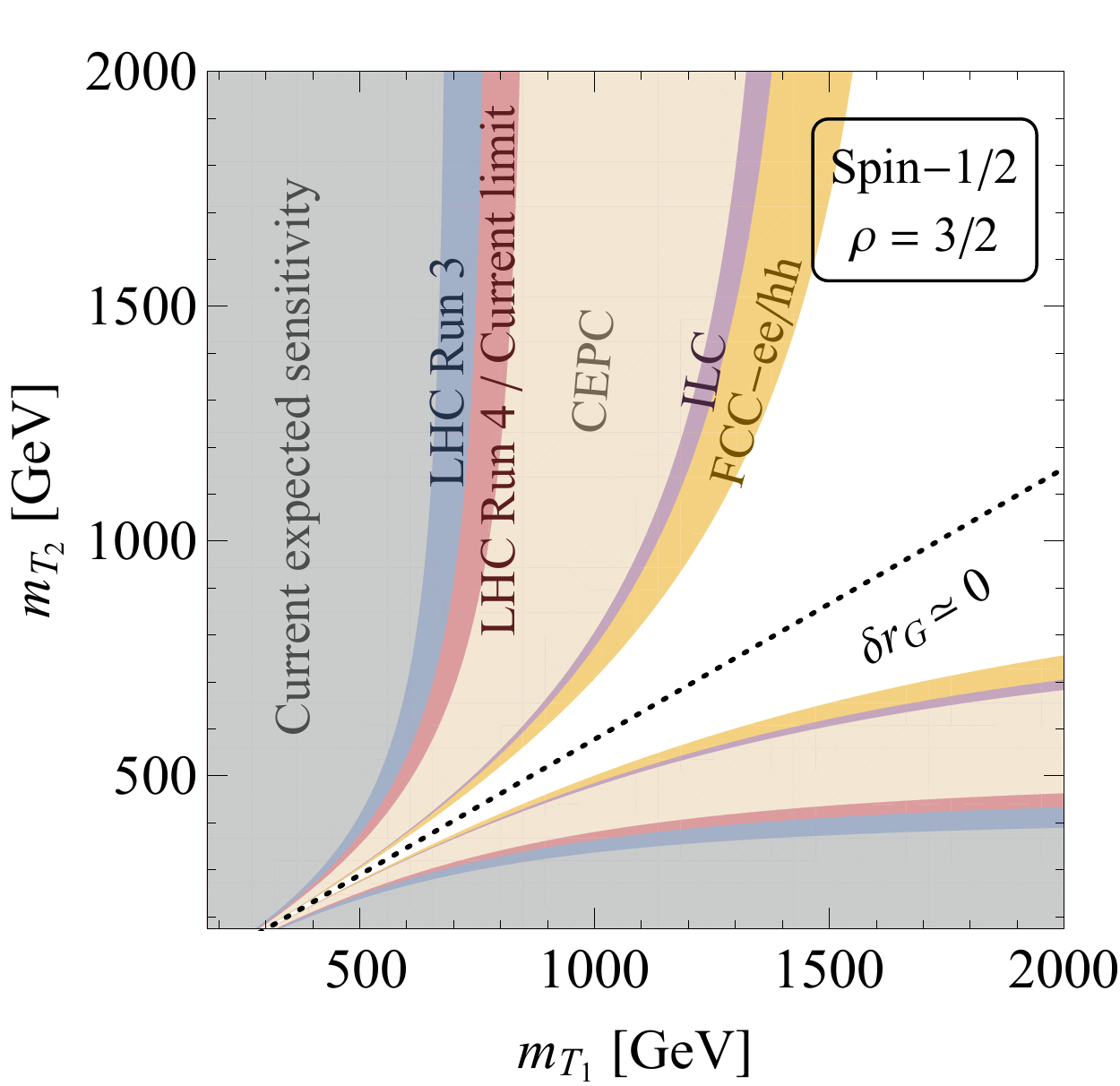} 
   \caption{ 
      Excluded parameter space and expected sensitivities at the $2\sigma$ C.L.~of current and future data (various colors) 
   for two spin-1/2 top partners in the $m_{T_2}$ versus $m_{T_1}$ plane. 
   The left plot shows the case in which both spin-1/2 top partners contribute equally to canceling the Higgs-mass contribution of the 
   top-quark loop, i.e.~$\rho=1/2$, where $\rho$ is defined in \eqref{2toprg}.  
   In the right plot, $T_2$ contributes with the same sign as the top-quark to the Higgs mass, but both contributions are cancelled by $T_1$, 
   $\rho=3/2$.  The latter allows for a ``stealth'' limit (black dotted line), in which Higgs precision measurements are not sensitive to the presence of spin-1/2 top-partners. 
   }
   \label{fig:figrhoboth}
\end{figure}

While the 2HDM presents one concrete way to reduce the sensitivity to spin-1/2 partners from Higgs precision measurements, 
we can study further improvements by introducing multiple top partners, as discussed in Section~\ref{subsec:better-models}. 
When $\rho \le 1$ in \eqref{2toprg}, the sensitivity to top partners remains unchanged, see \figref{figrhoboth} (left), but 
there is an extra fine-tuning penalty for moving away from the diagonal degenerate region as measured by 
\begin{equation}
\delta m_h^2=\frac{3}{8 \pi ^2}\left[|\rho| m_{T_1}^2   \log \left(\frac{\Lambda_\text{Strong}
   ^2}{m_{T_1}^2}\right) +|1-\rho| m_{T_2}^2  \log
   \left(\frac{\Lambda_\text{Strong}
   ^2}{m_{T_2}^2}\right)
   \right]\,.
\end{equation} 
However, $\rho \geq 1$ allows for a ``stealth'' scenario, as shown by the gray dashed line in \figref{figrhoboth} (right) for a fixed choice of $\rho$. 
In this case, there is an accidental cancellation to the $hgg$ and $h\gamma\gamma$ loop from the two top partners, which is not  
constrained by the Higgs precision observables under consideration.  However, it is probed by a complementary Higgs precision measurement
-- the $Zh$ cross section -- at future lepton colliders, as we discuss in Appendix A.  
Nevertheless, the current bounds on this scenario are weak and could thus provide a promising direction for model 
building, perhaps one in which a symmetry allows for the presence of the ``stealth" region with minimal fine-tuning. 

\section{Conclusions}
\label{s.conclusion}

In this paper, we performed a model-agnostic investigation of the limits from Higgs-precision data alone to probe naturalness from the 
presence of colored top partners.  There are many other complementary probes of naturalness, such as direct collider searches, electroweak precision constraints and rare decays.  However, while any given test may be avoided in principle through model-building tricks that allow for a ``natural" model, it is useful to understand how well any given probe can test colored naturalness.  Higgs precision tests are quite robust even on their own, since the couplings involved are inherently tied to the very question of naturalness itself. 

We find that with Higgs precision measurements alone, the HL-LHC can constrain spin-0 and spin-1/2 top partners almost to $\mathcal{O}(500)$~GeV in theories where there is only one spin-1/2 top partner or there is minimal mixing between the states.  With proposed future lepton and hadron colliders this can be extended to the TeV scale.   Spin-1 top partners are generically excluded to the multi-TeV regime.  However, we have also identified a number of ``blind-spots" where top partners can still be light even if future colliders are realized.   In particular, if there is a hierarchy between multiple top partners from mixing of the states, the standard probe using gluon-fusion can be avoided.  However, there are still bounds from 
Higgs precision measurements that are complementary to what is probed by gluon fusion.  For instance, in the case of spin-0 top partners, in the extreme limit where one eigenvalue becomes lighter than $m_h/2$, constraints from gluon-fusion Higgs production can still be avoided but there are strong bounds instead from the new contribution to the total width of the Higgs.  Nevertheless, there still exist particular points in parameter space that can avoid all Higgs precision measurements, similar to the light-sbottom window~\cite{Janot:2004cy,Batell:2013psa}.  While these blind-spots were known for spin-0 cases, we have extended them to lower masses and included decays, and demonstrated that they can also occur in fermionic top partner models as well.   This provides an interesting model building direction, since minimal fermionic top-partner models, such as in Little Higgs theories, are generically in more tension with Higgs precision constraints than their spin-0 counterparts.  There are other orthogonal probes that are only relevant in the future, such as radiative shifts in the $Zh$ cross-section, which we discuss in Appendix~A.  
It would also be interesting for very light masses to study the interplay between contributions to the well-measured $Z$-boson-width and 
the Higgs properties.  We leave this for future work.

Another potential avenue for light colored top partners comes from changes in the couplings of other SM particles or the introduction of new states that couple to the Higgs and affect its width.  We have categorized different changes to find what is the best way to still have light colored top partners in light of the current and future data.  
We find that the most promising direction to relax current constraints through the modification of SM-Higgs couplings is to modify the $h t \bar{t}$ coupling, 
which can be done in models with extended Higgs sectors.
However, the upcoming LHC Runs 3 and 4 and the proposed ILC and FCC-hh projects, which can measure the $ht\bar{t}$ coupling down to $1\%$ level, 
will extensively probe this possibility. 

It remains unclear whether the EW scale is natural.  
We have shown that future colliders will allow for a robust probe of the parameter space of 
natural models using Higgs properties alone.  


\subsection*{Acknowledgements}
We thank A. Cohen, J. Fan, T. Han, M. Low, G. Marques-Tavares, M. McCullough, M. Perelstein, M. Peskin, A. Pierce, M. Reece, M. Schmaltz, D. Shih, and T. Stefaniak.
R.E.~is supported by the DoE Early Career research program DESC0008061 and through a Sloan Foundation Research Fellowship. 
The work of P.M.~was supported in part by NSF CAREER Award NSF-PHY-1056833 and in part by NSF grant NSF-PHY-1620628.  
The work of H.R.~was supported in part by NSF grant PHY-1316617 and in part by DoE grant DESC0008061. 
The work of Y.Z.~was supported by DoE grant DESC0015845. 
 


\appendix

\section{Future Complementary Higgs Precision Probes}
\label{blindspot}

In this appendix, we explore two complementary Higgs precision probes, $\sigma(e^+e^-\to Zh)$ and (briefly) $gg\to hh$.  
These will only be measured with sufficiency precision at future colliders, but could potentially probe different regions of parameter space than 
$\mathcal{N}_{\hat{t}}$ and $\Gamma_{\hat{t}\hat{t}}$.  As was noted in earlier sections, the typically dominant effect on Higgs precision from colored top partners, $\mathcal{N}_{\hat{t}}$,  can have ``blind spots" if there is more than one top partner.  These ``blind-spots" have been noticed before, for instance in \cite{Craig:2014una, Fan:2014axa} with reference to colored stops in the natural SUSY paradigm. 

The inherent reason for the blind spots comes from the linear dependence on couplings in $\mathcal{N}_{\hat{t}}$, and occurs when \beq \frac{g_{h\t t_1 \t t_1}}{m_{\t t_1}^2}=-\frac{g_{h \t t_2 \t t_2}}{m_{\t t_2}^2}\eeq for stops and \beq \frac{\rho}{m^2_{T_1}}=-\frac{1-\rho}{m^2_{T_2}}\eeq for multiple fermionic top partners. In both cases, there is a destructive interference between the top-partner loops arising from the relative negative sign between the couplings of the top partners to the Higgs. 

Processes that depend quadratically on the Higgs coupling can potentially avoid this blind spot.  For instance the contribution to $\Gamma_{\hat{t}\hat{t}}$ for stops discussed in Section~\ref{s.spin-0} excludes a different region of parameter space than the ``blind-spot".  
However, as we discussed, $\Gamma_{\hat{t}\hat{t}}$ vanishes if one stop is sufficiently heavy, which leads to a new  
blind-spot.  
Nevertheless, there are other processes that depend quadraticallly on multiple Higgs couplings and thereby can fill in the $\mathcal{N}_{\hat{t}}$ ``blind-spot" more robustly if sufficient precision is achieved. One such example is Higgs WFR~\cite{Craig:2014una} from BSM states that cause deviations in tree-level Higgs couplings after canonical normalization.  These loop corrections to tree-level couplings are hard to measure, and the first opportunity to detect such deviations  would be at future lepton colliders where the $Zh$ coupling can be measured with great precision. The inclusive Higgstrahlung cross section $\sigma(e^+ e^- \to Zh)$ can itself be precisely measured to $< 1\%$ level~\cite{Dawson:2013bba,Peskin:2012we}, and as a result the $Zh$ coupling can be measured to an order of magnitude better than other Higgs couplings.

The quadratic dependence on Higgs couplings is straightforward to see.  For stops in natural SUSY the full WFR including stop-mixing was discussed in~\cite{Fan:2014axa}. The deviation in the $Zh$ coupling from WFR contributions is
\beq
\delta \sigma_{\text{Zh}}[\text{WFR}] =\frac{3}{16\pi^2} (g^2_{h_{11}}+g^2_{h_{22}}+2g^2_{h_{12}})I(m_h^2,m^2_{\tilde{t}_1},m^2_{\tilde{t}_2})\,, 
\eeq
where $I$ is the loop function defined in \cite{Fan:2014axa}, potentially covering the $\mathcal{N}_{\hat{t}}$ ``blind-spot".  
However, in~\cite{Craig:2014una} it was pointed out that the WFR contribution can dramatically mis-estimate $\delta \sigma_{Zh}$
and that a full one-loop calculation is instead required.  
Since other one-loop diagrams contain only one stop-Higgs vertex ($h\t t_1 \t t_1$ or $h\t t_2 \t t_2$, but not $h\t t_1 \t t_2$), there is destructive interference for particular choices of parameters.  
In particular, \cite{Craig:2014una} considered degenerate soft masses and observed that Higgstrahlung limits overlapped with 
$gg \to h$ limits. 

Here we explore Higgstrahlung probes in the most general stop mass plane (without restricting to degenerate soft masses).  
We follow the treatment of~\cite{Craig:2014una, Xiang:2017yfs} to calculate $\sigma(e^+e^- \rightarrow Zh)$  to include complete one-loop corrections using the \texttt{FeynArts 3.9} -- \texttt{FormCalc 9.4} -- \texttt{LoopTools 2.13} suite~\cite{Hahn:2000kx}.\footnote{We thank Matthew McCullough for providing his implementation of Natural SUSY in the FeynArts suite.}  We fix the $\sqrt s = 240\gev$ for all the results displayed here to match with the CEPC/FCC-ee setting. 
We also test the result with various center-of-mass energies, $\sqrt s = 240\gev$, $250 \gev$, $350 \gev$, and $500 \gev$, and find that the resulting excluded region is the same at the level of a few percent. 

\begin{figure}[t!]
   \centering
   \includegraphics[width=0.48\textwidth]{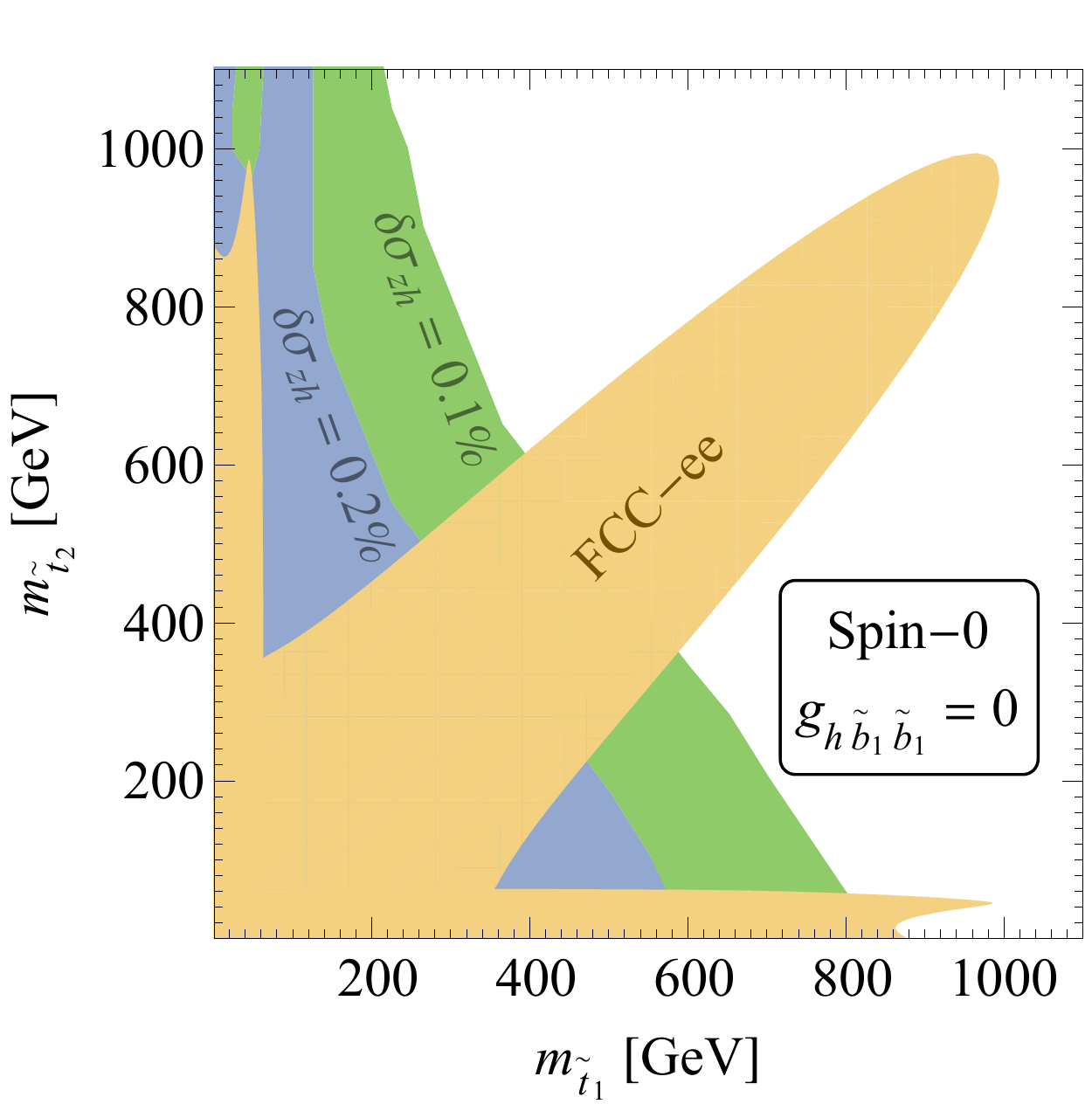}
~~~    \includegraphics[width=0.48\textwidth]{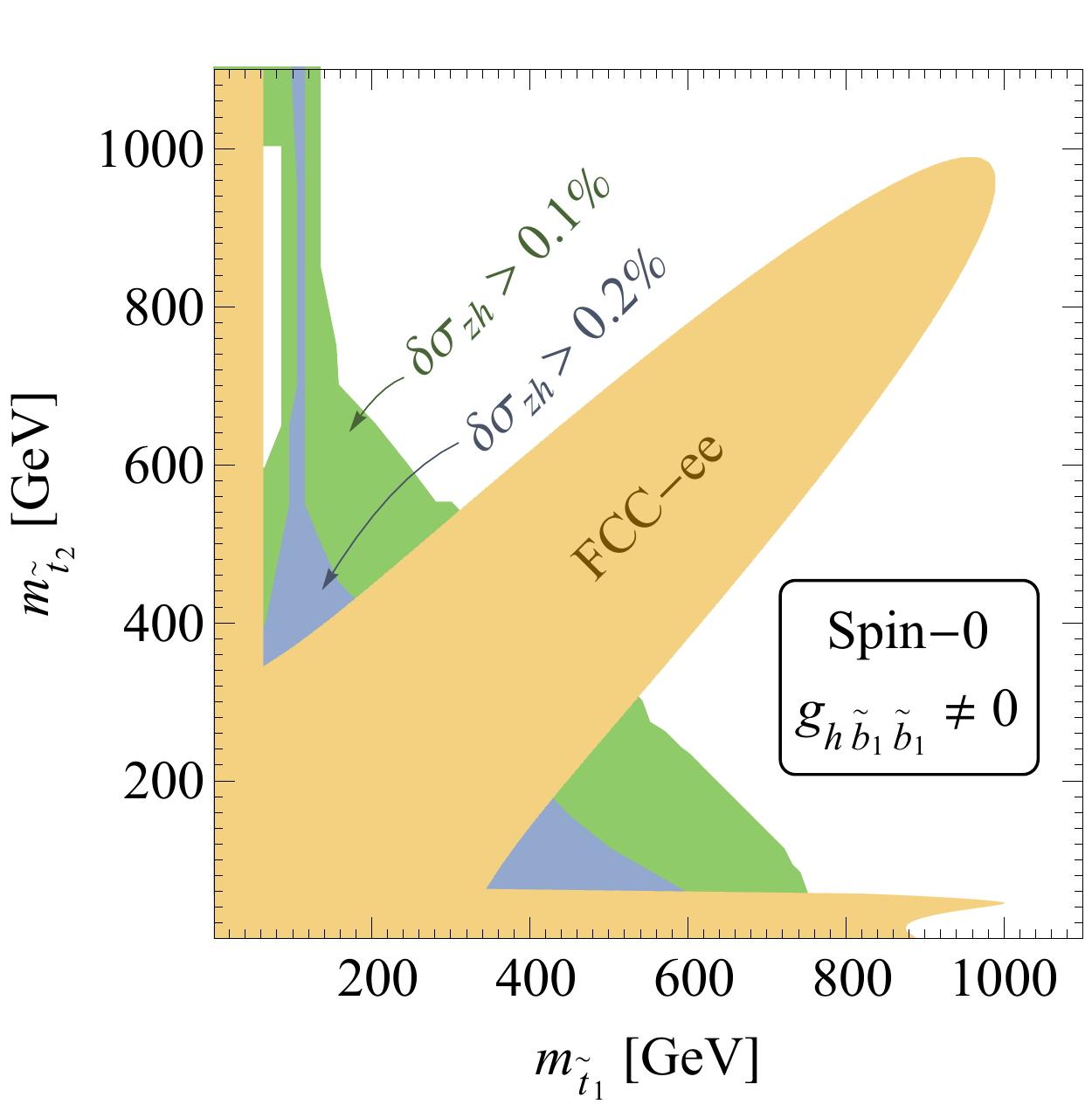}
   \caption{Expected sensitivities at the $2\sigma$ C.L.~of FCC-ee for spin-0 models with additional constraints from  $\delta \sigma_{Zh}$. 
    In the {\it left} plot, we assume $\tan \beta \simeq 1$ and $h \t b_1 \t b_1$ coupling vanishes (\eqref{gnb1b1=0}), while in the {\it right} plot, 
   $\tan \beta$ is large to maximize the $D$-term contributions in the stop and sbottom sector (\eqref{gnb1b1-D-term}). 
}
   \label{fig:stopstopZh}
\end{figure}
 
In~\figref{stopstopZh}, we show the additional constraints obtained from $\delta \sigma_{Zh}$ compared to the Higgs precision probes 
included in \figref{stopstop}. 
We find that with $\delta \sigma_{Zh} \geq 0.2\%$  (projected for ILC, CEPC and FCC-ee~\cite{Fujii:2015jha, Dawson:2013bba, CEPC-SPPCStudyGroup:2015csa}), we observe additional constraints in the non-degenerate region when $\tan \beta \rightarrow 1$. 
As seen in \figref{stopstopZh}, less additional parameter space is constrained when $\tan \beta$ is large. 
If we were to increase the statistics of the future lepton colliders and improve the measurement on $\delta \sigma_{Zh}$ to $0.1\%$, we start to probe more of the non-degenerate region in both cases. With $0.1\%$ of data, we can also robustly rule out $m_{\t t_1} \le 150 \gev$ in both cases.  However, one should note that this is tied to the ansatz that $\t t_1$ is mostly left-handed in our setup, which fixes the $\tilde{b}_1$ mass.  This is also the reason why the limits are not symmetric under the interchange of $\t t_1$ and $\t t_2$.  
It would be interesting to study fully the large-mixing region of small stop- and sbottom-masses in the MSSM to find 
robust lower bounds.

For fermionic top partners, which we consider to be part of an EFT, we do not implement a full one-loop analysis, as there can be additional dimension-six operators generated at the UV scale that could also contribute to the Higgstrahlung cross-section.   However, we can still make a conservative estimate of the contribution to the $Zh$ cross section from the top-partners using WFR in the EFT with the assumption that there are no large cancellations between the loop-effects and higher-dimension operators.  With this assumption, the deviation in the $Zh$ cross-section,  from the the finite contributions to Higgs WFR in the multiple fermionic top partner model in Section~\ref{subsec:better-models}, is given by,
\beq
\delta \sigma_{Zh}=-\frac{m_t^2 }{
 8  \pi^2}\left[\frac{\rho^2}{m_{T_1}^2}
+ \frac{(1-\rho)^2}{m_{T_2}^2}\right].
\eeq

\begin{figure}[t!]
   \centering
   \includegraphics[width=0.45\textwidth]{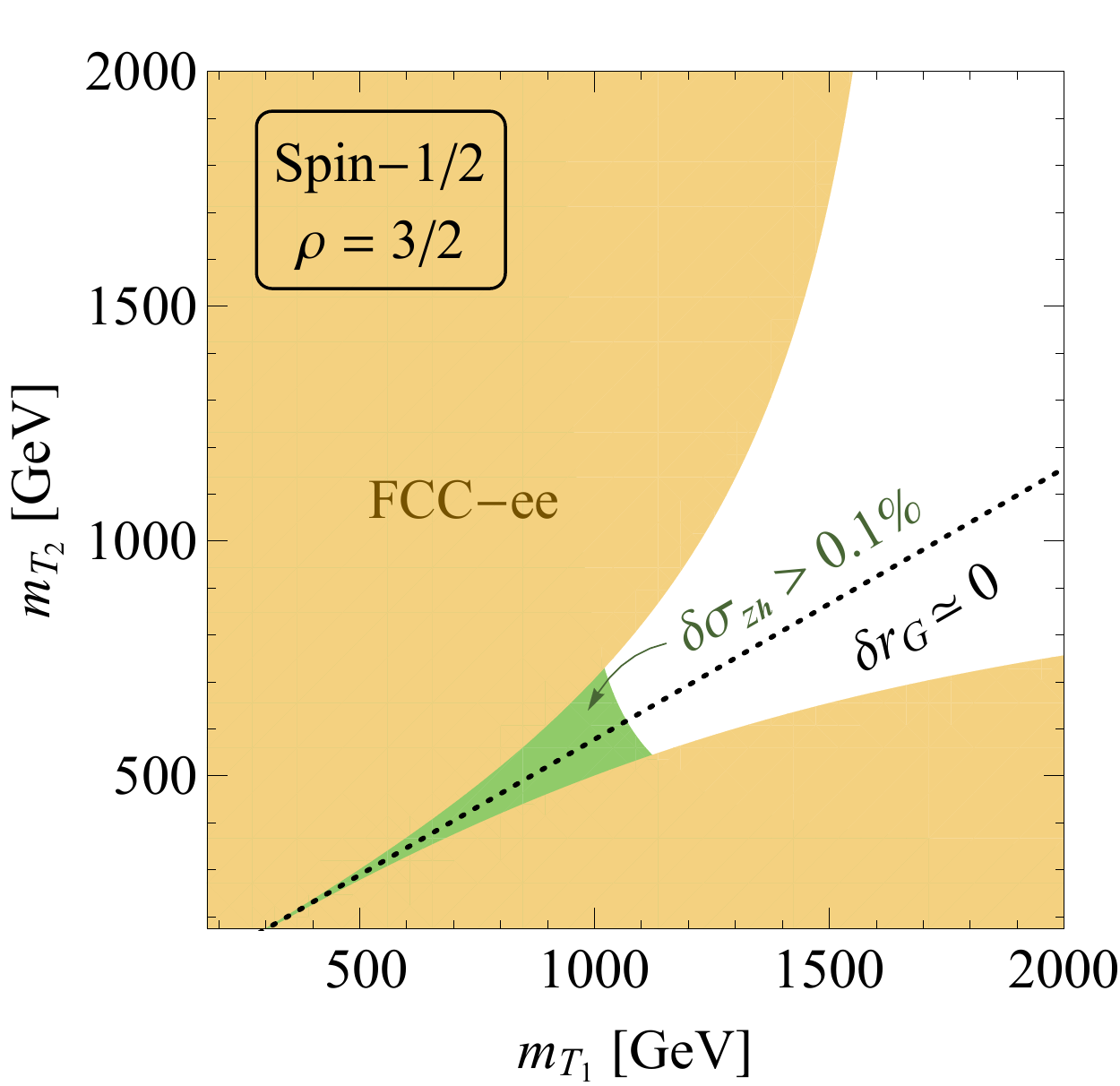} 
   \caption{ Expected sensitivities at the $2\sigma$ C.L.~of FCC-ee for two spin-1/2 top-partner model with additional constraints from $\delta \sigma_{Zh}$. 
   The projected sensitivity from FCC-ee is taken from \figref{figrhoboth}.}
   \label{fig:figrhobothZh}
\end{figure}

We find that the stealth region in the right panel of \figref{figrhoboth} can be covered to the TeV scale by the measurement of $\delta \sigma_{Zh}$ from future lepton colliders as shown in  \figref{figrhobothZh}.  
This is a conservative estimate, and the effects would in general be larger unless there was a symmetry or additional tuning of different 
contributions at the UV scale. 

Finally, we briefly comment on di-Higgs production, which also is quadratically sensitive to the Higgs-top-partner coupling.  Similar to $gg\to h$, colored top partners contribute to the double Higgs production process, $gg\to hh$, at the loop-level\cite{Batell:2015koa}.  It contains two Higgs vertices, which can spoil this cancellation, and naively we would expect some coverage of the ``blind-spots" by measuring the deviation of $\sigma({gg\to hh})$ from its SM prediction. However, even at future colliders, the total cross section for double-Higgs production is much smaller than single-Higgs production making this a difficult measurement without much discriminating power. 

At a 100 TeV hadron collider, with 30 ab$^{-1}$ of data, we can measure this cross section to 1.6$\%$ accuracy~\cite{Contino:2016spe}. 
However, even so, it is notoriously hard to differentiate between new colored particles in the loop and a change in the triple Higgs coupling~\cite{Dawson:2015oha}. 
We leave for future work a calculation of the constraints that includes a shape analysis of the $m^2_{hh}$ spectrum 
near the light top-partner mass threshold. 

\section{Loop-induced Higgs Couplings}
\label{loopcoupling}
\begin{figure}[t]
   \centering
   \includegraphics[width=0.48\textwidth]{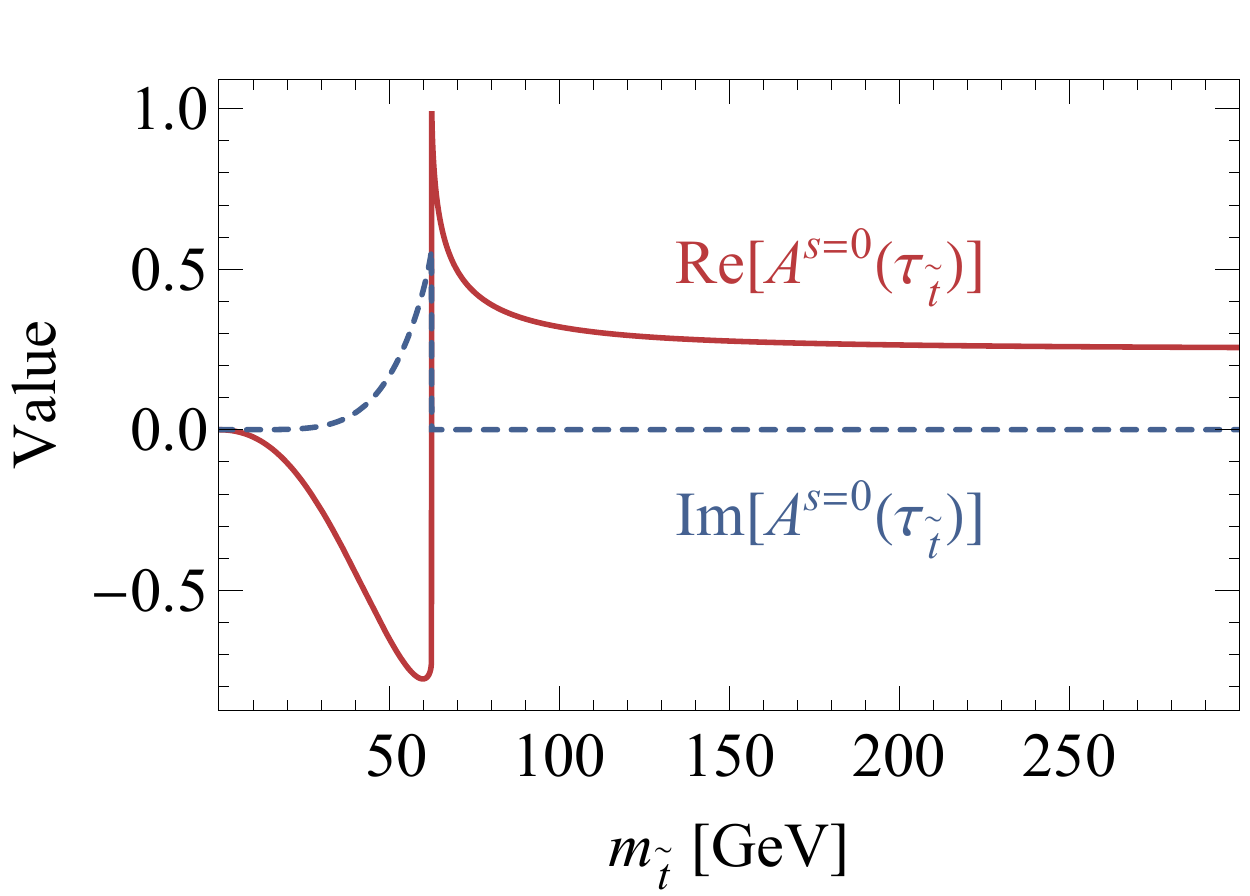} 
   \caption{The real and imaginary part of the pre-factor  of ${\mathcal A}^{\text{s}=0}(\tau_{\t t})$.}
   \label{fig:ReIm}
\end{figure}

The loop functions introduced in Section~\ref{s.precision} depend on the spin $s$ of the particle and are given by 
\begin{align}
{\mathcal A}^{\text{s}=0}(\tau) ={}&\f{3}{4} \f{1}{\tau^2} \left[ -\tau+ f(\tau)\right],\\
{\mathcal A}^{\text{s}=1/2}(\tau) ={}&\f{3}{2} \f{1}{\tau^2} \left[\tau + (\tau-1) f(\tau)\right],\\
{\mathcal A}^{\text{s}=1} (\tau)={}&\f{3}{4} \f{1}{\tau^2} \left[ -2\tau^2 -3\tau+3\left(1-2\tau\right)f(\tau)\right],
\end{align}
where  $\tau = m_h^2/4m_i^2$, and $m_i$ is the mass of loop particles. $f(\tau)$ is given by
\begin{equation}
f(\tau) = \begin{cases}
\left(\sin^{-1}{\sqrt{\tau}}\right)^2& \text{if } \tau \leq 1 \\
-\frac{1}{4}\left[\ln\frac{1+\sqrt{1-\tau^{-1}}}{1-\sqrt{1-\tau^{-1}}}-i\pi\right]^2 & \text{if } \tau > 1\,. 
\end{cases}
\end{equation}
When $m_i \to \infty$,
\begin{align}
{\mathcal A}^{\text{s} = 0}(\tau\rightarrow 0)& =\frac{1}{4},\\
{\mathcal A}^{\text{s}=1/2}(\tau\rightarrow 0)& =1,\\
{\mathcal A}^{\text{s}=1} (\tau\rightarrow 0)&=-\frac{21}{4}\,.
\end{align}
The loop functions become complex when $m_i <m_h/2$. In \figref{ReIm}, we take ${\mathcal A}^{\text{s}=0}$ as an example and show its real and imaginary parts as a function of $m_{\t t}$.  ${\mathcal A}^{\text{s}=1/2}$ and ${\mathcal A}^{\text{s}=1}$ have similar behaviors as ${\mathcal A}^{\text{s}=0}$.

\section{Cross-check with HiggsSignals}
\label{validity}

We use \texttt{HiggsSignals 1.4.0} to cross-check the results from our fitting method described in 
Sec.~\ref{s.data}.\footnote{We thank Tim Stefaniak for help in setting up \texttt{HiggsSignals}.}
 \texttt{HiggsSignals 1.4.0} contains the available data up to the summer of 2015.  
To make the comparison, we use the data sets with a single decay channel that are included in the 
\texttt{HiggsSignals 1.4.0} package.  
We choose to compare the constraints on \{$r_\gamma$, $r_G$\}, setting all other couplings to be their SM values.

\begin{figure}[t]
   \centering
   \includegraphics[width=0.96\textwidth]{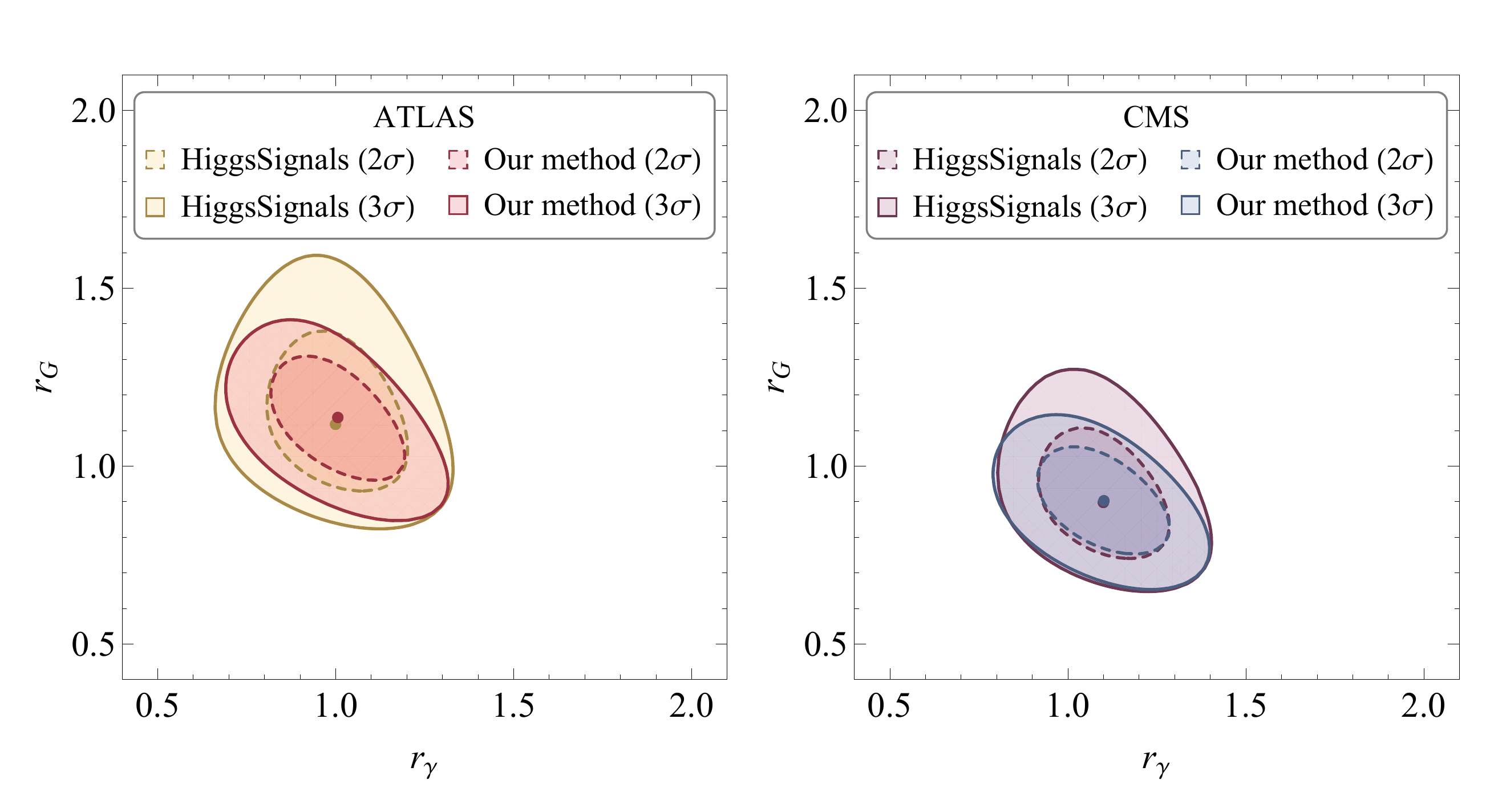} 
   \caption{A comparison of $r_G-r_\gamma$ joint fits from HiggsSignals and our fitting method described in Sec.~\ref{s.data} for ATLAS (left) and CMS (right) data (see text for more details).}
   \label{fig:validity}
\end{figure}

We show a comparison between our method and the \texttt{HiggsSignals} fit in Fig.~\ref{fig:validity}, finding reasonable 
agreement, especially for the $2\sigma$-CL exclusion contours.  
The main difference between the two methods is that 
\texttt{HiggsSignals} correlates the theoretical uncertainties in the SM production cross sections and branching ratios (see also~\cite{Arbey:2016kqi}), and also correlates 
the error on the individual production and decay with the BSM modifications to the couplings.  
If we set the theoretical uncertainties of SM cross sections and branching ratios to zero, our results agree with \texttt{HiggsSignals}. 
We note, however, that \texttt{HiggsSignals} also does not capture the correlations among the systematic uncertainties in the data, 
which is not public information. 

\section{Data Tables}
\label{datasets}
We here provide tables of the current and future Higgs-precision data used in our analyses.  Current data for LHC and Tevatron are listed in \tabref{muone}, \tabref{mutwo}, and \tabref{muthree}. The signal strength (the fourth column) is expressed in the format $\mu^{+\sigma_\text{up}}_{-\sigma_\text{down}}$, where $\sigma_\text{up}$ and $\sigma_\text{down}$ are the (asymmetric) 1$\sigma$-error bars of the observed signal strength $\mu$.  \tabref{muone} lists all searches with single decay and multiple productions. The signal strength~\eqref{mu} can be expressed as
\beq
\mu_{f}=\left( |r_G|^2 \xi_{G} + |r_V|^2 \xi_{V}+|r_t|^2\xi_{t}\right) \frac{|r_f|^2}{r_h+r_\text{inv}}\,, 
\eeq
where  $\xi_G$, $\xi_V$, and $\xi_t$ are the weights in the Higgs production for gluon fusion, vector-boson fusion plus associated production, and associated production with tops, respectively. The signal strength and weights given in \tabref{muthree} are defined in a similar manner, 
\begin{equation}
\mu_{\text{inv}}=\left( |r_G|^2 \xi_{G} + |r_V|^2 \xi_{V}+|r_t|^2\xi_{t}\right) \f{r_\text{inv}}{r_h+r_\text{inv}}\,.
\end{equation}
\tabref{mutwo} lists all searches with multiple decays and single production. The signal strength \eqref{mu} is instead expressed as
\beq
\mu_{f}=\frac{|r_t^2|}{r_h+r_\text{inv}}\left(|r_V^2|\zeta_{VV}+|r_b^2|\zeta_{bb}+|r_\tau|^2\zeta_{\tau\tau}\right),
\eeq
where $\zeta_{VV}$, $\zeta_{bb}$, and $\zeta_{\tau\tau}$ stand for weights in the Higgs decays into $WW$ plus $ZZ$, $bb$, 
and $\tau^+\tau^-$, respectively. 

Future ATLAS  Run 3 and Run 4 data are listed in \tabref{mufour}. We assume CMS will perform identical searches and effectively double the data listed in  \tabref{mufour} for our projections. Sensitivities for ILC (250 GeV, 2 ab$^{-1}$ $\oplus$ 350 GeV, 200 fb$^{-1}$ $\oplus$ 550 GeV, 4 ab$^{-1}$), CEPC (240 GeV, 10 ab$^{-1}$),  FCC-ee (240 GeV, 10 ab$^{-1}$ $\oplus$ 350 GeV, 2.6 ab$^{-1}$), and FCC-hh (100 TeV, 30 ab$^{-1}$) are listed in \tabref{future}. 

\begin{longtable}{@{} lcccrrrrc @{}} 
 \multicolumn{1}{l}{Channel} & \multicolumn{1}{l}{Analysis} &  \multicolumn{1}{l}{$\sqrt{s}$} & \multicolumn{1}{c}{$L$} & \multicolumn{1}{c}{$\mu$} & \multicolumn{1}{c}{ $\xi_G$} & \multicolumn{1}{c}{$\xi_V$} &  \multicolumn{1}{c}{$\xi_t$} &  \multicolumn{1}{c}{Ref.}  \\
 \multicolumn{1}{l}{} & \multicolumn{1}{l}{} &  \multicolumn{1}{l}{(TeV)} & \multicolumn{1}{l}{($\text{fb}^{-1}$)} & \multicolumn{1}{c}{} & \multicolumn{1}{c}{  (\%)} & \multicolumn{1}{c}{ (\%)} &  \multicolumn{1}{c}{(\%)} &  \multicolumn{1}{c}{}  \\
 \hline
\text{$\gamma \gamma $ (ttH; multijet)}	&	CMS	&	8	& full	&	$1.24^{+4.24}_{-2.70}$	&	3	&	3	&	94	&	\cite{Khachatryan:2014ira}	\\
\text{$\gamma \gamma $ (ttH; lepton)}	&	CMS	&	8	&full	&	$3.52^{+3.89}_{-2.45}$	&	0	&	4	&	96	&	\cite{Khachatryan:2014ira}	\\
\text{$\gamma \gamma $ (ttH)}	&	CMS	&	7	&full	&	$0.71^{+6.20}_{-3.56}$	&	3	&	5	&	92	&	\cite{Khachatryan:2014ira}	\\
\text{$\gamma \gamma $ (untagged 0)}	&	CMS	&	7	&full	&	$1.97^{+1.50}_{-1.25}$	&	80	&	19	&	1	&	\cite{Khachatryan:2014ira}	\\
\text{$\gamma \gamma $ (untagged 0)}	&	CMS	&	8	&full	&	$0.13^{+1.09}_{-0.74}$	&	71	&	27	&	2	&	\cite{Khachatryan:2014ira}	\\
\text{$\gamma \gamma $ (untagged 1)}	&	CMS	&	7	&full	&	$1.23^{+0.98}_{-0.88}$	&	92	&	8	&	0	&	\cite{Khachatryan:2014ira}	\\
\text{$\gamma \gamma $ (untagged 1)}	&	CMS	&	8	&full	&	$0.92^{+0.57}_{-0.49}$	&	82	&	18	&	1	&	\cite{Khachatryan:2014ira}	\\
\text{$\gamma \gamma $ (untagged 2)}	&	CMS	&	7	&full	&	$1.60^{+1.25}_{-1.17}$	&	92	&	8	&	0	&	\cite{Khachatryan:2014ira}	\\
\text{$\gamma \gamma $ (untagged 2)}	&	CMS	&	8	&full	&	$1.10^{+0.48}_{-0.44}$	&	89	&	11	&	0	&	\cite{Khachatryan:2014ira}	\\
\text{$\gamma \gamma $ (untagged 3)}	&	CMS	&	7	&full	&	$2.61^{+1.74}_{-1.65}$	&	92	&	8	&	0	&	\cite{Khachatryan:2014ira}	\\
\text{$\gamma \gamma $ (untagged 3)}	&	CMS	&	8	&full	&	$0.65^{+0.65}_{-0.89}$	&	89	&	10	&	0	&	\cite{Khachatryan:2014ira}	\\
\text{$\gamma \gamma $ (untagged 4)}	&	CMS	&	8	&full	&	$1.46^{+1.29}_{-1.24}$	&	91	&	8	&	0	&	\cite{Khachatryan:2014ira}	\\
\text{$\gamma \gamma $ (VBF; dijet 0)}	&	CMS	&	7	&full	&	$4.85^{+2.17}_{-1.76}$	&	19	&	81	&	0	&	\cite{Khachatryan:2014ira}	\\
\text{$\gamma \gamma $ (VBF; dijet 0)}	&	CMS	&	8	&full	&	$0.82^{+0.75}_{-0.58}$	&	14	&	85	&	0	&	\cite{Khachatryan:2014ira}	\\
\text{$\gamma \gamma $ (VBF; dijet 1)}	&	CMS	&	7	&full	&	$2.60^{+2.16}_{-1.76}$	&	38	&	61	&	0	&	\cite{Khachatryan:2014ira}	\\
\text{$\gamma \gamma $ (VBF; dijet 1)}	&	CMS	&	8	&full	&	$-0.21^{+0.75}_{-0.69}$	&	24	&	76	&	0	&	\cite{Khachatryan:2014ira}	\\
\text{$\gamma \gamma $ (VBF; dijet 2)}	&	CMS	&	8	&full	&	$2.60^{+1.33}_{-0.99}$	&	38	&	61	&	1	&	\cite{Khachatryan:2014ira}	\\
\text{$\gamma \gamma $ (VH; dijet)}	&	CMS	&	7	&full	&	$7.86^{+8.86}_{-6.40}$	&	27	&	71	&	2	&	\cite{Khachatryan:2014ira}	\\
\text{$\gamma \gamma $ (VH; dijet)}	&	CMS	&	8	&full	&	$0.39^{+2.16}_{-1.48}$	&	25	&	72	&	3	&	\cite{Khachatryan:2014ira}	\\
\text{$\gamma \gamma $ (VH; MET)}	&	CMS	&	7	&full	&	$4.32^{+6.72}_{-4.15}$	&	5	&	87	&	8	&	\cite{Khachatryan:2014ira}	\\
\text{$\gamma \gamma $ (VH; MET)}	&	CMS	&	8	&full	&	$0.08^{+1.86}_{-1.28}$	&	13	&	75	&	12	&	\cite{Khachatryan:2014ira}	\\
\text{$\gamma \gamma $ (VH; loose)}	&	CMS	&	7	&full	&	$3.10^{+8.29}_{-5.34}$	&	4	&	95	&	1	&	\cite{Khachatryan:2014ira}	\\
\text{$\gamma \gamma $ (VH; loose)}	&	CMS	&	8	&full	&	$1.24^{+3.69}_{-2.62}$	&	2	&	96	&	2	&	\cite{Khachatryan:2014ira}	\\
\text{$\gamma \gamma $ (VH; tight)}	&	CMS	&	8	&full	&	$-0.34^{+1.30}_{-0.63}$	&	0	&	96	&	4	&	\cite{Khachatryan:2014ira}	\\
\text{$WW$ (3$l$3$\nu $)}	&	CMS	&	7+8	&full	&	$0.56^{+1.27}_{-0.95}$	&	0	&	100	&	0	&	\cite{Chatrchyan:2013iaa}	\\
\text{$WW$ (0/1$j$)}	&	CMS	&	7+8	&full	&	$0.74^{+0.22}_{-0.20}$	&	82	&	18	&	0	&	\cite{Chatrchyan:2013iaa}	\\
\text{$WW$ (VBF; 2$j$)}	&	CMS	&	7+8	&full	&	$0.60^{+0.57}_{-0.46}$	&	20	&	80	&	0	&	\cite{Chatrchyan:2013iaa}	\\
\text{$WW$ (VH; 2$l$2$\nu $)}	&	CMS	&	7+8	&full	&	$0.39^{+1.97}_{-1.87}$	&	54	&	46	&	0	&	\cite{Chatrchyan:2013iaa}	\\
\text{$ZZ$ (0/1$j$)}	&	CMS	&	7+8	&full	&	$0.88^{+0.34}_{-0.27}$	&	90	&	10	&	0	&	\cite{Chatrchyan:2013mxa}	\\
\text{$ZZ$ (2$j$)}	&	CMS	&	7+8	&full	&	$1.55^{+0.95}_{-0.66}$	&	58	&	42	&	0	&	\cite{Chatrchyan:2013mxa}	\\
\text{$bb$ (VBF)}	&	CMS	&	7+8	&full	&	$2.80^{+1.60}_{-1.40}$	&	0	&	100	&	0	&	\cite{Khachatryan:2015bnx}	\\
\text{$bb$ (VH)}	&	CMS	&	7+8	&full	&	$0.89^{+0.43}_{-0.43}$	&	0	&	100	&	0	&	\cite{Khachatryan:2015bnx}	\\
\text{$bb$ (ttH)}	&	CMS	&	7+8	&full	&	$0.70^{+1.80}_{-1.80}$	&	0	&	0	&	100	&	\cite{Khachatryan:2015bnx}	\\
\text{$\tau \tau$ (ttH)}	&	CMS	&	7+8	&full	&	$-1.30^{+6.30}_{-5.50}$	&	0	&	0	&	100	&	\cite{Khachatryan:2014qaa}	\\
\text{$\tau \tau$ (0$j$)}	&	CMS	&	7+8	&full	&	$0.34^{+1.09}_{-1.09}$	&	98	&	2	&	0	&	\cite{Chatrchyan:2014nva}	\\
\text{$\tau \tau$ (1$j$)}	&	CMS	&	7+8	&full	&	$1.07^{+0.46}_{-0.46}$	&	77	&	23	&	0	&	\cite{Chatrchyan:2014nva}	\\
\text{$\tau \tau$ (VBF; 2$j$)}	&	CMS	&	7+8	&full	&	$0.94^{+0.41}_{-0.41}$	&	19	&	81	&	0	&	\cite{Chatrchyan:2014nva}	\\
\text{$\tau \tau$ (VH)}	&	CMS	&	7+8	&full	&	$-0.33^{+1.02}_{-1.02}$	&	0	&	100	&	0	&	\cite{Chatrchyan:2014nva}	\\
\text{$\gamma \gamma $ (central low $p_{T_t}$)}	&	ATLAS	&	7+8	&full	&	$0.62^{+0.42}_{-0.40}$	&	91	&	8	&	0	&	\cite{Aad:2014eha}	\\
\text{$\gamma \gamma $ (central high $p_{T_t}$)}	&	ATLAS	&	7+8	&full	&	$1.62^{+1.00}_{-0.83}$	&	67	&	31	&	2	&	\cite{Aad:2014eha}	\\
\text{$\gamma \gamma$ (forward low $p_{T_t}$)}	&	ATLAS	&	7+8	&full	&	$2.03^{+0.57}_{-0.53}$	&	91	&	9	&	0	&	\cite{Aad:2014eha}	\\
\text{$\gamma \gamma $ (forward high $p_{T_t}$)}	&	ATLAS	&	7+8	&full	&	$1.73^{+1.34}_{-1.18}$	&	66	&	33	&	1	&	\cite{Aad:2014eha}	\\
\text{$\gamma \gamma $ (VBF; loose)}	&	ATLAS	&	7+8	&full	&	$1.33^{+0.92}_{-0.77}$	&	33	&	67	&	0	&	\cite{Aad:2014eha}	\\
\text{$\gamma \gamma $ (VBF; tight)}	&	ATLAS	&	7+8	&full	&	$0.68^{+0.67}_{-0.51}$	&	15	&	85	&	0	&	\cite{Aad:2014eha}	\\
\text{$\gamma \gamma $ (VH; di-jet)}	&	ATLAS	&	7+8	&full	&	$0.23^{+1.67}_{-1.39}$	&	39	&	61	&	0	&	\cite{Aad:2014eha}	\\
\text{$\gamma \gamma $ (VH; MET)}	&	ATLAS	&	7+8	&full	&	$3.51^{+3.31}_{-2.42}$	&	7	&	86	&	7	&	\cite{Aad:2014eha}	\\
\text{$\gamma \gamma $ (VH; 1$l$)}	&	ATLAS	&	7+8	&full	&	$0.41^{+1.43}_{-1.06}$	&	0	&	98	&	2	&	\cite{Aad:2014eha}	\\
\text{$\gamma \gamma $ (ttH; hadronic)}	&	ATLAS	&	7+8	&full	&	$-0.84^{+3.23}_{-1.25}$	&	12	&	4	&	84	&	\cite{Aad:2014eha}	\\
\text{$\gamma \gamma $ (ttH; leptonic)}	&	ATLAS	&	7+8	&full	&	$2.42^{+3.21}_{-2.07}$	&	7	&	19	&	74	&	\cite{Aad:2014eha}	\\
\text{$WW$ ($e \mu $,$l_2=\mu $,$n_j=0$)}	&	ATLAS	&	7+8	&full	&	$1.08^{+0.41}_{-0.36}$	&	98	&	2	&	0	&	\cite{ATLAS:2014aga}	\\
\text{$WW$ ($e \mu $,$l_2=e$,$n_j=0$)}	&	ATLAS	&	7+8	&full	&	$1.40^{+0.49}_{-0.44}$	&	98	&	2	&	0	&	\cite{ATLAS:2014aga}	\\
\text{$WW$ ($e e \mu \mu $,$n_j=0$)}	&	ATLAS	&	7+8	&full	&	$0.47^{+0.74}_{-0.70}$	&	97	&	3	&	0	&	\cite{ATLAS:2014aga}	\\
\text{$WW$ ($e \mu $,$n_j=1$)}	&	ATLAS	&	7+8	&full	&	$1.16^{+0.51}_{-0.42}$	&	85	&	15	&	0	&	\cite{ATLAS:2014aga}	\\
\text{$WW$ ($e e \mu \mu $,$n_j$=1)}	&	ATLAS	&	7+8	&full	&	$0.19^{+1.12}_{-0.98}$	&	85	&	15	&	0	&	\cite{ATLAS:2014aga}	\\
\text{$WW$ (ggF; $e \mu $,$n_j\ge2$)}	&	ATLAS	&	7+8	&full	&	$1.20^{+0.96}_{-0.83}$	&	74	&	26	&	0	&	\cite{ATLAS:2014aga}	\\
\text{$WW$ (VBF; $e \mu $,$n_j\ge2$)}	&	ATLAS	&	7+8	&full	&	$0.98^{+0.48}_{-0.40}$	&	27	&	73	&	0	&	\cite{ATLAS:2014aga}	\\
\text{$WW$(VBF; $e e \mu \mu $,$n_j\ge2$)}	&	ATLAS	&	7+8	&full	&	$1.98^{+0.84}_{-0.67}$	&	25	&	75	&	0	&	\cite{ATLAS:2014aga}	\\

\text{$WW$(VH; 4$l$,2SFOS)}	&	ATLAS	&	7+8	&full	&	$-5.9^{+6.8}_{-4.1}$	&	0	&	100	&	0	&	\cite{Aad:2015ona}	\\
\text{$WW$(VH; 4$l$,1SFOS)}	&	ATLAS	&	7+8	&full	&	$9.6^{+8.1}_{-5.4}$	&	0	&	100	&	0	&	\cite{Aad:2015ona}	\\
\text{$WW$(VH; 3$l$,1SFOS/3SF)}	&	ATLAS	&	7+8	&full	&	$-2.9^{+2.7}_{-2.1}$	&	6	&	94	&	0	&	\cite{Aad:2015ona}	\\
\text{$WW$(VH; 3$l$,0SFOS)}	&	ATLAS	&	7+8	&full	&	$1.7^{+1.9}_{-1.4}$	&	0	&	100	&	0	&	\cite{Aad:2015ona}	\\
\text{$WW$(VH; 2$l$,DFOS) }	&	ATLAS	&	7+8	&full	&	$2.2^{+2.0}_{-1.9}$	&	51	&	49	&	0	&	\cite{Aad:2015ona}	\\
\text{$WW$(VH; 2$l$,SS2jet)}	&	ATLAS	&	7+8	&full	&	$7.6^{+6.0}_{-5.4}$	&	0	&	100	&	0	&	\cite{Aad:2015ona}	\\
\text{$WW$(VH; 2$l$,SS1jet)}	&	ATLAS	&	7+8	&full	&	$8.4^{+4.3}_{-3.8}$	&	0	&	100	&	0	&	\cite{Aad:2015ona}	\\

\text{$ZZ$ (ggF+ttH)}	&	ATLAS	&	7+8	&full	&	$1.66^{+0.51}_{-0.44}$	&	89	&	10	&	0	&	\cite{Aad:2014eva}	\\
\text{$ZZ$ (VBF+VH)}	&	ATLAS	&	7+8	&full	&	$0.26^{+1.64}_{-0.94}$	&	32	&	68	&	0	&	\cite{Aad:2014eva}	\\
\text{$bb$ (VH; 0$l$)}	&	ATLAS	&	8	&full	&	$-0.35^{+0.55}_{-0.52}$	&	0	&	100	&	0	&	\cite{Aad:2014xzb}	\\
\text{$bb$ (VH; 1$l$)}	&	ATLAS	&	8	&full	&	$1.17^{+0.66}_{-0.60}$	&	0	&	100	&	0	&	\cite{Aad:2014xzb}	\\
\text{$bb$ (VH; 2$l$)}	&	ATLAS	&	8	&full	&	$0.94^{+0.88}_{-0.79}$	&	0	&	100	&	0	&	\cite{Aad:2014xzb}	\\
\text{$bb$ (VBF; 2$j$)}	&	ATLAS	&	8	&full	&	$-0.80^{+2.30}_{-2.30}$	&	28	&	72	&	0	&	\cite{Aaboud:2016cns}	\\
\text{$bb$ (ttH; 2$l$)}	&	ATLAS	&	8	&full	&	$2.80^{+2.00}_{-2.00}$	&	0	&	0	&	100	&	\cite{Aad:2015gra}	\\
\text{$bb$ (ttH; lepton+jets)}	&	ATLAS	&	8	&full	&	$1.20^{+1.30}_{-1.30}$	&	0	&	0	&	100	&	\cite{Aad:2015gra}	\\
\text{$bb$ (ttH; hadronic)}	&	ATLAS	&	8	&full	&	$1.60^{+2.60}_{-2.60}$	&	0	&	0	&	100	&	\cite{Aad:2016zqi}	\\
$\tau _l\tau _l$ \text{(VBF)}	&	ATLAS	&	7+8	&full	&	$1.70^{+1.00}_{-0.90}$	&	42	&	58	&	0	&	\cite{Aad:2015vsa}	\\
$\tau _l\tau _l$ \text{(boosted)}	&	ATLAS	&	7+8	&full	&	$3.00^{+2.00}_{-1.70}$	&	65	&	35	&	0	&	\cite{Aad:2015vsa}	\\
$\tau _l\tau _h$ \text{(VBF)}	&	ATLAS	&	7+8	&full	&	$1.00^{+0.60}_{-0.50}$	&	34	&	66	&	0	&	\cite{Aad:2015vsa}	\\
$\tau _l\tau _h$ \text{(boosted)}	&	ATLAS	&	7+8	&full	&	$0.90^{+1.00}_{-0.90}$	&	67	&	33	&	0	&	\cite{Aad:2015vsa}	\\
$\tau _h\tau _h$ \text{(VBF)}	&	ATLAS	&	7+8	&full	&	$1.40^{+0.90}_{-0.70}$	&	39	&	61	&	0	&	\cite{Aad:2015vsa}	\\
$\tau _h\tau _h$ \text{(boosted)}	&	ATLAS	&	7+8	&full	&	$3.60^{+2.00}_{-1.60}$	&	62	&	38	&	0	&	\cite{Aad:2015vsa}	\\
\text{$\tau _h\tau _h$ (ZH)}	&	ATLAS	&	8	&full	&	$4.60^{+3.20}_{-3.20}$	&	0	&	100	&	0	&	\cite{Aad:2015zrx}	\\
\text{$\tau _l\tau _h$ (ZH)}	&	ATLAS	&	8	&full	&	$1.00^{+3.50}_{-3.50}$	&	0	&	100	&	0	&	\cite{Aad:2015zrx}	\\
\text{$\tau _h\tau _h$ (WH)}	&	ATLAS	&	8	&full	&	$1.80^{+3.10}_{-3.10}$	&	0	&	100	&	0	&	\cite{Aad:2015zrx}	\\
\text{$\tau _l\tau _h$ (WH)}	&	ATLAS	&	8	&full	&	$1.30^{+2.80}_{-2.80}$	&	0	&	100	&	0	&	\cite{Aad:2015zrx}	\\

\text{$\gamma \gamma $ (ggF)}	&	CMS	&	13	& 35.9 &	$1.11^{+0.19}_{-1.18}$	&	100	&	0	&	0	&	\cite{CMS:2017rli}	\\

\text{$\gamma \gamma $ (VBF)}	&	CMS	&	13	& 35.9 &	$0.5^{+0.6}_{-0.5}$	&	0	&	100	&	0	&	\cite{CMS:2017rli}	\\

\text{$\gamma \gamma $ (ttH)}	&	CMS	&	13	& 35.9 &	$2.2^{+0.9}_{-0.8}$	&	0	&	0	&	100	&	\cite{CMS:2017rli}	\\

\text{$\gamma \gamma $ (VH)}	&	CMS	&	13	& 35.9 &	$2.3^{+1.1}_{-1.0}$	&	0	&	100	&	0	&	\cite{CMS:2017rli}	\\

\text{$WW$ (0$j$)}	&	CMS	&	13	&15.2 & $0.9^{+0.4}_{-0.3}$	&	97	&	3	&	0	&	\cite{CMS-PAS-HIG-16-021}	\\
\text{$WW$ (1$j$)}	&	CMS	&	13	&15.2 &$1.1^{+0.4}_{-0.4}$	&	85	&	15	&	0	&	\cite{CMS-PAS-HIG-16-021}	\\
\text{$WW$ (2$j$)}	&	CMS	&	13	&15.2 &$1.3^{+1.0}_{-1.0}$	&	74	&	26	&	0	&	\cite{CMS-PAS-HIG-16-021}	\\

\text{$WW$ (VBF; 2$j$)}	&	CMS	&	13	&15.2 & $1.4^{+0.8}_{-0.8}$	&	38	&	62	&	0	&	\cite{CMS-PAS-HIG-16-021}	\\
\text{$WW$ (VH; 2$j$)}	&	CMS	&	13	&15.2 &$2.1^{+2.3}_{-2.2}$	&	54	&	46	&	0	&	\cite{CMS-PAS-HIG-16-021}	\\
\text{$WW$ (WH; 3$l$)}	&	CMS	&	13	&15.2 &$-1.4^{+1.5}_{-1.5}$	&	4	&	96	&	0	&	\cite{CMS-PAS-HIG-16-021}	\\

\text{$ZZ$ (untagged)}	&	CMS	&	13	&35.9 &	$1.17^{+0.23}_{-0.21}$	&	95	&	5	&	0	&	\cite{CMS:2017jkd}	\\
\text{$ZZ$ (VBF; $1j$)}	&	CMS	&	13	&35.9 &	$0.97^{+0.40}_{-0.32}$	&	86	&	14	&	0	&	\cite{CMS:2017jkd}	\\
\text{$ZZ$ (VBF; $2j$)}	&	CMS	&	13	&35.9 &	$0.63^{+0.51}_{-0.34}$	&	46	&	54	&	0	&	\cite{CMS:2017jkd}	\\
\text{$ZZ$ (VH; hadronic)}	&	CMS	&	13	&35.9 &	$0.76^{+0.78}_{-0.48}$	&	68	&	30	&	2	&	\cite{CMS:2017jkd}	\\
\text{$bb$ (ttH; lepton+jets)}	&	CMS	&	13	&12.9	&	$-0.43^{+1.02}_{-1.02}$	&	0	&	0	&	100	&	\cite{CMS:2016zbb}	\\
\text{$bb$ (ttH; $2l$)}	&	CMS	&	13	&12.9&	$-0.04^{+1.50}_{-1.39}$	&	0	&	0	&	100	&	\cite{CMS:2016zbb}	\\

\text{$bb$ (VBF)}	&	CMS	&	13	&2.3	&	$-3.70^{+2.40}_{-2.50}$	&	28	&	72	&	0	&	\cite{CMS:2016mmc}	\\

\text{$bb$ (boosted)}	&	CMS	&	13	&35.9 &	$2.30^{+1.80}_{-1.60}$	&	54	&	31	&	15	&	\cite{CMS-PAS-HIG-17-010}	\\

\text{$\tau \tau$ ($e \mu$)}	&	CMS	&	13	&35.9 &	$0.66^{+0.61}_{-0.59}$	&	17	&	83	&	0	&	\cite{CMS:2017wyg}	\\
\text{$\tau \tau$ ($e \tau_h$)}	&	CMS	&	13	&35.9 &	$0.56^{+0.58}_{-0.56}$	&	48	&	52	&	0	&	\cite{CMS:2017wyg}	\\
\text{$\tau \tau$ ($\mu \tau_h$)}	&	CMS	&	13	&35.9 &	$1.09^{+0.41}_{-0.41}$	&	32	&	68	&	0	&	\cite{CMS:2017wyg}	\\
\text{$\tau \tau$ ($\tau_h \tau_h$)}	&	CMS	&	13	&35.9 &	$1.30^{+0.37}_{-0.33}$	&	59	&	41	&	0	&	\cite{CMS:2017wyg}	\\

\text{$\gamma \gamma $ (ggF)}	&	ATLAS	&	13	&36.1&	$0.8^{+0.19}_{-0.18}$	&	93	&	7	&	0	&	\cite{ATLAS:2017gamgam}	\\

\text{$\gamma \gamma $ (VBF)}	&	ATLAS	&	13	&36.1&	$2.1^{+0.60}_{-0.60}$	&	42	&	58	&	0	&	\cite{ATLAS:2017gamgam}	\\

\text{$\gamma \gamma $ (VH)}	&	ATLAS	&	13	&36.1&	$0.7^{+0.9}_{-0.8}$	&	53	&	44	&	3	&	\cite{ATLAS:2017gamgam}	\\

\text{$\gamma \gamma $ (ttH)}	&	ATLAS	&	13	&36.1&	$0.5^{+0.6}_{-0.6}$	&	11	&	7	&	82	&	\cite{ATLAS:2017gamgam}	\\

\text{$WW$ (VBF; $e\mu $,$n_j\geq 2$)}	&	ATLAS	&	13	&5.8	&	$1.70^{+1.10}_{-0.90}$	&	38	&	62	&	0	&	\cite{ATLAS:2016gld}	\\
\text{$WW$ (WH; 3$l$+MET)}	&	ATLAS	&	13	&5.8	&	$3.20^{+4.40}_{-4.20}$	&	4	&	96	&	0	&	\cite{ATLAS:2016gld}	\\

\text{$ZZ$ (ggF; 0$j$)}&	ATLAS	&	13	&36.1&	$1.22^{+0.34}_{-0.29}$	&	98	&	2	&	0	&	\cite{ATLAS:2017ZZ}	\\
\text{$ZZ$  (ggF;1$j$;$p_T^H$ low) }&	ATLAS	&	13	&36.1&	$0.5^{+0.85}_{-0.76}$	&	92	&	8	&	0	&	\cite{ATLAS:2017ZZ}	\\

\text{$ZZ$ (ggF;1$j$;$p_T^H$ med)}&	ATLAS	&	13	&36.1&	$1.3^{+0.98}_{-0.73}$	&	84	&	16	&	0	&	\cite{ATLAS:2017ZZ}	\\
\text{$ZZ$ (ggF;1$j$;$p_T^H$ high) }&	ATLAS	&	13	&36.1&	$1.3^{+2.4}_{-1.7}$	&	75	&	25	&	0	&	\cite{ATLAS:2017ZZ}	\\

\text{$ZZ$  (ggF;2$j$)}&	ATLAS	&	13	&36.1&	$1.5^{+1.4}_{-1.0}$	&	69.5	&	29.5	&	1	&	\cite{ATLAS:2017ZZ}	\\
\text{$ZZ$  (VBF;$p_T^j$ low)  }&	ATLAS	&	13	&36.1&	$2.9^{+2.0}_{-1.6}$	& 62	&	37	&	1	&	\cite{ATLAS:2017ZZ}	\\

\text{$ZZ$ (VBF;$p_T^j$ high)}&	ATLAS	&	13	&36.1&	$13^{+12.0}_{-8.0}$	&	57	&	40	&	3	&	\cite{ATLAS:2017ZZ}	\\

\text{$bb$ (VH; $2l$)}	&	ATLAS	&	13	&36.1&$1.9^{+0.78}_{-0.64}$	&	0	&	100	&	0	&	\cite{ATLAS:2017bb}	\\
\text{$bb$ (VH; $1l$)}	&	ATLAS	&	13	&36.1&	$1.43^{+0.69}_{-0.59}$	&	0	&	100	&	0	&	\cite{ATLAS:2017bb}	\\
\text{$bb$ (VH; $0l$)}	&	ATLAS	&	13	&36.1&	$0.45^{+0.53}_{-0.51}$	&	0	&	100	&	0	&	\cite{ATLAS:2017bb}	\\

\text{$bb$ (VBF; 2$j$+$\gamma$)}	&	ATLAS	&	13	&12.6	&	$-3.90^{+2.80}_{-2.70}$	&	0	&	100	&	0	&	\cite{ATLAS:2016lgh}	\\
\text{$bb$ (ttH; $2l$)}	&	ATLAS	&	13	&13.2	&	$4.60^{+2.90}_{-2.30}$	&	0	&	0	&	100	&	\cite{ATLAS:2016awy}	\\
\text{$bb$ (ttH; $1l$)}	&	ATLAS	&	13	&13.2	&	$1.60^{+1.10}_{-1.10}$	&	0	&	0	&	100	&	\cite{ATLAS:2016awy}	\\
\hline
$\gamma \gamma$	&	Tevatron	&	1.96	&full	&	$5.79^{+3.39}_{-3.12}$	&	78	&	22	&	0	&	\cite{Aaltonen:2013ioz}	\\
\text{$WW$}	&	Tevatron	&	1.96	&full	&	$0.94^{+0.85}_{-0.83}$	&	78	&	22	&	0	&	\cite{Aaltonen:2013ioz}	\\
\text{$bb$}	&	Tevatron	&	1.96	&full	&	$1.59^{+0.69}_{-0.72}$	&	0	&	100	&	0	&	\cite{Aaltonen:2013ioz}	\\
$\tau \tau$	&	Tevatron	&	1.96	&full	&	$1.68^{+2.28}_{-1.68}$	&	50	&	50	&	0	&	\cite{Aaltonen:2013ioz}	\\
\caption{Signal strength $\mu_{f}$ for current LHC and Tevatron data. The central value, sup-script and subscript for $\mu$ represents observed signal strength, $1\sigma$ upper error bar, and $1\sigma$ down error bar respectively.  $\xi_G$, $\xi_V$, and $\xi_t$ indicate for weights in the Higgs production for gluon fusion, vector boson fusion plus associated production, and associated production with tops, respectively. Official values for weights are used when available, otherwise estimates are made.}
   \label{tab:muone}
\end{longtable}

\begin{table}[htbp]
   \centering
   \begin{tabular}{@{} lccrrrrrc @{}} 
    \multicolumn{1}{l}{Channel} & \multicolumn{1}{l}{Analysis} &  \multicolumn{1}{l}{$\sqrt{s}$} &  \multicolumn{1}{c}{$L$} & \multicolumn{1}{c}{$\mu$} & \multicolumn{1}{c}{ $\zeta_{VV}$ } & \multicolumn{1}{c}{$\zeta_{bb}$} &  \multicolumn{1}{c}{$\zeta_{\tau\tau}$} &  \multicolumn{1}{c}{Ref.}  \\
    
      \multicolumn{1}{l}{} & \multicolumn{1}{l}{} &  \multicolumn{1}{l}{ (TeV)} &   \multicolumn{1}{l}{($\text{fb}^{-1}$)} & \multicolumn{1}{c}{} & \multicolumn{1}{c}{  (\%)} & \multicolumn{1}{c}{ (\%)} &  \multicolumn{1}{c}{ (\%)} &  \multicolumn{1}{c}{}  \\

\hline
\text{ttH ($4l$)}	&	CMS	&	8	&full&	$-4.70^{+5.00}_{-1.30}$	&	72	&	0	&	28	&	\cite{Khachatryan:2014qaa}	\\
\text{ttH ($3l$)}	&	CMS	&	8	&full&	$3.10^{+2.40}_{-2.00}$	&	77	&	0	&	23	&	\cite{Khachatryan:2014qaa}	\\
\text{ttH ($2l_{ss}$)}	&	CMS	&	8	&full&	$5.30^{+2.10}_{-1.80}$	&	77	&	0	&	23	&	\cite{Khachatryan:2014qaa}	\\
\text{ttH ($1l2\tau_h$)}	&	CMS	&	13	&35.9&	$-1.20^{+1.50}_{-1.47}$	&	3	&	0	&	97	&	\cite{CMS:2017lgc}	\\
\text{ttH ($2l_{ss}1\tau_h$)}	&	CMS	&	13	&35.9&	$0.86^{+0.79}_{-0.66}$	&	42	&	0	&	58	&	\cite{CMS:2017lgc}	\\
\text{ttH ($3l1\tau_h$)}	&	CMS	&	13	&35.9&	$1.22^{+1.34}_{-1.00}$	&	43	&	0	&	57	&	\cite{CMS:2017lgc}	\\
\text{ttH ($4l$)}	&	CMS	&	13	&35.9&	$0.90^{+2.30}_{-1.60}$	&	72	&	0	&	28	&	\cite{CMS-PAS-HIG-17-004}	\\
\text{ttH ($3l$)}	&	CMS	&	13	&	35.9&$1.00^{+0.80}_{-0.70}$	&	79	&	0	&	21	&	\cite{CMS-PAS-HIG-17-004}	\\
\text{ttH ($2l_{ss}$)}	&	CMS	&	13	&35.9&	$1.70^{+0.60}_{-0.50}$	&	80	&	0	&	20	&	\cite{CMS-PAS-HIG-17-004}	\\

\text{ttH ($4l$)}	&	ATLAS	&	8	&full&	$1.80^{+6.90}_{-6.90}$	&	82	&	4	&	14	&	\cite{ATLAS:2015aea}	\\
\text{ttH ($3l$)}	&	ATLAS	&	8	&full&	$2.80^{+2.20}_{-1.80}$	&	81	&	4	&	15	&	\cite{ATLAS:2015aea}	\\
\text{ttH ($2l1\tau_h$)}	&	ATLAS	&	8&full	&	$-0.90^{+3.10}_{-2.00}$	&	37	&	1	&	62	&	\cite{ATLAS:2015aea}	\\
\text{ttH ($2l0\tau_h$)}	&	ATLAS	&	8&full	&	$2.80^{+2.10}_{-1.90}$	&	83	&	2	&	15	&	\cite{ATLAS:2015aea}	\\
\text{ttH ($1l2\tau_h$)}	&	ATLAS	&	8&full	&	$-9.60^{+9.60}_{-9.70}$	&	4	&	3	&	93	&	\cite{ATLAS:2015aea}	\\
\text{ttH ($3l$)}	&	ATLAS	&	13	&13.2&	$0.50^{+1.70}_{-1.60}$	&	78	&	2	&	20	&	\cite{ATLAS:2016ldo}	\\
\text{ttH ($2l0 \tau_h$})	&	ATLAS	&	13	&13.2&	$4.00^{+2.10}_{-1.70}$	&	80	&	3	&	17	&	\cite{ATLAS:2016ldo}	\\
\text{ttH ($2l1\tau_h$})	&	ATLAS	&	13	&13.2&	$6.20^{+3.60}_{-2.70}$	&	48	&	1	&	51	&	\cite{ATLAS:2016ldo}	\\
   \end{tabular}
   \caption{Signal strength for $ttH$ multi-lepton searches $\mu_{f}$ for current LHC data. $\zeta_{VV}$, $\zeta_{bb}$, and $\zeta_{\tau\tau}$ indicate weights in the Higgs decays into $WW$ plus $ZZ$, $b\bar{b}$, and $\tau^+\tau^-$, respectively. Official values for weights are used when given. ``$2l_{ss}$" stands for two same-sign di-leptons.}
   \label{tab:mutwo}
\end{table}

\begin{table}[htbp]
   \centering
   \begin{tabular}{@{} lccrrrrrc @{}} 
 \multicolumn{1}{l}{Channel} & \multicolumn{1}{l}{Analysis} &  \multicolumn{1}{l}{$\sqrt{s}$ } & \multicolumn{1}{c}{$L$} & \multicolumn{1}{c}{$\mu$} & \multicolumn{1}{c}{ $\xi_G$} & \multicolumn{1}{c}{$\xi_V$} &  \multicolumn{1}{c}{$\xi_t$} &  \multicolumn{1}{c}{Ref.}  \\
  \multicolumn{1}{l}{} & \multicolumn{1}{l}{} &  \multicolumn{1}{l}{(TeV)} & \multicolumn{1}{c}{($\text{fb}^{-1}$)} & \multicolumn{1}{c}{} & \multicolumn{1}{c}{(\%)} & \multicolumn{1}{c}{ (\%)} &  \multicolumn{1}{c}{ (\%)} &  \multicolumn{1}{c}{}  \\

\hline
\multirow{ 2}{*}{\text{Invisible (VBF)}} 	&	\multirow{ 2}{*}{CMS} 	&	7+8	& full&	\multirow{ 2}{*}{$0.16^{+0.15}_{-0.15}$}	&	\multirow{ 2}{*}{9} 	&	\multirow{ 2}{*}{91} 	&	\multirow{ 2}{*}{0} 	&	\multirow{ 2}{*}{\cite{CMS:2016rfr}}\\
	&	&	+13	& 2.3&	&		&		&		&	\\

\multirow{ 2}{*}{\text{Invisible (VH)}}  	&	\multirow{ 2}{*}{CMS}	&	7+8	&full&	\multirow{ 2}{*}{$0.00^{+0.12}_{-0.12}$}	&	\multirow{ 2}{*}{0}	&	\multirow{ 2}{*}{100}	&	\multirow{ 2}{*}{0}	&	\multirow{ 2}{*}{\cite{CMS:2016rfr}}	\\	&	&	+13	&2.3&		&	&	&		&	\\

\multirow{ 2}{*}{\text{Invisible (ggF)}} 	&	\multirow{ 2}{*}{CMS}	&	7+8	&full&	\multirow{ 2}{*}{$0.00^{+0.27}_{-0.27}$}	&	\multirow{ 2}{*}{70}	&	\multirow{ 2}{*}{30}	&	\multirow{ 2}{*}{0}	&	\multirow{ 2}{*}{\cite{CMS:2016rfr}}	\\
	&	&	+13	&2.3&		&		&		&	&	\\

\text{Invisible (VH; mono-$j$)}	&	CMS	&	13	&35.9&	$0.00^{+0.38}_{-0.38}$	&	73	&	27	&	0	&	\cite{CMS:2017tbk}	\\
\text{Invisible (VH; mono-$V$)}	&	CMS	&	13	&35.9&	$0.00^{+0.25}_{-0.25}$	&	39	&	61	&	0	&	\cite{CMS:2017tbk}	\\
\text{Invisible (ZH)}	&	CMS	&	13	&35.9&	$0.00^{+0.20}_{-0.20}$	&	0	&	100	&	0	&	\cite{CMS:2017gbj}	\\
\text{Invisible (ZH)}	&	ATLAS	&	7+8	&full&	$0.00^{+0.38}_{-0.38}$	&	0	&	100	&	0	&	\cite{Chatrchyan:2014tja}	\\
\text{Invisible (VBF)}	&	ATLAS	&	8	&full&	$0.00^{+0.14}_{-0.14}$	&	6	&	94	&	0	&	\cite{Aad:2015txa}	\\
\text{Invisible (VH)}	&	ATLAS	&	8	&full&	$0.00^{+0.40}_{-0.40}$	&	39	&	61	&	0	&	\cite{Aad:2015uga}	\\
\text{Invisible (1$j$+MET)}	&	ATLAS	&	8	&full&	$0.00^{+0.81}_{-0.81}$	&	52	&	48	&	0	&	\cite{Aad:2015zva}	\\
\text{Invisible (ZH; $ee$)}	&	ATLAS	&	13	&36.1&	$0.00^{+0.31}_{-0.31}$	&	0	&	100	&	0	&	\cite{ATLAS:2017inv}	\\
\text{Invisible (ZH; $\mu\mu$)}	&	ATLAS	&	13	&36.1&	$0.00^{+0.50}_{-0.50}$	&	0	&	100	&	0	&	\cite{ATLAS:2017inv}	\\
   \end{tabular}
   \caption{Signal strengths for Higgs invisible searches $\mu_{\text{inv}}$ for current LHC data. Official signal strengths and $1\sigma$ error bars are used if the likelihood curve is provided. Otherwise we assume the observed signal strength is 0 and translate the $95\%$ upper limits on signal strength, $\sigma^{95\%}_\text{inv}$, into $1\sigma$ error bars ($\sigma^\text{up/down}_\text{inv}=\sigma^{95\%}_\text{inv}/\sqrt{3.84}$). 
   Official values for weights are used when given.}
   \label{tab:muthree}
\end{table}

\begin{table}[htbp]
   \centering
     \begin{tabular}{clrrrrc}
    \multicolumn{1}{c}{$L$}  & \multicolumn{1}{l}{Channel} & \multicolumn{1}{c}{$\mu$} & \multicolumn{1}{c}{ $\xi_G$ (\%)} & \multicolumn{1}{c}{$\xi_V$ (\%)} &  \multicolumn{1}{c}{$\xi_t$ (\%)} &  \multicolumn{1}{c}{Ref.}  \\
\hline
   \multirow{ 16}{*}{300 fb$^{-1}$} &  \text{$\gamma \gamma $ ($0j$)} & $ 1.0^{+0.19}_{-0.19} $ & 92 & 8 & 0  &  \cite{atlas2014:pro}\\
   & \text{$\gamma \gamma $ ($1j$) } & $ 1.0^{+0.27}_{-0.27} $ & 82 & 18 & 0  &   \cite{atlas2014:pro}\\
& \text{$\gamma \gamma $ (VBF) } & $ 1.0^{+0.47}_{-0.47} $ & 39 & 61 & 0  &     \cite{atlas2014:pro}\\
& \text{$\gamma \gamma $ (WH-like) } & $ 1.0^{+0.48}_{-0.48} $ & 2 & 79 & 19  &  \cite{atlas2014:pro}\\
& \text{$\gamma \gamma $ (ZH-like)} & $ 1.0^{+0.85}_{-0.85} $ & 2 & 79 & 19  &  \cite{atlas2014:pro}\\
& \text{$\gamma \gamma $ (ttH-like) } & $ 1.0^{+0.38}_{-0.38} $ & 0 & 0 & 100  &  \cite{atlas2014:pro}\\
& \text{$WW$ ($0j$)} & $ 1.0^{+0.18}_{-0.18} $ & 98 & 2 & 0  &  \cite{atlas2014:pro}\\
& \text{$WW$ ($1j$)} & $ 1.0^{+0.30}_{-0.30} $ & 88 & 12 & 0  &  \cite{atlas2014:pro}\\
& \text{$WW$ (VBF-like) } & $ 1.0^{+0.21}_{-0.21} $ & 8 & 92 & 0  &  \cite{atlas2014:pro}\\
& \text{$ZZ$ (VH-like)} & $ 1.0^{+0.35}_{-0.35} $ & 30 & 56 & 14  &  \cite{atlas2014:pro}\\ 
& \text{$ZZ$ (ttH-like)} & $ 1.0^{+0.49}_{-0.49} $ & 9 & 6 & 85  &  \cite{atlas2014:pro}\\
& \text{$ZZ$ (VBF-like)} & $ 1.0^{+0.36}_{-0.36} $ & 45 & 54 & 1  &  \cite{atlas2014:pro}\\
& \text{$ZZ$ (ggF-like)} & $ 1.0^{+0.12}_{-0.12} $ & 89 & 10 & 1  &  \cite{atlas2014:pro}\\
& \text{$bb$ (WH)} & $ 1.0^{+0.57}_{-0.57} $ & 0 & 100 & 0  &  \cite{atlas2014:pro}\\
& \text{$bb$ (ZH)} & $ 1.0^{+0.29}_{-0.29} $ & 0 & 100 & 0  &  \cite{atlas2014:pro}\\
& \text{$\tau \tau $ (VBF-like)} & $ 1.0^{+0.21}_{-0.21} $ & 20 & 80 & 0  &  \cite{atlas2014:pro}\\
& Invisible (VBF) & $0.0^{+0.11}_{-0.11}$ & 0 & 100 & 0 & \cite{CMS-DP-2016-064}\\
 \hline
 
 \multirow{ 16}{*}{3 ab$^{-1}$} &
 \text{$\gamma \gamma $ ($0j$) } & $ 1.0^{+0.16}_{-0.16} $ & 92 & 8 & 0  &    \cite{atlas2014:pro}\\
& \text{$\gamma \gamma $ ($1j$) } & $ 1.0^{+0.23}_{-0.23} $ & 82 & 18 & 0  &    \cite{atlas2014:pro}\\ 
& \text{$\gamma \gamma $ (VBF) } & $ 1.0^{+0.22}_{-0.22} $ & 39 & 61 & 0  &   \cite{atlas2014:pro}\\  
& \text{$\gamma \gamma $ (WH-like) } & $ 1.0^{+0.19}_{-0.19} $ & 2 & 79 & 19  &   \cite{atlas2014:pro}\\  
& \text{$\gamma \gamma $ (ZH-like) } & $ 1.0^{+0.28}_{-0.28} $ & 2 & 79 & 19  &   \cite{atlas2014:pro}\\  
& \text{$\gamma \gamma $ (ttH-like)} & $ 1.0^{+0.17}_{-0.17} $ & 0 & 0 & 100  &   \cite{atlas2014:pro}\\ 
& \text{$WW$ ($0j$) } & $ 1.0^{+0.16}_{-0.16} $ & 98 & 2 & 0  &   \cite{atlas2014:pro}\\  
& \text{$WW$ ($1j$) } & $ 1.0^{+0.26}_{-0.26} $ & 88 & 12 & 0  &  \cite{atlas2014:pro}\\ 
& \text{$WW$ (VBF-like) } & $ 1.0^{+0.15}_{-0.15} $ & 8 & 92 & 0  &  \cite{atlas2014:pro}\\ 
& \text{$ZZ$ (VH-like) } & $ 1.0^{+0.13}_{-0.13} $ & 30 & 56 & 14  &   \cite{atlas2014:pro}\\  
& \text{$ZZ$ (ttH-like) } & $ 1.0^{+0.20}_{-0.20} $ & 9 & 6 & 85  &   \cite{atlas2014:pro}\\  
& \text{$ZZ$ (VBF-like) } & $ 1.0^{+0.21}_{-0.21} $ & 45 & 54 & 1  &   \cite{atlas2014:pro}\\ 
& \text{$ZZ$ (ggF-like) } & $ 1.0^{+0.11}_{-0.11} $ & 89 & 10 & 1  &   \cite{atlas2014:pro}\\ 
& \text{$bb$ (WH) } & $ 1.0^{+0.37}_{-0.37} $ & 0 & 100 & 0  &  \cite{atlas2014:pro}\\ 
& \text{$bb$ (ZH) } & $ 1.0^{+0.14}_{-0.14} $ & 0 & 100 & 0  &   \cite{atlas2014:pro}\\  
& \text{$\tau \tau $ (VBF-like) } & $ 1.0^{+0.19}_{-0.19} $ & 20 & 80 & 0  &   \cite{atlas2014:pro}\\ 
& Invisible (VBF) & $0.0^{+0.10}_{-0.10}$ & 0 & 100 & 0 & \cite{CMS-DP-2016-064}\\

\end{tabular}      
\caption{Projected ATLAS (CMS) Run 3 and Run 4 data used in fits from~\cite{atlas2014:pro}. Data with integrated luminosity 300 fb$^{-1}$ and 3 ab$^{-1}$ are listed in the upper and lower blocks respectively.  The center of  energy is assumed to be 14~TeV. The weights for production channels are taken from~\cite{Bechtle:2014ewa}. In our projections, we assume CMS can achieve  the same Higgs precision measurement as ATLAS. Effectively we double the data listed above for LHC Run 3 and Run 4 projections. 
}
   \label{tab:mufour}
\end{table}

\begin{table}[!htbp]
   \centering
   \begin{tabular}{@{} crrrr @{}} 
      
       & ILC & CEPC & FCC-ee & FCC-hh\\
      \hline
$\sigma_{\Gamma_h}$ & 1.8\% & 1.9\% & 1\% & -- \\
$\sigma_{r_b}$ & 0.7\%& 0.92\% & 0.42\% & -- \\
$\sigma_{r_c}$ & 1.2 \% & 1.2\%& 0.71\% & -- \\
$\sigma_{r_G}$ & 1\% & 1.1\%& 0.8\% & -- \\
$\sigma_{r_W}$ & 0.42\% & 0.87\%& 0.19\% & -- \\
$\sigma_{r_\tau}$ & 0.9\% & 1\%&0.54\% & -- \\
$\sigma_{r_Z}$ & 0.32\%& 0.18\%& 0.15\% &  -- \\
$\sigma_{r_\gamma}$ & 3.4\% & 3.3\%&1.5\% &  -- \\
$\sigma_{r_\mu}$ & 9.2\% & 6.1\% & 6.2\% & -- \\
$\sigma_{r_t}$ & 3\% & -- & 13\% & 1\%\\
$B_\text{inv}$ & $0.29\%$ & $0.2\%$ & $0.19\%$ & -- \\
   \end{tabular}
   \caption{Constraints on sensitivities for ILC (250 GeV, 2 ab$^{-1}$ $\oplus$ 350 GeV, 200 fb$^{-1}$ $\oplus$ 550 GeV, 4 ab$^{-1}$)~\cite{Fujii:2015jha}, CEPC (240 GeV, 10 ab$^{-1}$)~\cite{CEPC-SPPCStudyGroup:2015csa},  FCC-ee (240 GeV, 10 ab$^{-1}$ $\oplus$ 350 GeV, 2.6 ab$^{-1}$) from~\cite{Dawson:2013bba, FCChh:2017}, and FCC-hh (100 TeV, 30 ab$^{-1}$)~\cite{FCChh:2017}. Note that since most of sensitivities of FCC-hh are still under study (except for $r_t$)~\cite{FCChh:2017}, we use the corresponding values from FCC-ee for our projections.  $B_\text{inv}$ in the last row are the upper limits with 95\% CL.}
   \label{tab:future}
\end{table}

\clearpage

\bibliographystyle{JHEP} 
\bibliography{color.bib}

\end{document}